\documentclass[fleqn,10pt]{wlscirep}

\usepackage[utf8]{inputenc}
\usepackage[T1]{fontenc}

\usepackage{graphicx}
\usepackage{subcaption}

\usepackage{amsmath}
\usepackage{amssymb}
\usepackage{mathtools}

\usepackage{booktabs}
\usepackage{array}
\usepackage{multirow}
\usepackage{tabularx}

\usepackage{tikz}
\usetikzlibrary{
arrows.meta,
positioning,
calc,
shapes.geometric,
fit,
shadows,
matrix,
decorations.pathreplacing,
shadows.blur,
patterns
}

\usepackage{enumitem}

\usepackage[section]{placeins}

\usepackage[superscript,nomove]{cite}

\usepackage{microtype}

\usepackage{lineno}

\usepackage{hyperref}

\hypersetup{
colorlinks=true,
linkcolor=black,
citecolor=black,
urlcolor=black
}
\usepackage{xcolor}
\usepackage{soul}
\sethlcolor{yellow}
\renewcommand{\hl}[1]{#1}

\usepackage{caption}

\newcommand{\rv}[1]{\mathrm{#1}}
\makeatletter
\newcommand*{\rom}[1]{\expandafter\@slowromancap\romannumeral #1@}
\makeatother

\newcommand{\beginsupplement}{
\setcounter{section}{0}
\setcounter{figure}{0}
\setcounter{table}{0}
\setcounter{equation}{0}
\renewcommand{\thesection}{S\arabic{section}}
\renewcommand{\thefigure}{S\arabic{figure}}
\renewcommand{\thetable}{S\arabic{table}}
\renewcommand{\theequation}{S\arabic{equation}}
}

\newcommand{\supplementcaptions}{
\captionsetup[figure]{name={Supplementary Figure}}
\captionsetup[table]{name={Supplementary Table}}
}

\newcommand{\suptitle}[1]{
\begin{center}
{\Huge\bfseries #1\par}
\end{center}\vspace{0.5em}
}

\newcommand{\supauthors}[1]{
\begin{center}
#1
\end{center}\vspace{2em}
}

\title{DNA-MGC+: A versatile codec for reliable and resource-efficient data storage on synthetic DNA}

\author[1]{Ramy Khabbaz}
\author[1,2]{Jérémy Mateos}
\author[1,2]{Marc Antonini}
\author[1,*]{Serge Kas Hanna}
\affil[1]{Côte d’Azur University, CNRS, I3S, Sophia Antipolis, France}
\affil[2]{Pearcode, Sophia Antipolis, France}
\affil[*]{serge.kas-hanna@cnrs.fr}

\begin{abstract}
\hl{The biochemical processes underlying DNA data storage, including synthesis, amplification, and sequencing, are inherently noisy, leading to insertion, deletion, and substitution errors as well as sequence dropouts that challenge reliable data retrieval. Here we introduce DNA-MGC+, a DNA storage codec designed for reliable and resource-efficient retrieval under diverse operating conditions. We evaluate DNA-MGC+ across various \emph{in silico} and \emph{in vitro} settings. In comparison with selected representative codecs, the results show that DNA-MGC+ consistently improves key performance metrics under explicit reliability constraints, including required sequencing depth, read cost, decoding time, storage density, and error-correction capability. \emph{In silico} experiments show that DNA-MGC+ supports reliable decoding at error rates up to 24\% for a 15-KB file, while separate scalability experiments demonstrate reliable decoding of files up to 200~MB at a 5\% error rate. \emph{In vitro} experiments involving a 24-KB file and electrochemical synthesis show that different DNA-MGC+ configurations achieve minimum sequencing depths below~3\texttimes~and read costs below 3.5~nts/bit under both Illumina and Nanopore sequencing.}
\end{abstract}

\begin{document}
\flushbottom
\maketitle
\thispagestyle{empty}

\section*{Introduction}
The exponential growth of digital information has made conventional storage technologies increasingly unsustainable~\cite{rydning2022worldwide}, motivating research into alternative storage media, including the use of biological molecules. DNA has emerged as a particularly promising medium owing to its high density and durability~\cite{church2012, grass2015robust}. One gram of DNA can theoretically store exabytes of data, and DNA molecules can remain stable for centuries at room temperature under suitable conditions~\cite{church2012,grass2015robust}, enabling compact and sustainable long-term storage. The DNA storage pipeline consists of a writing process, a storage phase, and a reading process. During writing, binary information is digitally encoded into quaternary sequences over the four DNA nucleotides Adenine ($\mathsf{A}$), Guanine ($\mathsf{G}$), Cytosine ($\mathsf{C}$), and Thymine ($\mathsf{T}$), which are then synthesized as short DNA molecules known as oligonucleotides. The encoded sequences must be short to adhere to the length limitations imposed by current DNA synthesis technologies. During reading, the stored oligonucleotides are amplified, sequenced, and the resulting reads are digitally processed and decoded to retrieve the original information.

While several proof-of-concept experiments have already demonstrated the feasibility of this pipeline, scalability remains a central challenge. The main obstacles are related to reliability, speed, and cost. DNA synthesis, amplification, and sequencing are inherently noisy biochemical processes that introduce {\em errors} and {\em biases} throughout the pipeline~\cite{heckel2019characterization, gimpel2023digital}. Existing prototypes predominantly rely on high-fidelity synthesis and sequencing technologies, such as material deposition-based synthesis (Twist Bioscience) and Illumina sequencing, to mitigate these effects. Such technologies, however, are slow and expensive, limiting current systems to small-scale demonstrations that store megabytes of data rather than the exabytes envisioned. Faster and lower-cost alternatives, such as photolithographic synthesis, offer more scalable solutions but exhibit significantly higher error rates~\cite{antkowiak2020low, lietard2021chemical}. A key step toward scalability is therefore to address these errors and biases algorithmically rather than preventing them biochemically through high-fidelity technologies, thereby enabling the use of faster, cheaper, yet more error-prone technologies. Achieving this goal requires the development of efficient encoders and decoders (codecs) specifically adapted to the DNA storage channel, whose distinctive noise characteristics differ fundamentally from those encountered in conventional communication and storage systems~\cite{heckel2019characterization}.

Errors in DNA storage manifest as insertion, deletion, and substitution (IDS) errors at the base level, affecting individual nucleotides~\cite{heckel2019characterization}. In addition, biases introduced along the storage pipeline, such as those arising during PCR amplification, together with the stochastic nature of sequencing, can lead to sequence dropouts, i.e., the complete loss of encoded sequences~\cite{gimpel2023digital}. A natural algorithmic approach to mitigate both errors and dropouts is the use of error-correcting codes (ECCs)~\cite{milenkovic2024dna, sabary2024survey_coding}. ECCs introduce structured logical redundancy into the stored data to enable reliable information retrieval under noisy conditions. A common design strategy is to employ a two-layer coding architecture in which the inner code adds redundancy within each encoded sequence to correct base-level errors, while the outer code introduces redundant sequences to compensate for dropout events~\cite{shomorony2022information}.

ECCs provide several key benefits that enhance DNA storage systems. First, they improve data retrieval reliability by reducing the probability of decoding failures in the presence of errors and dropouts~\cite{hanna2025reliability}. Second, they reduce the number of sequencing reads required for successful decoding, which in turn decreases both the data retrieval time and the overall sequencing effort~\cite{erlich2017dna}. Third, in addition to decreasing sequencing depth, ECCs also allow reducing the number of physical molecular copies required for reliable data retrieval, thereby enabling higher storage densities~\cite{gimpel2025comparison}. Fourth, they enable the correction of synthesis errors that may appear systematically across reads and therefore cannot be mitigated simply by increasing sequencing depth~\cite{press2020hedges}.

Due to the established benefits of ECCs, the design of codecs for DNA storage with robust error-correction capabilities has been the subject of extensive research. The ECCs adopted in existing DNA storage codecs vary widely in scope. In particular, they differ in whether they are used for error detection or error correction, whether they address base-level errors and/or dropouts, and in the types of base-level errors they can handle. Existing designs include codecs that combine dropout recovery with error detection~\cite{goldman2013towards,grass2015robust,bornholt2016dna,blawat2016forward,erlich2017dna,organick2018random,appuswamy2019oligoarchive}, codecs that jointly address dropouts and substitution errors~\cite{chandak2019improved, antkowiak2020low, meiser2020reading, jpegdnaSPIE2025}, codecs whose ECCs support dropout recovery together with correction of all three types of IDS errors~\cite{press2020hedges, song2022robust, welzel2023dna, yan2025dna}, and codecs that focus on correcting IDS errors within individual sequences without providing explicit dropout recovery~\cite{zhang2026gungnir}. Most of these codecs also enforce content-specific constraints on the encoded DNA sequences, typically by limiting homopolymer length and balancing GC content, with some additionally avoiding specific sequence motifs~\cite{welzel2023dna}. Such constrained coding strategies follow the general principle of preventing or reducing errors and biases by avoiding sequence patterns believed to be problematic for the underlying biochemical processes. Some works focus exclusively on constrained coding and therefore design codecs without explicit ECCs~\cite{church2012,tabatabaei2015rewritable,yazdi2017portable,ping2022towards}, including approaches that additionally account for thermodynamic constraints~\cite{tabatabaei2015rewritable,ping2022towards}. A recent study suggests that addressing errors through ECCs is more efficient than attempting to avoid them through constrained coding alone~\cite{11164904}.

Beyond conventional coding approaches over the standard four-nucleotide alphabet, alternative encoding strategies based on degenerate bases and composite DNA letters have also been explored~\cite{anavy2019data,choi2019high,zhao2024composite}. Recent works have further investigated deep-learning-based reconstruction~\cite{bar2025scalable}, neural decoding of polar codes over synchronization-error channels~\cite{aharoni2025neural}, and joint probabilistic processing tailored to Nanopore reads~\cite{welter2026end}. These approaches can exploit complex empirical error patterns and soft information from multiple reads, potentially improving robustness and reducing coding redundancy or decoding complexity. However, they typically require representative training data, accurate channel models, or specialized computational procedures, which can limit their portability across experimental workflows and complicate direct system-level comparisons.

In this work, we focus on complete DNA storage codecs based on the standard four-nucleotide alphabet and introduce DNA-MGC+ (Marker Guess \& Check Plus). \hl{DNA-MGC+ provides efficient protection against both IDS errors and sequence dropouts, while maintaining low sequencing, decoding, and storage requirements.} Through a broad range of \emph{in silico} and \emph{in vitro} settings, including varying error and bias profiles as well as experimental results obtained with both Illumina and Nanopore sequencing, we show that DNA-MGC+ consistently outperforms other selected representative codecs. Specifically, DNA-MGC+ achieves simultaneous gains across several key performance metrics under explicit reliability constraints, including the minimum required sequencing depth, read cost, decoding time, maximal error-correction capability, and storage density.

\hl{\emph{In silico} simulations show that DNA-MGC+ supports reliable decoding at error rates up to 24\% for a 15-KB file and, in separate scalability experiments at a 5\% error rate, for files up to 200~MB. Moreover, when evaluated with experimentally derived error and bias profiles using the DT4DDS}~\cite{gimpel2023digital} \hl{simulator, DNA-MGC+ achieves a storage density of approximately 57~EB/g under low-fidelity conditions. We also validate DNA-MGC+ \emph{in vitro} using an experimental workflow that combines electrochemical DNA synthesis (GenScript) with both Illumina and Nanopore sequencing. Notably, this experimental setting is more error-prone than those used for codec evaluation in most prior studies, which typically rely on high-fidelity material-deposition-based synthesis (e.g., Twist Bioscience) and/or Illumina sequencing. For a 24-KB file, the \emph{in vitro} results show that different DNA-MGC+ configurations achieve minimum sequencing depths below~3\texttimes~and read costs below 3.5~nts/bit under both sequencing technologies. Collectively, these results highlight the reliability, resource efficiency, and versatility of DNA-MGC+, reflecting its ability to combine robust data retrieval with low read requirements across diverse operating conditions.}

\hl{Beyond these quantitative performance gains, our experimental study reveals additional findings of broader interest. First, we observe that comparable data retrieval performance can be achieved under both Illumina and Nanopore sequencing when using state-of-the-art Nanopore basecalling methods and robust codecs, despite the higher error rates associated with Nanopore. Second, we find that codecs equipped with strong error-correcting codes perform similarly with and without content-specific constraints imposed on encoded sequences. Finally, among the sequence-level characteristics examined in our experiments, we identify Gibbs free energy as particularly important for efficient data retrieval, given its stronger association with coverage variability than commonly constrained characteristics such as homopolymer length and GC content.}

\section*{Results}

\subsection*{Conceptual design of the DNA-MGC+ codec}
We outline below the main steps of the DNA-MGC+ codec, describing how a binary file is encoded into a collection of DNA sequences and subsequently decoded from sequencing data. This high-level overview is intended to convey the structure and main operating principles of the codec, while the full mathematical and algorithmic details are deferred to the Methods section.

\begin{enumerate}[leftmargin=*]
    \item {\em Fragmentation:} The input binary file is partitioned into non-overlapping short fragments. The fragment length is determined as a function of the desired target DNA sequence length and user-defined codec parameters.

    \item {\em Outer encoding:} The collection of binary fragments is encoded using a Reed–Solomon (RS) code~\cite{reed1960polynomial}, which introduces \emph{inter-sequence} redundancy by generating additional binary fragments. The RS code acts as an \emph{outer code}, enabling recovery from dropouts (lost sequences) as well as correction of residual errors from other decoding stages. 
    
    \item {\em Indexing and inner encoding:} A unique binary index is appended to each fragment, after which the indexed fragments are encoded using the Marker Guess \& Check Plus (MGC+) code~\cite{hanna2024GC, MGCP}. The MGC+ code acts as an {\em inner code} that introduces {\em intra-sequence} redundancy through structured encoding applied separately to each indexed fragment. This redundancy enables the correction of IDS errors within individual DNA sequences. The MGC+ encoding process includes an unconstrained binary-to-quaternary mapping ($00 \leftrightarrow \mathsf{A}$, $01 \leftrightarrow \mathsf{T}$, $10 \leftrightarrow \mathsf{C}$, $11 \leftrightarrow \mathsf{G}$) and therefore produces coded DNA sequences without content-specific constraints.

    \item {\em Filtering:} The codec optionally supports filtering of the resulting DNA sequences based on \emph{arbitrary} content-specific constraints. These may include, but are not limited to, homopolymer length, GC content, sequence motifs, or thermodynamic properties. This capability is enabled by the outer RS code, which allows the generation of an excess pool of candidate sequences from which any subset can be selected to represent the file through puncturing.
    
    \item {\em Decoding:} Decoding operates on noisy, unordered DNA sequences, which may correspond either to raw sequencing reads or to consensus sequences obtained after clustering. The sequences provided to the decoder are not required to match the reference sequence length, as insertions and deletions can be handled by the inner MGC+ code. Decoding proceeds by applying the inner MGC+ decoder, extracting fragment indices, decoding the outer RS code, and concatenating the recovered fragments to reconstruct the original file.
\end{enumerate}

The main motivation behind this design is to combine strong base-level IDS error-correction capability at the inner code with sequence-level erasure \emph{and} substitution correction at the outer code. The inner-code component is provided by MGC+, which offers more efficient correction of IDS errors than the inner codes typically used in existing DNA storage codecs, while also communicating decoding information to the outer code. The outer-code component is provided by the RS code, which can treat detected inner-decoding failures as sequence-level erasures (dropouts), thereby reducing the amount of redundancy required for recovery, while also correcting residual (undetected) substitution errors. Another distinguishing feature compared with existing designs is the ability to support arbitrary sequence constraints through filtering while retaining the erasure- and substitution- correction capability of RS codes. In contrast, previous filtering-based designs have typically relied on outer fountain codes, whose correction capability is largely limited to erasures. Finally, the decoding procedure is naturally parallelizable at both the inner and outer decoding stages, which improves scalability to larger files.

\subsection*{Overview of the evaluation framework}
We evaluate the performance of the proposed DNA-MGC+ codec in comparison with selected codecs under three settings: (1)~fully synthetic channel models, (2)~experimentally derived {\em in silico} error and bias profiles, and (3)~{\em in vitro} wet-lab experiments. In all settings, evaluation follows an end-to-end pipeline in which a binary input file is encoded into DNA sequences using the codec under consideration, processed through the associated channel model or experimental workflow, and decoded back into binary form.

In the first \emph{in silico} setting, we consider a randomly generated binary file of fixed size 15~KB and evaluate the performance of DNA-MGC+ and the comparison codecs under multiple synthetic channel scenarios that capture different error and bias regimes. The following six codecs were selected for comparison based on their prominence in the literature, the availability of wet-lab validation, and outcomes of previous benchmarking efforts~\cite{gimpel2025comparison}: DNA-Aeon~\cite{welzel2023dna}, HEDGES~\cite{press2020hedges}, DNA-Fountain~\cite{erlich2017dna}, DNA-RS~\cite{antkowiak2020low,meiser2020reading}, DNA-StairLoop~\cite{yan2025dna}, and an LDPC-based scheme~\cite{chandak2019improved}. Across all evaluated codecs, we consider a total of 16 configurations spanning three code-rate classes, with target rates of approximately 0.50~(low), 1.00~(medium), and 1.50~(high) bits per nucleotide (bits/nt). The design length of the encoded DNA sequences is chosen to be as close as possible to 150 nucleotides. We then increase the input file size beyond 15~KB to assess the system-level scalability of the best-performing codec configurations.

In the second \emph{in silico} setting, we again consider a randomly generated binary file of size 15~KB and compare DNA-MGC+ against the same six representative codecs. In this case, however, the evaluation uses experimentally derived error and bias profiles based on the DT4DDS~\cite{gimpel2023digital} digital twin for DNA storage. Finally, in the \emph{in vitro} setting, we evaluate DNA-MGC+ using wet-lab experiments performed on a file of size approximately 24~KB. This evaluation includes a comparison with the two best-performing codec configurations identified from the preceding \emph{in silico} evaluations. The corresponding experimental workflow and codec configurations are described in the dedicated sections.

Performance is assessed using three metrics, reported consistently across all evaluation settings. First, decoding reliability is measured through the minimum {\em coverage depth} required to retrieve the file with an exact match, subject to a reliability constraint that requires successful decoding in 50 out of 50 independent trials. Second, we report the {\em read cost}, defined as the ratio of the minimum required coverage depth to the code rate. Normalizing by the code rate penalizes lower-rate configurations and removes the bias that would otherwise arise when comparing only the coverage depth across codecs storing different amounts of information per nucleotide. This metric therefore enables a fair comparison between codecs operating at different code rates by quantifying the minimum number of nucleotides that must be read per stored information bit (nts/bit). Third, we report the average {\em decoding time} measured at the minimum reliable coverage depth. Decoding time excludes clustering, alignment, and consensus formation, which are handled separately to isolate the computational cost of the codec decoding algorithms themselves. All computations are performed on a machine equipped with an AMD Ryzen™ Threadripper™ PRO 7985WX CPU, 256 GB of available memory, and 64 CPU cores. The number of cores used is specified for each experiment, and a two-hour decoding time limit is imposed in all experiments. Unless otherwise stated, all results are obtained using a common preprocessing pipeline prior to decoding, in which noisy reads are clustered using CD-HIT~\cite{cd-hit}, followed by multiple sequence alignment using Kalign~\cite{kalign} and consensus calling.

\subsection*{In silico performance under synthetic error and bias models}
\begin{figure}[t!]
    \centering

    \begin{subfigure}{0.32\textwidth}
        \centering
        \includegraphics[width=\linewidth]{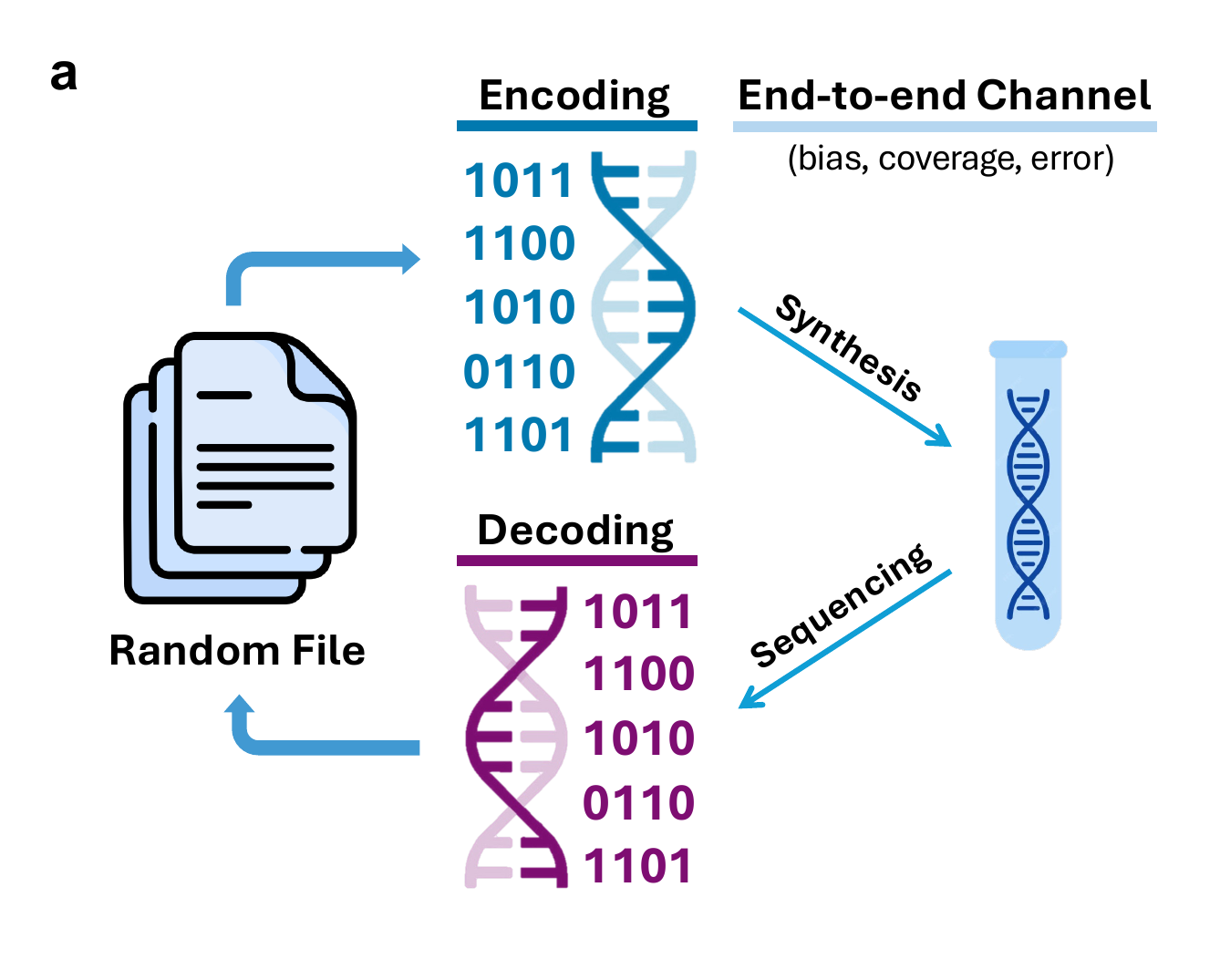}
    \end{subfigure}
    \hfill
    \begin{subfigure}{0.32\textwidth}
        \centering
        \includegraphics[width=\linewidth]{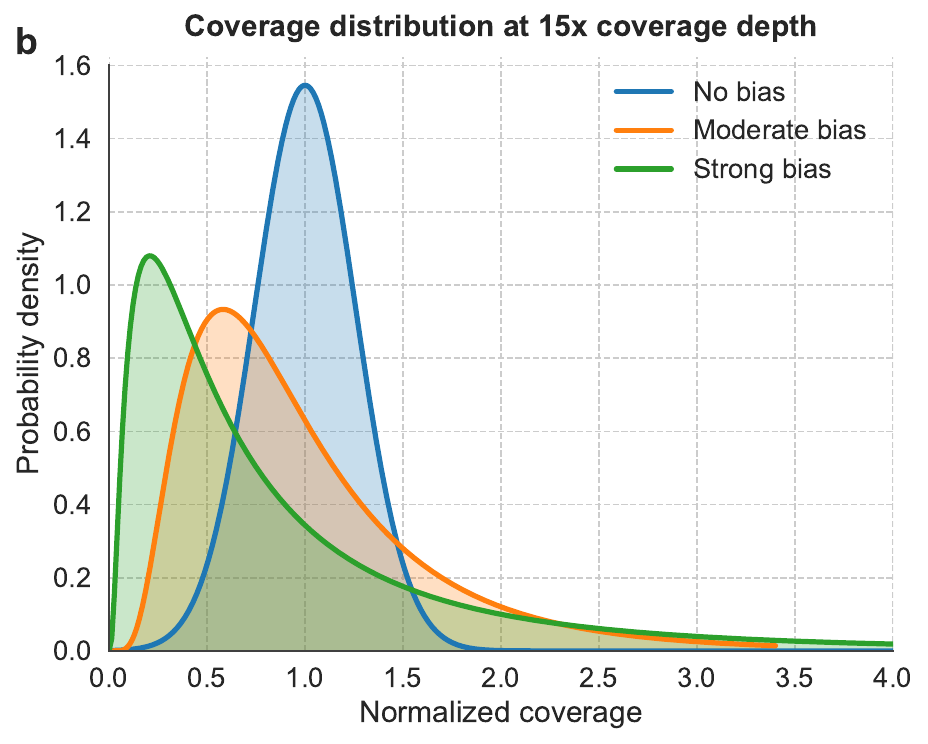}
    \end{subfigure}
    \hfill
    \begin{subfigure}{0.32\textwidth}
        \centering
        \includegraphics[width=\linewidth]{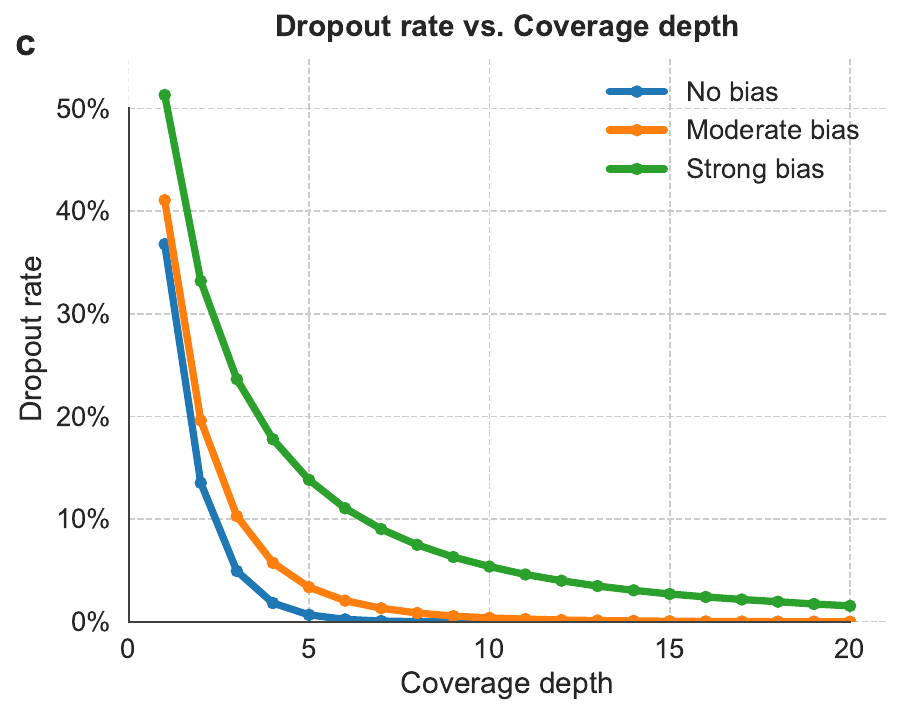}
    \end{subfigure}

    \vspace{0.2cm}

    \begin{subfigure}{\textwidth}
        \centering
        \includegraphics[width=\linewidth]{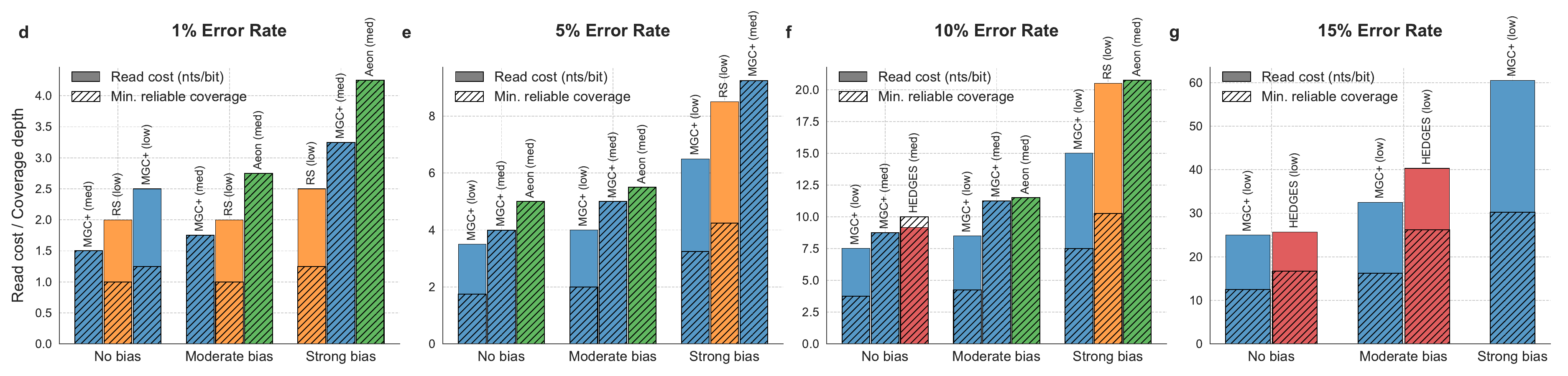}
    \end{subfigure}

    \vspace{0.4cm}

    \begin{subfigure}{0.32\textwidth}
        \centering
        \includegraphics[width=\linewidth]{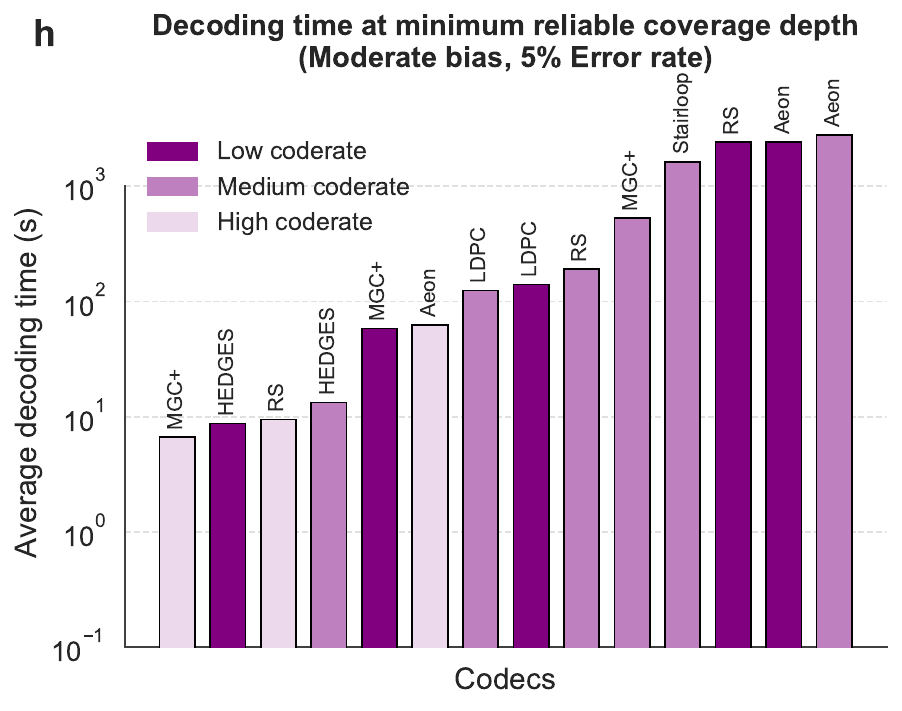}
    \end{subfigure}
    \hfill
    \begin{subfigure}{0.32\textwidth}
        \centering
        \includegraphics[width=\linewidth]{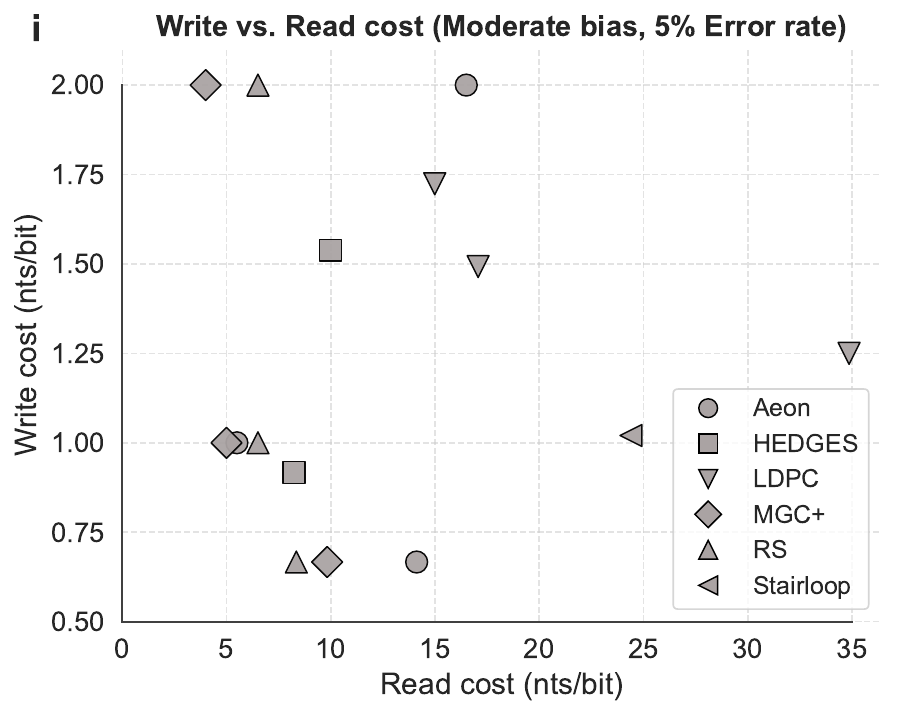}
    \end{subfigure}
    \hfill
    \begin{subfigure}{0.32\textwidth}
        \centering
        \includegraphics[width=\linewidth]{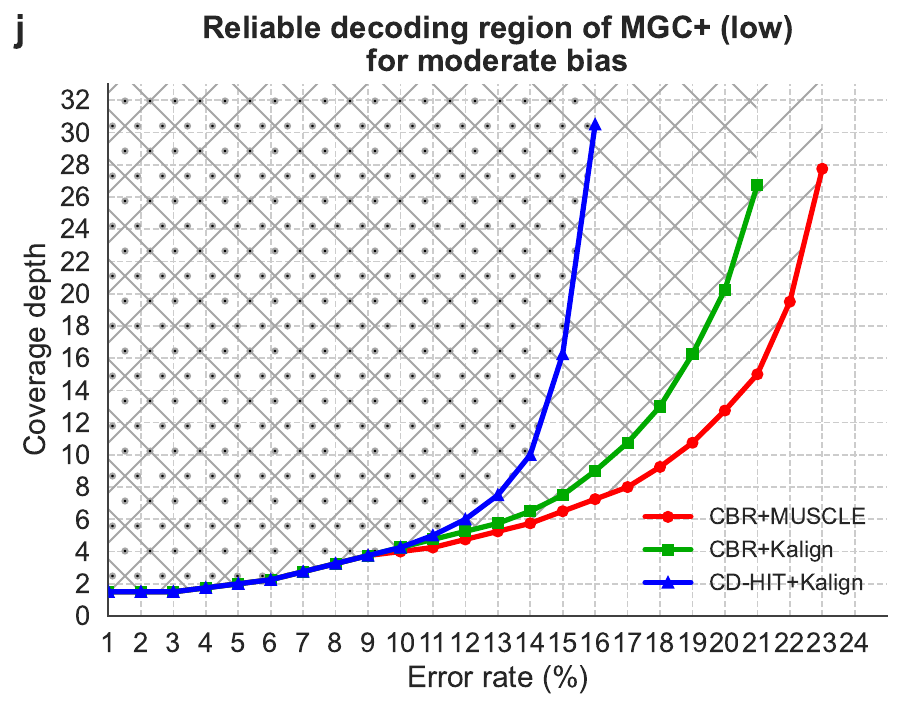}
    \end{subfigure}

\caption{\textbf{In silico codec performance for a file of size 15KB under synthetic error and bias models.} (\textbf{a})~Schematic of the evaluation pipeline, in which a randomly generated file is encoded into DNA sequences and decoded after processing through a synthetic channel modeling the end-to-end DNA storage process, parameterized by bias, coverage depth, and error rate. (\textbf{b})~Coverage distributions for the three considered bias regimes, based on a lognormal model with unit mean and standard deviations $\sigma = 0$ (no bias), $\sigma = 0.5$ (moderate bias), and $\sigma = 1$ (strong bias). Coverage distributions are normalized by their mean, which equals the coverage depth, set to 15\texttimes~in this plot. \textbf{(c)}~Dropout rates as a function of coverage depth across the three bias regimes, showing the expected fraction of reference sequences receiving zero reads. \textbf{(d)--(g)} Minimum coverage depth and associated read cost achieved by the three best-performing codecs across different bias--error combinations. \textbf{(h)}~Average decoding times for the moderate bias and 5\% error rate scenario, measured at the minimum coverage depth required for reliable decoding using a single CPU core. \textbf{(i)}~Trade-off between write cost and read cost for the moderate bias and 5\% error rate scenario. \textbf{(j)}~Reliable decoding region of the DNA-MGC+ codec (low-rate configuration) for the moderate-bias case, highlighting the error rate and coverage depth values for which reliable decoding is achievable under different clustering--alignment combinations.}
\label{fig:silico1}
\end{figure}

We start our performance analysis by evaluating the DNA-MGC+ codec in comparison with the other aforementioned codecs under a synthetic end-to-end channel model. This model captures key sources of variability in the DNA storage pipeline, including base-level errors, coverage bias across reference DNA sequences, and the stochastic nature of the sequencing process. The channel is parameterized by three independent quantities: a bias parameter, a coverage depth, and an error rate (Fig.~\ref{fig:silico1}a). Given a set of reference DNA sequences produced by the encoder, the channel generates a collection of noisy reads through three successive stages:
\begin{enumerate}[leftmargin=*]
    \item {\em Bias:}  Each reference sequence is assigned a multiplicative weight drawn from a lognormal distribution with unit mean and standard deviation governed by the channel bias parameter. These weights are normalized to form a probability vector that defines the sequence-dependent sampling probabilities to be used in the next stage.

    \item {\em Coverage:} Reads are generated by random sampling with replacement according to the probability vector obtained in the previous stage, where the total number of reads is equal to the number of reference sequences multiplied by the channel coverage depth parameter.
    
    \item {\em Error:} Base-level errors are introduced randomly and independently into each read according to the channel error rate, with substitutions, deletions, and insertions occurring in fixed proportions of 53\%, 45\%, and 2\% of the total error rate, respectively, consistent with ratios reported in the literature~\cite{gimpel2023digital}.
\end{enumerate}

The coverage depth parameter fixes the average number of reads per reference sequence, while the bias parameter $\sigma\geq0$ influences how unevenly these reads are distributed across sequences. The adopted bias model is motivated by experimental observations showing that the coverage distribution is positively skewed and is well-approximated by a lognormal distribution~\cite{gimpel2023digital, chen2020quantifying}.
In our simulations we consider three representative bias regimes: $\sigma = 0$ (no bias), $\sigma = 0.5$ (moderate bias), and $\sigma = 1$ (strong bias). When $\sigma = 0$, all sequences are sampled with equal probability, leading to a symmetric coverage distribution, whereas for $\sigma > 0$ the increasing standard deviation induces progressively skewed coverage distributions~(see Fig.~\ref{fig:silico1}b). An important consequence of skewed coverage is an increased occurrence of sequence dropouts, whereby some reference sequences receive zero reads at the channel output. The dropout rate grows with increasing bias strength~(see Fig.~\ref{fig:silico1}c). 

\subsubsection*{Performance across multiple channel scenarios}
We first consider a fixed file size of 15~KB and evaluate DNA-MGC+ and the comparison codecs across 12 different synthetic channel scenarios, obtained by combining the three bias regimes with four error rates (1\%, 5\%, 10\%, and 15\%). For each scenario, we determine the minimum coverage depth required for reliable decoding for all 16 configurations of the comparison codecs, in addition to three configurations of DNA-MGC+ corresponding to target code rates of 0.50, 1.00, and 1.50~bits/nt. This is done by performing a binary search over coverage depths in the range~[1, 32] with a resolution of 0.25, under the reliability constraint of 50 successful decodings (with an exact match) out of 50 trials. The corresponding read cost is then computed by normalizing the minimum coverage depth by the associated code rate. Figs.~\ref{fig:silico1}d–g report, for each of the 12 scenarios, the three best-performing codecs in terms of read cost, together with their corresponding coverage depths. 

The results in Figs.~\ref{fig:silico1}d--g show that the strongest performers across the synthetic channel conditions are DNA-MGC+, HEDGES, DNA-Aeon, and DNA-RS. In particular, DNA-MGC+ achieves the lowest read cost in 11 out of the 12 bias--error combinations, with the only exception occurring under strong bias at 1\% error rate, where the low-rate configuration of DNA-RS attains a lower read cost. Beyond read cost, DNA-MGC+ also achieves the lowest coverage depth among all codecs for error rates of 5\%, 10\%, and 15\% across all bias regimes, demonstrating that its reliability advantage persists as both dropouts and errors increase. Notably, under the highest error rate of 15\%, only 2 out of the 19 tested codec configurations meet the reliability constraint within the simulated coverage depth range of [1, 32]. These two cases correspond to the low-rate configurations of HEDGES and DNA-MGC+ (see Fig.~\ref{fig:silico1}g). Both succeed in the no-bias and moderate-bias regimes, while DNA-MGC+ is the only codec that achieves reliable decoding under strong bias at 15\% error rate.

In Figs.~\ref{fig:silico1}h--i, we highlight results for the moderate-bias, 5\% error-rate scenario in terms of average decoding time and the relation between write cost and read cost. The average decoding times reported in Fig.~\ref{fig:silico1}h are measured at the minimum coverage depth required by each codec for reliable decoding, with the corresponding decoding computations restricted to the use of a single CPU core for all codecs. The results show that DNA-MGC+ consistently ranks among the three fastest codecs across all three code-rate classes. Fig.~\ref{fig:silico1}i illustrates the trade-off between write cost and read cost for the same scenario, where the write cost is defined as the reciprocal of the code rate expressed in~nts/bit. Codecs with lower write cost (that is, higher code rate) generally require higher read cost to achieve reliable decoding, reflecting the classical trade-off between logical redundancy added during encoding and redundancy supplied through coverage depth. DNA-MGC+ achieves a favorable balance in this trade-off, as it lies near the lower envelope of the cloud of points, combining relatively low write cost with some of the lowest read costs among the tested codecs.

We further examined the robustness of the low-rate configuration of DNA-MGC+, which, as shown in Fig.~\ref{fig:silico1}g, supports reliable decoding at error rates up to 15\% even under strong bias. Building on this observation, we assessed how far the error rate can be increased while still satisfying the reliable decoding constraint within the coverage depth range of [1, 32]. Using the same preprocessing pipeline applied across all codecs (CD-HIT + Kalign), we found that reliable decoding remains achievable up to an error rate of 16\% under moderate bias. Fig.~\ref{fig:silico1}j shows the reliable decoding region for the moderate-bias regime. Similar figures for the no-bias and strong-bias regimes are provided in Supplementary Figs. S5 and S6, showing that the maximum achievable error rate is also 16\% in the no-bias case, and 15\% in the strong-bias case. Our analysis further reveals that at these high error rates, clustering becomes the limiting factor as most reads end up being unclustered. This in turn renders the subsequent alignment stage ineffective since the majority of clusters contain only a single read. Adjusting the parameters of CD-HIT did not help mitigate this effect.

To address the clustering bottleneck, we tested an alternative clustering algorithm introduced by Rashtchian et al.~\cite{rashtchian2017clustering}, referred to here as ``Clustering Billions of Reads'' (CBR), combined with the same alignment method Kalign. This approach extends the reliable decoding region to 21\% under moderate bias, as illustrated in Fig.~\ref{fig:silico1}j. For error rates of 10\% or less, CBR offers no measurable advantage over CD-HIT, and the improvement becomes apparent only as the error rate approaches the regime where CD-HIT no longer clusters effectively. We also tested pairing CBR with a more computationally intensive alignment algorithm, MUSCLE~\cite{muscle5}, which further extends the reliable decoding region to 23\% under moderate bias. Beyond this point, even CBR becomes ineffective as almost all reads remain unclustered. For the no-bias and strong-bias regimes, the results provided in Supplementary Figs. S5 and S6 indicate that by using CBR and MUSCLE, DNA-MGC+ achieves reliable decoding up to an error rate of 24\% in the no-bias case, and 21\% in the strong-bias case. \hl{More broadly, further extending the reliable decoding region in high-error-rate and large-scale settings while keeping the same code parameters may require alternative clustering strategies specifically designed for such scenarios, such as hash-sketch fuzzy clustering}~\cite{shao2025hsfc}\hl{, or index-based clustering combined with carefully designed index sets}.

Additional results obtained under the same synthetic channel model are provided in the Supplementary Information. These include the encoding time required by each codec for the 15-KB file (Supplementary Fig. S4), results similar to those in Fig.~\ref{fig:silico1} under equal insertion, deletion, and substitution proportions (Supplementary Figs. S1--S3), a comparison between DNA-MGC+ and the Gungnir codec~\cite{zhang2026gungnir} (Supplementary Note S4), and full numerical results for all 19 codec configurations across the 12 channel scenarios (Supplementary Tables S6--S8).

\subsubsection*{Scalability across file sizes}
While the preceding results evaluate codec performance for a fixed 15-KB file across different channel conditions, we next assess the system-level scalability of selected codecs by varying the input file size from a few kilobytes to hundreds of megabytes. For this experiment, we focus on the moderate-bias, 5\% error-rate scenario of the synthetic channel model and consider the four codecs that achieved the best performance in terms of read cost under this scenario, according to the results in Fig.~\ref{fig:silico1} and Supplementary Table~S7, namely DNA-MGC+ (low-rate), DNA-Aeon (medium-rate), DNA-RS (medium-rate), and HEDGES (medium-rate). The encoded sequence length is kept fixed at approximately 150~nts while the file size is varied. For DNA-MGC+, the redundancy allocation of the low-rate configuration is slightly adjusted for files larger than 50~KB by reducing the inner-code redundancy, allowing the encoded sequences to accommodate the additional packet index required for larger files while maintaining the same sequence length, as described in the Methods section and Supplementary Table~S1.

\begin{figure}[t!]
    \centering
    \includegraphics[width=\textwidth]{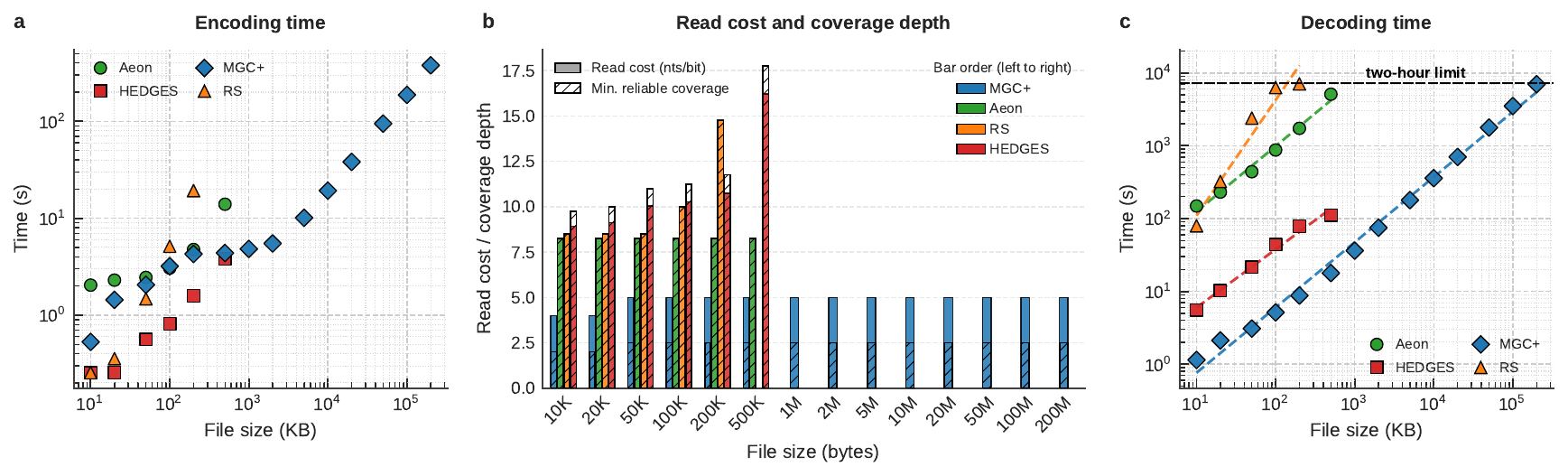} 
    \caption{\textbf{In silico codec performance across varying file sizes under the 5\% error and moderate-bias scenario of the synthetic channel model.} (\textbf{a})~Encoding time as a function of file size for DNA-Aeon (medium-rate), HEDGES (medium-rate), DNA-RS (medium-rate), and DNA-MGC+ (low-rate). (\textbf{b})~Minimum coverage depth required by each codec for reliable decoding and corresponding read cost (nts/bit) as a function of file size. (\textbf{c})~Average decoding time as a function of file size, measured at the minimum coverage depth required for reliable decoding.}
    \label{silico11}
\end{figure}

Since the objective of this experiment is to evaluate system-level scalability, we allow each codec to use the full 64 CPU cores whenever supported by its implementation, unlike the preceding experiments, in which computations were intentionally limited to a single CPU core. The two-hour decoding time limit per file is nevertheless maintained. To avoid making clustering the limiting factor in this larger-scale setting, all codecs are evaluated using the CBR clustering algorithm, which scales more favorably to large numbers of reads than CD-HIT, followed by Kalign alignment and consensus calling. For each file size and codec configuration, we report the encoding time (Fig.~\ref{silico11}a), the minimum coverage depth required for reliable decoding under the same 50-out-of-50 reliability criterion used throughout this work and the associated read cost (Fig.~\ref{silico11}b), and the average decoding time measured at the minimum coverage depth (Fig.~\ref{silico11}c). \hl{Rather than repeating the full binary search over [1,32] at every file size, we use the minimum coverage depth found for the smallest file size as a starting point and refine it locally (in steps of 0.25) for larger file sizes, applying the same 50-out-of-50 decoding criterion at each tested depth}. When a codec reaches an implementation-specific limitation or exceeds the two-hour decoding time limit, larger file sizes are not evaluated for that codec.

The results in Fig.~\ref{silico11} show that DNA-MGC+ scales more favorably to larger files than the comparison codecs while maintaining both reliable and efficient decoding. In particular, DNA-MGC+ successfully decodes files of up to 200~MB within the imposed two-hour decoding time limit while attaining a read cost below 5~nts/bit throughout the evaluated range. Fig.~\ref{silico11}b further shows that the minimum coverage depth required for reliable decoding and the corresponding read cost remain nearly constant for DNA-MGC+ as the file size increases \hl{from 10~KB to 200~MB}. The only exception is a small increase for file sizes above 50~KB, which is expected since the reduction in inner-code redundancy required to accommodate the additional packet index slightly reduces the overall error-correction capability. Nevertheless, DNA-MGC+ consistently achieves the lowest read cost among the evaluated codecs across all tested file sizes. Fig.~\ref{silico11}c further indicates that the decoding time of DNA-MGC+ grows nearly linearly with file size, with a slope (growth rate) of approximately one. Both the decoding time and its growth rate are lower than those of the comparison codecs. In terms of encoding time (Fig.~\ref{silico11}a), DNA-MGC+ exhibits performance comparable to DNA-Aeon and DNA-RS, depending on the file size, whereas HEDGES provides the fastest encoding. \hl{Additional results for DNA-MGC+ across varying file sizes and higher error rates are provided in Supplementary Note~S5}.

The remaining codecs exhibit more limited scalability. DNA-Aeon maintains constant coverage requirements and read cost across the evaluated file sizes but reaches the two-hour decoding time limit for file sizes beyond 500~KB. DNA-RS shows a progressive increase in read cost as the file size grows from 50~KB to 500~KB, indicating that additional reading effort is required to maintain reliable decoding for larger files, before also reaching the decoding time limit. HEDGES exhibits a similar increase in read cost as the file size increases from 10~KB to 500~KB, beyond which no further experiments could be performed \hl{due to a design} limitation rather than decoding complexity. \hl{Specifically, the HEDGES codec design uses fixed-size packet and sequence indices, imposing a hard limit of 256 packets and 256 sequences per packet, which is also hard-coded in the available implementation and is therefore insufficient for larger files}.

\subsection*{In silico performance under experimentally derived error and bias profiles}
Our second {\em in silico} performance evaluation follows a standardized workflow based on DT4DDS~\cite{gimpel2023digital}, a digital twin for DNA data storage built on real experimental datasets and designed to reproduce the behavior of state-of-the-art storage pipelines. In this study, we focus on the “low-fidelity” workflow defined in Ref.~\cite{gimpel2025comparison}, which models a pipeline consisting of electrochemical synthesis~(e.g., GenScript), amplification using an error-prone polymerase (e.g., Taq), and short paired-end Illumina sequencing. This scenario is characterized by higher error rates and more pronounced bias compared with high-fidelity pipelines based on material deposition synthesis (e.g., Twist) and low-error polymerases (e.g., Q5). As such, it provides a particularly relevant setting for assessing the ability of the codecs to compensate for the error and bias introduced by low-fidelity technologies.

Compared with the synthetic model used previously, this experimentally derived model imposes a more nuanced error profile. While the overall end-to-end error rate \hl{is fixed at approximately 1.6\%, as measured by aligning the final simulated reads to their corresponding encoded sequences using minimap2}~\cite{li2018minimap2}, errors exhibit correlation across neighboring bases, position-dependent behavior with increased rates near sequence extremities, and asymmetric substitution patterns derived from empirical measurements. In addition, each stage of the workflow is modeled separately, such that the resulting error composition reflects the successive biochemical processes \hl{and inherent processing steps of the workflow} rather than a simple end-to-end mixture. For example, under the considered workflow, deletions predominantly originate during synthesis, whereas substitutions mainly arise during sequencing. More specifically, the error profile includes a 1.56\% deletion rate during synthesis, a 0.011\% substitution rate during amplification, and substitution rates of 0.11\% and 0.25\% in the forward and reverse Illumina reads, respectively, \hl{with no insertion errors introduced by the workflow}. \hl{The paired-end reads are} subsequently merged using NGmerge~\cite{gaspar2018ngmerge}.

The experimentally derived workflow also provides a more refined treatment of coverage. Instead of a single coverage depth parameter, the model enables separate parametrization of {\em physical redundancy} and {\em sequencing depth}. Physical redundancy is defined as the average number of physical DNA molecules per reference sequence during storage, whereas sequencing depth corresponds to the average number of sequencing reads obtained per encoded sequence. This distinction is important because physical redundancy directly constrains the achievable storage density (SD), which can be approximated as $$\text{SD} \approx \frac{\text{code rate}}{\text{physical redundancy}}\times 113.7,$$ where the rate of the codec is expressed in bits/nt and the storage density in exabytes per gram of DNA (EB/g)~\cite{gimpel2025comparison}. A codec that maintains reliable decoding at low physical redundancy therefore unlocks higher storage densities, since fewer molecular copies are required for data retrieval.

Using the low-fidelity workflow, we evaluate the same 19 codec configurations examined under the synthetic model on a randomly generated 15-KB file, together with an additional DNA-MGC+ configuration of code rate 0.50~bits/nt whose parametrization is optimized for this specific workflow. In the synthetic model evaluation, DNA-MGC+ was assessed using fixed parametrizations across 12 different error and bias conditions. In general, it is possible to optimize the parametrization of DNA-MGC+ according to the characteristics of the model under consideration. This corresponds to a design tradeoff between the redundancy allocated to the inner and outer codes, which we elaborate on in the Discussion section. The corresponding parameter settings for all evaluated configurations are provided in the Supplementary Information.

\begin{figure}[t!]
    \centering

    \begin{subfigure}{0.33\textwidth}
        \centering
        \includegraphics[width=\linewidth]{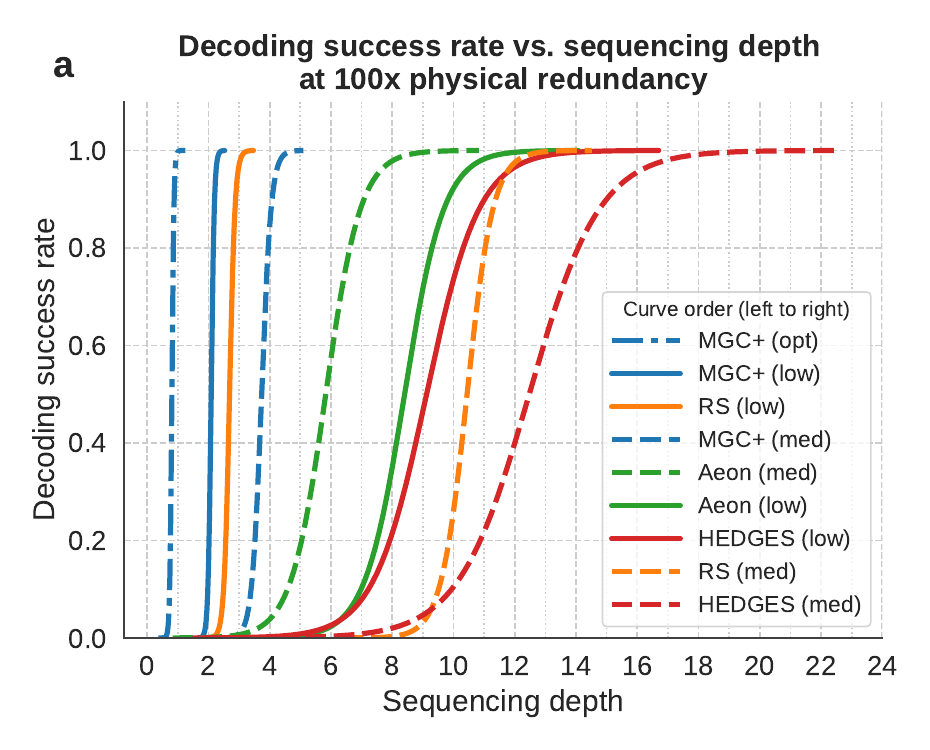}
    \end{subfigure}
    \hfill
    \begin{subfigure}{0.33\textwidth}
        \centering
        \includegraphics[width=\linewidth]{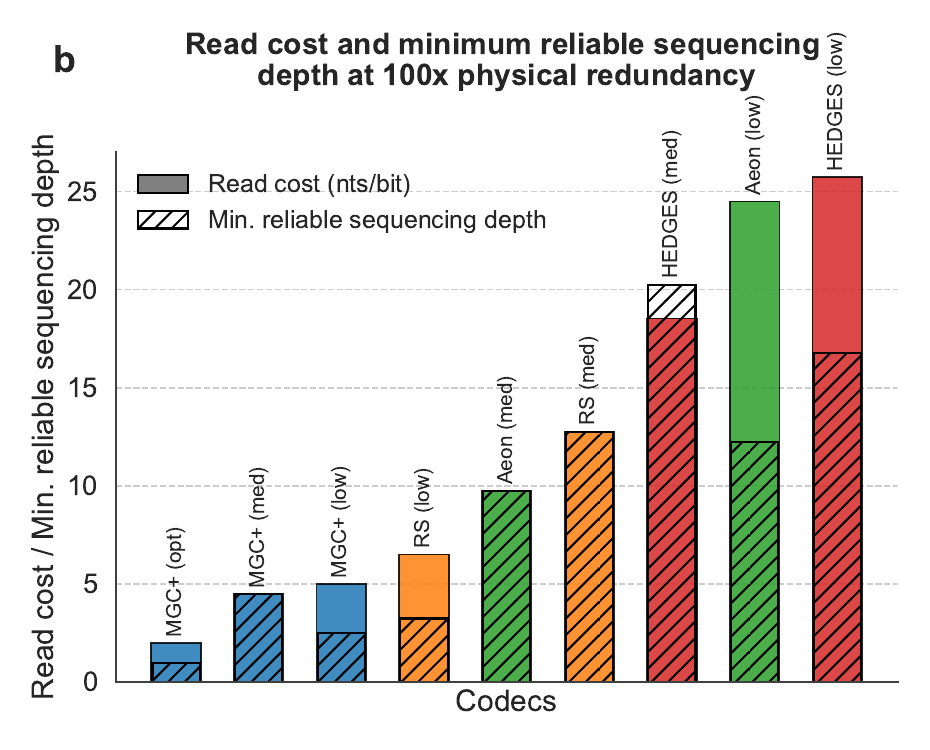}
    \end{subfigure}
    \hfill
    \begin{subfigure}{0.33\textwidth}
        \centering
        \includegraphics[width=\linewidth]{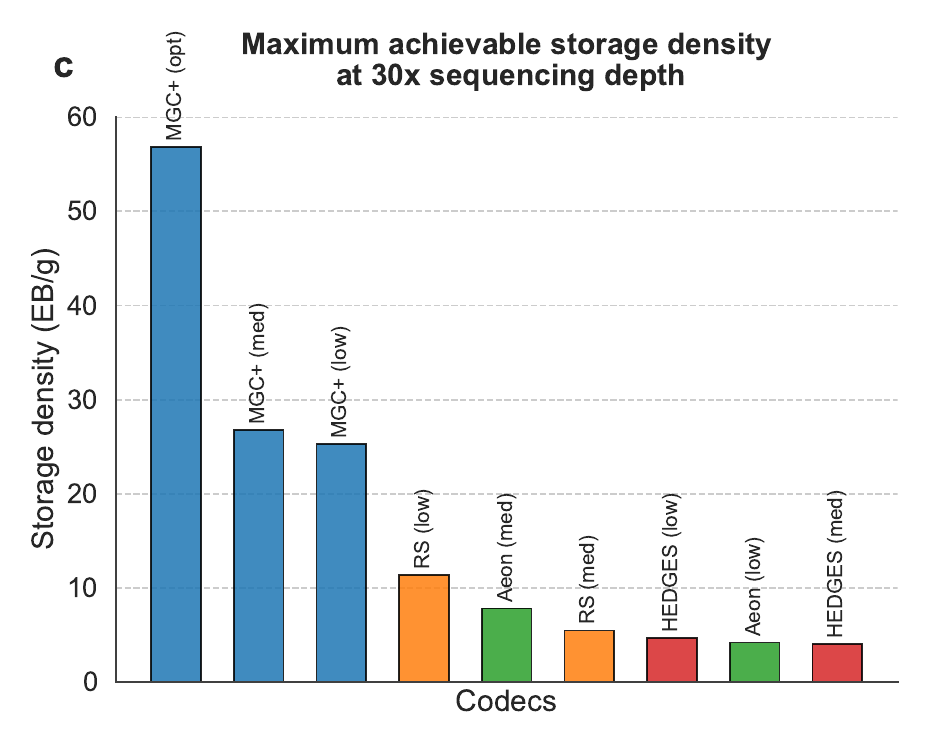}
    \end{subfigure}

\caption{\textbf{In silico codec performance for a file of size 15~KB under experimentally derived error and bias profiles.} (\textbf{a})~Decoding success rate as a function of sequencing depth at a fixed physical redundancy of 100\texttimes~under the DT4DDS low-fidelity workflow. Curves correspond to logistic regression fits of the empirical decoding outcomes; line style indicates the code rate (solid: low, dashed: medium, dash-dot: optimized) (\textbf{b})~Minimum sequencing depth required for reliable decoding at 100\texttimes~physical redundancy, together with the corresponding read cost. (\textbf{c})~Maximum achievable storage density, expressed in exabytes per gram of DNA, at a fixed sequencing depth of 30\texttimes, computed from the minimum physical redundancy required for reliable decoding.}
\label{fig:silico2}
\end{figure}

In one setting, we fix the physical redundancy to 100\texttimes~and measure the decoding success rate as a function of sequencing depth (Fig. \ref{fig:silico2}a), as well as the minimum sequencing depth and associated read cost required to satisfy the same 50-out-of-50 decoding reliability criterion used earlier (Fig. \ref{fig:silico2}b). Notably, only 9 out of the 20 tested codec configurations meet the reliability constraint within the simulated sequencing depth range of [1, 32]. Among these, the low-rate, medium-rate, and optimized DNA-MGC+ configurations achieve the lowest read costs. In particular, the optimized configuration enables reliable decoding at a sequencing depth as low as 1\texttimes, corresponding to a read cost of 2~nts/bit, while the two non-optimized DNA-MGC+ configurations also outperform all remaining codecs in terms of read cost. All numerical results for this setting are summarized in Supplementary Tables S9 and S10.

In another setting, we fix the sequencing depth to 30\texttimes~and determine the minimum physical redundancy required for reliable decoding. While reduced physical redundancy, like reduced sequencing depth, increases the occurrence of sequence dropouts, it also introduces a more fundamental challenge: the limited number of physical molecular copies reduces variability across reads and limits the effectiveness of clustering (see Discussion section for more details). Fig.~\ref{fig:silico2}c reports the maximum achievable storage density (SD) for each codec configuration, obtained by applying the above SD formula using the measured minimum physical redundancy. The results show that the three DNA-MGC+ configurations achieve the highest storage densities, with the optimized DNA-MGC+ configuration enabling reliable decoding at a physical redundancy of only 1\texttimes, corresponding to a storage density of approximately 57~EB/g.

\subsection*{In vitro experimental evaluation}
To evaluate the performance of DNA-MGC+ under {\em in vitro} experimental conditions, we encoded a 24-KB file obtained by zlib compression of a text document containing the Universal Declaration of Human Rights in four languages, using four different configurations of DNA-MGC+ and one configuration of each comparison codec, DNA-Aeon and HEDGES. DNA oligonucleotides corresponding to the six codec configurations were synthesized in a single pool via electrochemical synthesis performed by GenScript. The target design length of the encoded DNA sequences was set to 126~nts, below the maximum length of 170 imposed by GenScript. This choice allowed sufficient margin for the inclusion of 20-nt forward and reverse primer sequences, as well as a 4-nt unique tag to distinguish among the six configurations. Sequencing was performed using both Oxford Nanopore sequencing and paired-end Illumina sequencing. 

The four DNA-MGC+ configurations considered in this experimental evaluation are based on two distinct inner and outer code parametrizations with different redundancy levels. The first parametrization yields an overall code rate of 1.03 bits/nt and is denoted as design~$\mathrm{A}$, while the second introduces additional redundancy and results in a lower code rate of 0.71 bits/nt, denoted as design~$\mathrm{B}$. For each design, we consider an unfiltered variant, in which no constraints are imposed on the content of the encoded DNA sequences, as well as a filtered variant. 

In the filtered variants, an excess number of candidate sequences is generated using the codec, and only those satisfying a set of constraints are retained. These constraints include restrictions on homopolymer length, GC content, and the avoidance of selected sequence motifs, together with an additional filtering step based on thermodynamic properties (Gibbs free energy). By design of the DNA-MGC+ codec, the filtering step does not affect the code rate, so the filtered variants of designs $\mathrm{A}$ and $\mathrm{B}$ retain code rates of 1.03 and 0.71 bits/nt, respectively. These constraints, however, do impact encoding and decoding complexity. Full details of the filtering criteria and parameter selection are provided in the Methods section and in Supplementary Table~S1, respectively.

In addition to DNA-MGC+, one configuration of each of DNA-Aeon and HEDGES was included in the same experimental pool, with code rates of 1.0 and 0.61 bits/nt, respectively. These codecs were selected based on their performance in the {\em in silico} evaluations (Figs.~\ref{fig:silico1} and~\ref{fig:silico2}), where both achieved reliable decoding and consistently ranked among the strongest performers behind DNA-MGC+. In particular, DNA-Aeon (1.0 bit/nt) was chosen for its favorable read cost performance at medium rate across multiple {\em in silico} scenarios (see Figs.~\ref{fig:silico1}d--g and Fig.~\ref{fig:silico2}b). HEDGES (0.61 bits/nt), on the other hand, was selected for its strong robustness to IDS errors at low rate (see Fig.~\ref{fig:silico1}g). Although DNA-RS also exhibited competitive {\em in silico} performance, its main advantage lies in the outer RS code, whose benefits are shared with DNA-MGC+ by design. At the inner code level, however, DNA-MGC+ provides stronger capabilities for correcting insertion and deletion errors. DNA-RS was therefore not included in the {\em in vitro} comparison, since its principal strengths are inherited by DNA-MGC+ and can be further enhanced through optimized redundancy allocation, as observed in Fig.~\ref{fig:silico2}.

\begin{figure}[t!]
    \centering
    \includegraphics[width=0.49\textwidth]{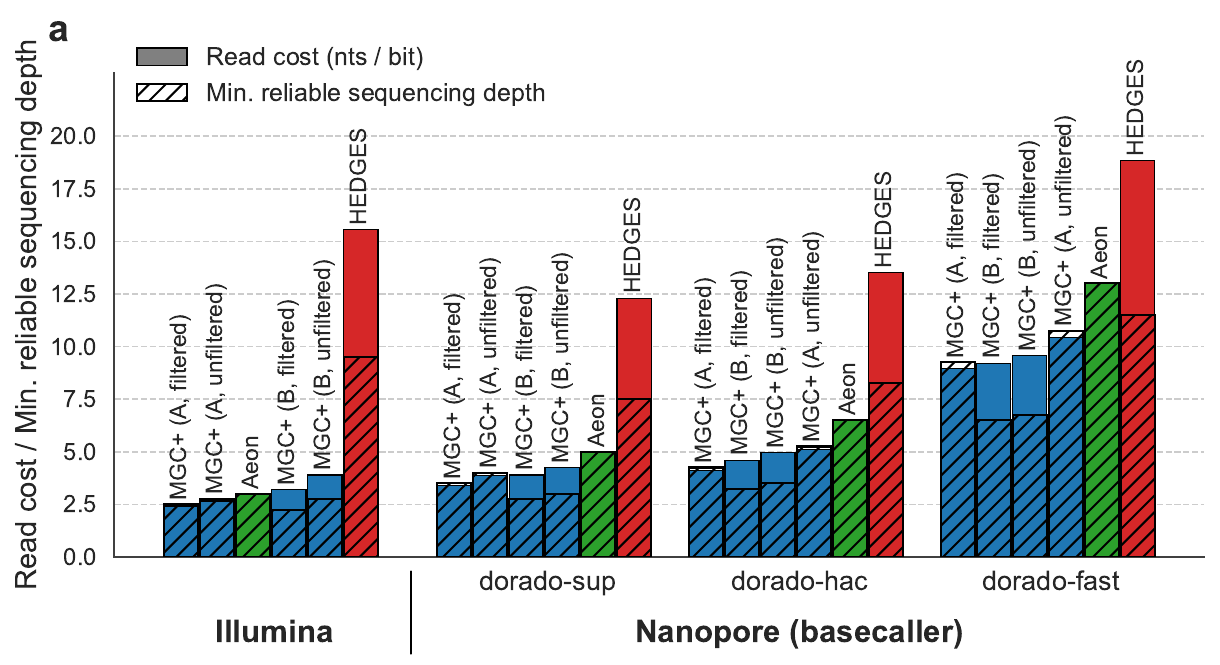}
    \includegraphics[width=0.49\textwidth]{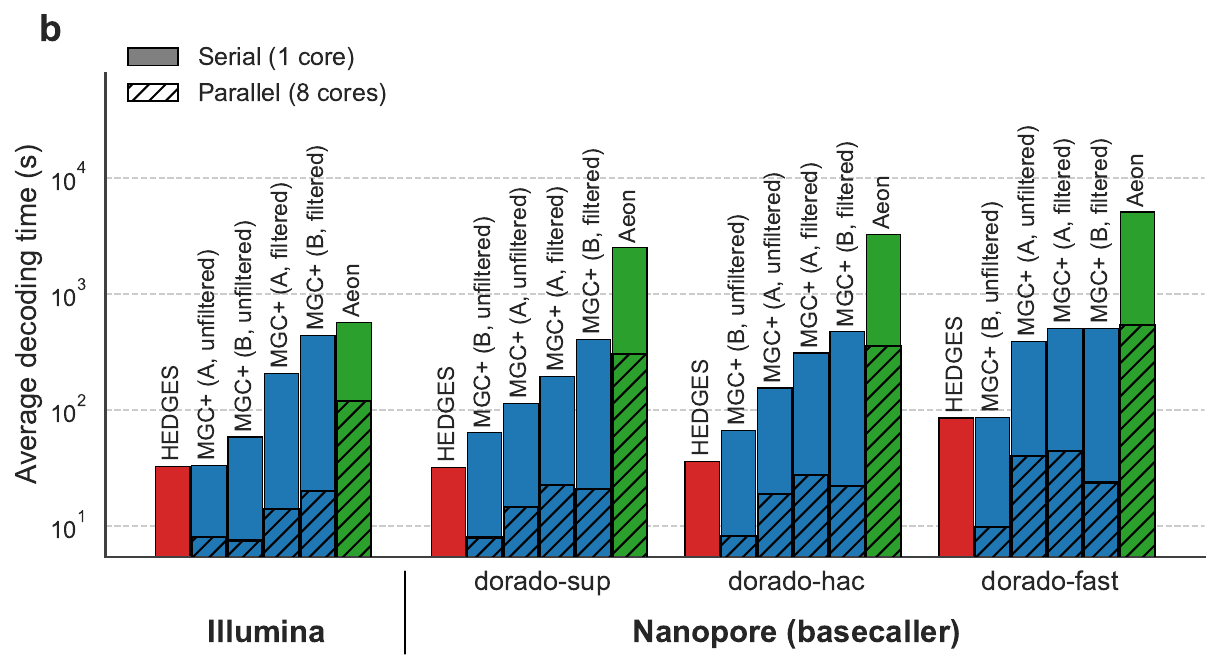}
    \caption{\textbf{In vitro performance of DNA-MGC+ and comparison codecs under Illumina and Oxford Nanopore sequencing.} (\textbf{a})~Minimum sequencing depth and corresponding read cost required for reliable decoding, obtained via progressive read downsampling, under Illumina and Nanopore sequencing. For Nanopore data, results are reported for multiple Dorado basecalling models with varying computational complexity. (\textbf{b})~Average decoding time required to recover the 24-KB stored file, measured at the minimum sequencing depth needed for reliable decoding for each codec configuration.}
    \label{fig:wetlab}
\end{figure}

The stored file was successfully decoded with an exact match under both Nanopore and Illumina sequencing, across all six codec configurations, albeit with quantitatively different performance outcomes. The performance comparison across multiple metrics is shown in Fig.~\ref{fig:wetlab}. For each codec configuration, we report the minimum sequencing depth required for reliable decoding, determined via progressive read downsampling under the same 50-out-of-50 decoding reliability criterion used in the {\em in silico} evaluations. The downsampling is applied after the raw reads are processed to trim primers, filter payloads by length ($\pm$10 nt of the design length), and demultiplex into per-configuration subsets using the 4-nt tags, as detailed in the Methods section. The corresponding read cost, expressed in nucleotides per information bit, is also reported. We further report the average decoding time for recovering the stored file, measured at the minimum sequencing depth required by each codec for reliable decoding. As in the previous results, decoding times exclude clustering, alignment, and consensus calling, which were performed using CD-HIT and Kalign.

For Nanopore sequencing, multiple Dorado~\cite{dorado2025} basecalling models were tested, revealing pronounced tradeoffs between basecalling accuracy and computational cost. The fast, high accuracy (hac), and super accuracy (sup) models exhibited basecalling runtimes differing by approximately one order of magnitude between successive models, with corresponding base-level error rates of 5.6 \textpm~0.2\%, 2.2 \textpm~0.15\%, and 1.4 \textpm~0.1\% (averaged across the six codec configurations), respectively. As expected, the results in Fig.~\ref{fig:wetlab}a show that reduced basecalling accuracy leads to a systematic increase in the minimum sequencing depth required for reliable decoding, and consequently in the associated read cost.  Across all three basecallers, all four configurations of DNA-MGC+ consistently require lower sequencing depth and read cost than both DNA-Aeon and HEDGES. In particular, under the dorado-sup basecaller, the filtered variant of design~$\mathrm{B}$ achieves the lowest minimum sequencing depth of 2.75\texttimes, while the filtered variant of design~$\mathrm{A}$ attains the lowest read cost of 3.4~nts/bit. Furthermore, the filtered variants of DNA-MGC+ provide modest but consistent improvements over their unfiltered counterparts across all cases.

Under Illumina sequencing, we observed an average base-level error rate of 0.34 \textpm~0.03\%, which is lower than that of Nanopore sequencing, as expected. Overall, the performance trends under Illumina closely mirror those obtained with Nanopore. In particular, DNA-MGC+ outperforms both DNA-Aeon and HEDGES, with the filtered variant of design~$\mathrm{B}$ achieving the lowest minimum sequencing depth of 2.25\texttimes, and the filtered variant of design~$\mathrm{A}$ attaining the lowest read cost of 2.4~nts/bit. A detailed performance comparison across Nanopore and Illumina platforms is provided in the Discussion section.

In terms of decoding time, Fig.~\ref{fig:wetlab}b reveals clear differences across codecs, with consistent trends observed under both Illumina and Nanopore sequencing.  When decoding is performed on a single core, HEDGES achieves the fastest decoding time, with the unfiltered variant of design~$\mathrm{B}$ of DNA-MGC+ following closely, while DNA-Aeon is approximately one order of magnitude slower. When parallelization across eight CPU cores is enabled, DNA-MGC+ achieves the fastest decoding performance overall, requiring less than 10 seconds on average for decoding. This improvement reflects the structure of the DNA-MGC+ decoder, which supports parallel processing at both the inner and outer code levels. The publicly available implementation of DNA-Aeon also supports parallel execution, and the corresponding parallel decoding results are therefore reported in Fig.~\ref{fig:wetlab}b, whereas the available implementation of HEDGES does not support parallel decoding and is thus reported under single-core execution only. Another observation worth noting is that the unfiltered variants of DNA-MGC+ consistently decode faster than their filtered counterparts. This difference stems from the codec design, discussed in more detail in the Methods section, whereby filtered variants rely on a more complex decoding algorithm at the outer RS code level.

\section*{Discussion}
The performance evaluations presented in this work demonstrate that DNA-MGC+ consistently outperforms the selected codecs across a broad range of \emph{in silico} frameworks and \emph{in vitro} experimental conditions. These evaluations span different error and bias profiles as well as multiple sequencing technologies with distinct basecalling models, highlighting the \emph{versatility} of DNA-MGC+ in delivering superior performance across diverse operating conditions. The \emph{reliability} and \emph{resource-efficiency} of DNA-MGC+ are reflected in several key performance metrics. Specifically, the codec achieves simultaneous gains in coverage depth requirements, read cost, decoding time, maximal error-correction capability, and storage density. In practice, these gains translate into improved data reliability, reduced sequencing effort, and faster data retrieval. Importantly, the results further indicate that DNA-MGC+ is efficient even under low-fidelity biochemical processes. Such processes are more cost-effective, and are therefore expected to play a central role in overcoming current scalability bottlenecks in DNA data storage in the future. These improvements are a direct consequence of the coding strategy implemented in DNA-MGC+, as discussed next.

The design of the DNA-MGC+ codec follows a two-layer architecture, in which logical redundancy is introduced at both the inner and outer coding stages with complementary roles. At the inner level, an MGC+ code~\cite{MGCP} protects against base-level errors by enabling the correction of insertions, deletions, and substitutions (IDS) within individual DNA sequences. At the outer level, a Reed-Solomon (RS) code~\cite{reed1960polynomial} mitigates the effects of coverage bias by enabling recovery from sequence dropouts, while also correcting residual errors that remain after inner decoding. Under a fixed target code rate, this separation of roles leads to a natural redundancy allocation tradeoff: allocating more redundancy to the inner layer than to the outer layer enhances IDS error-correction capability at the expense of reduced robustness to dropout events, and vice versa. When the underlying error and bias profile is known or estimated, this tradeoff can be optimized, as reflected by the improved performance of the optimized DNA-MGC+ configuration in Fig.~\ref{fig:silico2}. 

\begin{table}[h!]
\centering
\caption{\textbf{Inner and outer code components of the selected codecs and their scope of operation.}}
\label{tab:codec_architecture}
\begin{tabular}{lcccc}
\hline
\textbf{Codec} & \textbf{Inner code} & \textbf{Outer code} & \textbf{Inner code scope} & \textbf{Outer code scope} \\
\hline
DNA-MGC+      & MGC+              & RS        & IDS errors (correction)        & Dropouts and residual errors \\
DNA-Aeon~\cite{welzel2023dna}      & Arithmetic & Fountain  & IDS errors (correction)        & Dropouts \\
HEDGES~\cite{press2020hedges}        & Hedges            & RS        & IDS errors (correction)        & Dropouts and residual errors \\
DNA-Fountain~\cite{erlich2017dna}  & RS                & Fountain  & Substitution errors (detection) & Dropouts \\
DNA-RS~\cite{antkowiak2020low,meiser2020reading}        & RS                & RS        & Substitution errors (correction) & Dropouts and residual errors \\
\hline
\end{tabular}
\end{table}

Several of the comparison codecs considered in this work follow the same two-layer coding architecture as DNA-MGC+, including DNA-Aeon, HEDGES, DNA-Fountain, and DNA-RS, whose respective inner and outer components are summarized in Table~\ref{tab:codec_architecture}. As observed in Figs.~\ref{fig:silico1} and~\ref{fig:silico2}, among the codecs evaluated in this work, those based on this inner-outer architecture achieve better performance than DNA-StairLoop~\cite{yan2025dna} and the LDPC-based scheme~\cite{chandak2019improved}, which follow conceptually different design principles. DNA-Fountain is a notable exception, as its reduced performance stems from limitations of its inner code, whose functionality is deliberately restricted to detection rather than correction of substitution errors. The advantage of using an RS code as an outer code is well established in coding theory~\cite{bar-lev25, hanna2025reliability}, as RS codes are optimal in recovering from sequence dropouts with minimal redundancy, while also allowing correction of residual errors after inner decoding. Consistent with this theory, DNA-RS and DNA-MGC+ consistently emerge as the strongest performers in bias-dominated regimes such as Fig.~\ref{fig:silico1}d and Fig.~\ref{fig:silico2}. In parallel, strong performance under high-error conditions is most evident for codecs whose inner codes support correction of IDS errors. This explains why DNA-MGC+, HEDGES, and DNA-Aeon outperform schemes limited to substitution-only correction in Figs.~\ref{fig:silico1}f–g. Overall, the gains of DNA-MGC+ with respect to existing designs arise from the combination of \emph{(i)} enhanced IDS error-correction capability at the inner MGC+ layer; and \emph{(ii)} the use of an outer RS code that is both optimal for mitigating sequence dropouts and capable of correcting residual inner-decoding errors.

The \emph{in vitro} results in Fig.~\ref{fig:wetlab} further indicate how structural differences between codecs translate into distinct performance under Illumina and Nanopore sequencing. Achieving reliable decoding with limited reads is a dual challenge, as low coverage increases both sequence dropouts and the effective error rate in consensus sequences. Accordingly, the minimum sequencing depth required by a codec reflects both its outer-code robustness to dropouts and inner-code IDS correction capability. Notably, the results in Fig.~\ref{fig:wetlab} show that DNA-MGC+ maintains nearly identical sequencing depth requirements under Illumina and Nanopore (dorado-sup) sequencing, despite the higher base-level error rate of the latter. This stability is attributable to the strong IDS error-correction capability of the inner MGC+ code, which effectively compensates for elevated error rates and preserves overall decoding performance. In contrast, DNA-Aeon performs substantially better under Illumina and approaches the sequencing depth levels of DNA-MGC+, indicating comparable dropout robustness but weaker IDS correction capability. HEDGES exhibits similar performance across both platforms yet consistently requires higher sequencing depths than DNA-MGC+ and DNA-Aeon, suggesting that its primary limitation lies in dropout mitigation rather than in IDS correction.

Beyond the challenge of reliable decoding at low sequencing depths, the second studied \emph{in silico} framework highlights a distinct and more stringent challenge: reliable decoding at low physical redundancy, which enables higher storage densities~(Fig.~\ref{fig:silico2}c). While reduced sequencing depth limits the total number of reads, reduced physical redundancy limits the number of independent molecular copies of each reference sequence. Although both effects increase sequence dropouts, low physical redundancy introduces a more fundamental decoding challenge that cannot be resolved by simply increasing the number of reads. Specifically, when only a small number of physical copies of each DNA sequence are present, errors introduced during early stages of the storage pipeline, such as synthesis, tend to propagate across most subsequent reads corresponding to the same reference sequence. This results in reduced read variability and the emergence of systematic error patterns, which in turn limit the effectiveness of clustering and consensus calling. Consequently, a larger share of the error-correction burden must be handled by the codec itself rather than by upstream read-processing operations. The superior IDS error-correction capability of DNA-MGC+ at the inner-code level is reflected in the results shown in Fig.~\ref{fig:silico2}c, where reliable decoding is achieved at lower physical redundancies than comparison codecs, thereby enabling higher storage densities.

In addition to their role in dropout recovery and residual error correction, outer codes also influence the flexibility of constrained coding strategies. For this reason, despite the aforementioned advantages of RS outer codes, Fountain-based outer codes have also attracted significant attention in the literature. A key feature of Fountain codes is their ability to generate a virtually unlimited number of encoded DNA sequences. This enables constrained coding through simple filtering, where sequences that violate the desired constraints are discarded until a sufficient number of valid sequences is obtained~\cite{erlich2017dna}. In the proposed DNA-MGC+ codec, we introduce a design methodology that allows this filtering paradigm to be extended to outer RS codes, as detailed in the Methods section. As a result, the DNA-MGC+ codec supports filtering based on arbitrary content-specific constraints while retaining the error-correction advantages of outer RS codes.

The \emph{in vitro} results in Fig.~\ref{fig:wetlab}a show that modest but consistent performance gains are achieved when filtering is applied to DNA-MGC+. The filtering used for this experiment imposes several concurrent constraints. These include limiting the maximum homopolymer length to four, enforcing a GC content between 45\% and 55\%, avoiding motifs such as di- and trinucleotide repeats, and ranking candidate sequences by their predicted folding Gibbs free energy ($\Delta G$), with those whose $\Delta G$ values are closest to zero retained. The limited extent of the observed gains compared to the unfiltered case can be attributed to two main factors. First, the inherent robustness of DNA-MGC+ to errors and biases allows it to compensate for content-induced effects without the need to explicitly avoid them through constrained coding. Second, the impact of factors such as long homopolymers and high GC content on modern synthesis and sequencing technologies appears to be less detrimental than suggested in earlier works, which is in line with recent independent experimental findings~\cite{11164904}. 

Despite the limited magnitude of the gains, the consistent improvement observed with filtering indicates that some of the considered sequence properties do indeed influence the retrieval process. In particular, our statistical analysis of the experimental data reveals that $\Delta G$ is the dominant factor contributing to the observed gains. Specifically, we identify Spearman correlation coefficients of $+0.31$ for Nanopore (and $+0.26$ for Illumina) between the $\Delta G$ of unfiltered DNA-MGC+ sequences and their sequencing read counts, with p-values below $10^{-50}$ in both cases. These statistics support the hypothesis that sequences with $\Delta G$ closer to zero (less negative) receive more reads and are therefore easier to decode. In contrast, the corresponding correlation coefficients for GC content and maximum homopolymer length are weaker, at $-0.21$ and $+0.08$ for Nanopore, respectively. Partial Spearman correlations further confirm that the $\Delta G$--coverage association is largely independent of GC content or homopolymer length, whereas the marginal association between GC content and coverage largely disappears after controlling for $\Delta G$. \hl{These results therefore indicate that $\Delta G$ is the dominant sequence-level feature associated with coverage variability among those examined in our experiments.} Additional details and a more thorough statistical analysis are provided in Supplementary Note~S3.


\section*{Methods} \label{methods}
\subsection*{DNA-MGC+ codec design}

\paragraph{Encoding.} Consider an input binary file of size $B$ bits. The DNA-MGC+ encoder outputs $N$ reference DNA sequences that collectively represent the file, each of length $L_{\text{ref}}$ nucleotides. The encoding is carried out through a series of structured transformations that introduce logical redundancy in order to enable reliable data retrieval under error and dropout events. This process, illustrated in~Fig.~\ref{fig:encoding_pipeline}, comprises the following sequential steps.

\begin{enumerate}[label=(\arabic*), leftmargin=*]
    \item {\em Fragmentation: } The input binary file is partitioned into $K=\lceil B/k \rceil$ non-overlapping fragments, each of length $k$ bits.
    \item {\em Outer RS Code: } The $K$ fragments are encoded using an outer Reed-Solomon (RS) code over the finite field $\mathbb{F}_{2^{\ell_\mathrm{out}}}$ to introduce inter-sequence redundancy, where each RS symbol consists of $\ell_\mathrm{out}$ bits. The outer RS code generates $c_{\text{out}}$ redundant fragments, resulting in a total of $N = K + c_{\text{out}}$ encoded fragments and an outer code rate of $\rho_{\text{out}}=K/N$.
    \item {\em Indexing: } Each outer-encoded fragment is prepended with a unique binary index of length $\ell_{\text{out}}$ bits, forming indexed fragments of length $k+\ell_{\text{out}}$ bits. The index length $\ell_{\text{out}}$ is chosen to uniquely identify any position in the outer RS codeword.

    \item {\em Inner MGC+ Code: } 
    Each indexed fragment is individually encoded using the Marker Guess \& Check Plus~(MGC+) code to introduce intra-sequence redundancy. The encoding proceeds in two stages, where part of the redundancy is introduced in the binary domain, and another part is added in the DNA domain after binary-to-quaternary mapping.

    \begin{enumerate}[label=(\arabic{enumi}.\arabic*)]
        \item {\em Binary-domain redundancy: } The inner MGC+ code introduces $c_{\text{in}}+1$ parity symbols derived from an underlying RS code over $\mathbb{F}_{2^{\ell_{\text{in}}}}$, where each RS symbol consists of $\ell_{\text{in}}$ bits. This adds $(c_{\text{in}}+1)\ell_{\text{in}}$ redundant bits to each indexed fragment, resulting in binary sequences of total length $k+\ell_{\text{out}}+(c_{\text{in}}+1)\ell_{\text{in}}$~bits. The first $c_{\text{in}}$ parity symbols are referred to as \emph{guess} parities, while the final parity symbol is referred to as the \emph{check} parity, reflecting their distinct roles in the decoding procedure described later.
        
        \item {\em DNA-domain redundancy: } 
        Each resulting binary sequence is mapped into a quaternary DNA sequence according to $00 \mapsto \mathsf{A}$, $01 \mapsto \mathsf{T}$, $10 \mapsto \mathsf{C}$, and $11 \mapsto \mathsf{G}$. Optionally, periodic 2-nt markers ``$\mathsf{AC}$'' are inserted after every $\ell_{\text{in}}$ DNA symbols. The check parity symbol is further encoded by mapping it into a short DNA barcode of length~$\beta$, where the used barcodes are drawn from a predefined DNA codebook with minimum edit distance $d_{\text{min}}$. 
    \end{enumerate}
\end{enumerate}

\begin{figure}[t!]
    \centering
    \resizebox{\textwidth}{!}{%

\begin{tikzpicture}[
  font=\fontfamily{phv}\selectfont,
  >=Stealth,
  arrow/.style={-Stealth, thick, draw=gray!70},
  bitbox/.style={
    draw=gray!90, fill=gray!5, rounded corners=1pt, thick,
    inner sep=4pt, font=\fontfamily{pcr}\selectfont\footnotesize
  },
  rlabel/.style={font=\fontfamily{phv}\selectfont\scriptsize},
  bitbox_outer/.style={bitbox, fill=blue!15},
  bitbox_idx/.style={bitbox, fill=orange!15},
  bitbox_red/.style={bitbox, fill=red!15},
  bitbox_in/.style={bitbox, inner xsep=2pt} 
]


\coordinate (R1) at (0,0);
\coordinate (R2) at ([yshift=-6mm]R1);
\coordinate (RK) at ([yshift=-20mm]R1);  

\coordinate (RKp) at ([yshift=-6mm]RK);
\coordinate (RN)  at ([yshift=-16mm]RK);

\node[bitbox, anchor=west] (v11) at (R1) {01\ldots01};
\node[bitbox, right=0mm of v11] (v12) {10\ldots11};
\node[right=0mm of v12, font=\fontfamily{phv}\selectfont] (vell1) {$\cdots$};
\node[bitbox, right=0mm of vell1] (v1L) {00\ldots10};

\node[bitbox, anchor=west] (v21) at (R2) {11\ldots00};
\node[bitbox, right=0mm of v21] (v22) {01\ldots10};
\node[right=0mm of v22, font=\fontfamily{phv}\selectfont] (vell2) {$\cdots$};
\node[bitbox, right=0mm of vell2] (v2L) {11\ldots01};

\node[below=0mm of v21, font=\fontfamily{phv}\selectfont] {$\vdots$};
\node[below=0mm of v22, font=\fontfamily{phv}\selectfont] {$\vdots$};
\node[below=4mm of vell2, font=\fontfamily{phv}\selectfont] {$\cdots$};
\node[below=0mm of v2L, font=\fontfamily{phv}\selectfont] (anc1) {$\vdots$};

\node[bitbox, anchor=west] (vK1) at (RK) {10\ldots11};
\node[bitbox, right=0mm of vK1] (vK2) {00\ldots01};
\node[right=0mm of vK2, font=\fontfamily{phv}\selectfont] (vellK) {$\cdots$};
\node[bitbox, right=0mm of vellK] (vKL) {11\ldots10};

\node[bitbox_outer, anchor=west] (vKp1) at (RKp) {00\ldots10};
\node[bitbox_outer, right=0mm of vKp1] (vKp2) {11\ldots00};
\node[right=0mm of vKp2, font=\fontfamily{phv}\selectfont] (vellKp) {$\cdots$};
\node[bitbox_outer, right=0mm of vellKp] (vKpL) {00\ldots01};

\node[below=-2mm of vKp1, font=\fontfamily{phv}\selectfont] {$\vdots$};
\node[below=-2mm of vKp2, font=\fontfamily{phv}\selectfont] {$\vdots$};
\node[below=2mm of vellKp, font=\fontfamily{phv}\selectfont] {$\cdots$};
\node[below=-2mm of vKpL, font=\fontfamily{phv}\selectfont] {$\vdots$};

\node[bitbox_outer, anchor=west] (vN1) at (RN) {11\ldots01};
\node[bitbox_outer, right=0mm of vN1] (vN2) {00\ldots10};
\node[right=0mm of vN2, font=\fontfamily{phv}\selectfont] (vellN) {$\cdots$};
\node[bitbox_outer, right=0mm of vellN] (vNL) {10\ldots11};

\draw [decorate, decoration={brace, amplitude=5pt, mirror}] 
        ($(v11.north west)+(-0.1,0)$) -- ($(vK1.south west)+(-0.1,0)$) 
        node [midway, left=6pt, font=\fontfamily{phv}\selectfont\footnotesize] {$K$};

\draw [decorate, decoration={brace, amplitude=5pt, mirror}] 
        ($(vKp1.north west)+(-0.1,0)$) -- ($(vN1.south west)+(-0.1,0)$) 
        node [midway, left=4pt, font=\fontfamily{phv}\selectfont\footnotesize] {$c_{\text{out}}$};

\draw[<->, black]
  ([yshift=1.0mm]v11.north west) -- ([yshift=1.0mm]v11.north east)
  node[midway, above, font=\fontfamily{phv}\selectfont\footnotesize] {$\ell_{\text{out}}$};
\draw[<->, black]
  ([yshift=1.0mm]v1L.north west) -- ([yshift=1.0mm]v1L.north east)
  node[midway, above, font=\fontfamily{phv}\selectfont\footnotesize] {$\ell_{\text{out}}$};

\draw[draw=gray!90, densely dashed, thick, rounded corners=3pt] 
      (-0.77,1.22) rectangle (5.25,-2.3);

\draw[draw=gray!90, densely dashed, thick, rounded corners=3pt] 
      (-0.95,1.6) rectangle (5.43,-4.2);
      
\node[above] at (2.1,1.15) {\fontfamily{phv}\selectfont\scriptsize\color{gray} (1) Fragmentation};
\node[above] at (2.1,1.53) {\fontfamily{phv}\selectfont\scriptsize\color{gray} (2) Outer RS Code (inter-sequence redundancy)};

\draw [arrow, densely dotted, black, thin] ($(v11.north west)+(0.09,0)$ ) -- ($(vN1.south west)+(0.09,-0.3)$);
\draw [arrow, densely dotted, black, thin] ($(v1L.north west)+(0.09,0)$ ) -- ($(vNL.south west)+(0.09,-0.3)$);
\draw[<->, black]
  ([yshift=5.5mm]v11.north west) -- ([yshift=5.5mm]v1L.north east)
  node[midway, above, font=\fontfamily{phv}\selectfont\footnotesize] {$k$};


\coordinate (S1)  at ([xshift=12mm]v1L.east);
\coordinate (S2)  at ([xshift=12mm]v2L.east);
\coordinate (SK2) at ([xshift=12mm]vKL.east);
\coordinate (SKp2) at ([xshift=12mm]vKpL.east);
\coordinate (SN2) at ([xshift=12mm]vNL.east);

\tikzset{
  bitbox_in_reg/.style={bitbox_in},
  bitbox_in_out/.style={bitbox_outer, inner xsep=2pt},
  bitbox_red_in/.style={bitbox_red, inner xsep=2pt},
  bitbox_idx_in/.style={bitbox_idx, inner xsep=2pt}
}

\node[bitbox_idx_in, anchor=west] (w11a) at (S1) {01\ldots};
\node[bitbox_idx_in, right=0mm of w11a] (w11b) {\ldots10};
\node[bitbox_in_reg, right=0.3mm of w11b] (w12) {0\ldots0};
\node[bitbox_in_reg, right=0mm of w12] (w13) {0\ldots1};
\node[right=0mm of w13, font=\fontfamily{phv}\selectfont] (w1dotsA) {$\cdots$};
\node[bitbox_in_reg, right=0mm of w1dotsA] (w14) {1\ldots0};
\node[bitbox_red_in, right=0mm of w14] (w15r) {1\ldots1};
\node[right=0mm of w15r, font=\fontfamily{phv}\selectfont] (w1dotsR) {$\cdots$};
\node[bitbox_red_in, right=0mm of w1dotsR] (w1Lr) {0\ldots1};

\node[bitbox_idx_in, anchor=west] (w21a) at (S2) {11\ldots};
\node[bitbox_idx_in, right=0mm of w21a] (w21b) {\ldots00};
\node[bitbox_in_reg, right=0.3mm of w21b] (w22) {1\ldots0};
\node[bitbox_in_reg, right=0mm of w22] (w23) {1\ldots0};
\node[right=0mm of w23, font=\fontfamily{phv}\selectfont] (w2dotsA) {$\cdots$};
\node[bitbox_in_reg, right=0mm of w2dotsA] (w24) {0\ldots1};
\node[bitbox_red_in, right=0mm of w24] (w25r) {0\ldots1};
\node[right=0mm of w25r, font=\fontfamily{phv}\selectfont] (w2dotsR) {$\cdots$};
\node[bitbox_red_in, right=0mm of w2dotsR] (w2Lr) {1\ldots1};

\node[below=0mm of w21a, font=\fontfamily{phv}\selectfont] (anc2) {$\vdots$};
\node[below=0mm of w21b, font=\fontfamily{phv}\selectfont] {$\vdots$};
\node[below=0mm of w22,  font=\fontfamily{phv}\selectfont] {$\vdots$};
\node[below=0mm of w23,  font=\fontfamily{phv}\selectfont] {$\vdots$};
\node[below=4mm of w2dotsA, font=\fontfamily{phv}\selectfont] {$\cdots$};
\node[below=0mm of w24, font=\fontfamily{phv}\selectfont] {$\vdots$};
\node[below=0mm of w25r, font=\fontfamily{phv}\selectfont] {$\vdots$};
\node[below=4mm of w2dotsR, font=\fontfamily{phv}\selectfont] {$\cdots$};
\node[below=0mm of w2Lr, font=\fontfamily{phv}\selectfont] (anc3) {$\vdots$};

\node[bitbox_idx_in, anchor=west] (wK1a) at (SK2) {10\ldots};
\node[bitbox_idx_in, right=0mm of wK1a] (wK1b) {\ldots11};
\node[bitbox_in_reg, right=0.3mm of wK1b] (wK2) {1\ldots0};
\node[bitbox_in_reg, right=0mm of wK2] (wK3) {0\ldots1};
\node[right=0mm of wK3, font=\fontfamily{phv}\selectfont] (wKdotsA) {$\cdots$};
\node[bitbox_in_reg, right=0mm of wKdotsA] (wK4) {1\ldots1};
\node[bitbox_red_in, right=0mm of wK4] (wK5r) {0\ldots0};
\node[right=0mm of wK5r, font=\fontfamily{phv}\selectfont] (wKdotsR) {$\cdots$};
\node[bitbox_red_in, right=0mm of wKdotsR] (wKLr) {1\ldots0};

\node[bitbox_idx_in, anchor=west] (wKp1a) at (SKp2) {01\ldots};
\node[bitbox_idx_in, right=0mm of wKp1a] (wKp1b) {\ldots01};
\node[bitbox_in_out, right=0.3mm of wKp1b] (wKp2) {0\ldots1};
\node[bitbox_in_out, right=0mm of wKp2] (wKp3) {1\ldots0};
\node[right=0mm of wKp3, font=\fontfamily{phv}\selectfont] (wKpdotsA) {$\cdots$};
\node[bitbox_in_out, right=0mm of wKpdotsA] (wKp4) {0\ldots1};
\node[bitbox_red_in, right=0mm of wKp4] (wKp5r) {1\ldots0};
\node[right=0mm of wKp5r, font=\fontfamily{phv}\selectfont] (wKpdotsR) {$\cdots$};
\node[bitbox_red_in, right=0mm of wKpdotsR] (wKpLr) {0\ldots1};

\node[below=-2mm of wKp1a, font=\fontfamily{phv}\selectfont] {$\vdots$};
\node[below=-2mm of wKp1b, font=\fontfamily{phv}\selectfont] {$\vdots$};
\node[below=2mm of wKpdotsA, font=\fontfamily{phv}\selectfont] {$\cdots$};
\node[below=-2mm of wKp4, font=\fontfamily{phv}\selectfont] {$\vdots$};
\node[below=-2mm of wKp5r, font=\fontfamily{phv}\selectfont] {$\vdots$};
\node[below=2mm of wKpdotsR, font=\fontfamily{phv}\selectfont] {$\cdots$};
\node[below=-2mm of wKpLr, font=\fontfamily{phv}\selectfont] {$\vdots$};

\node[bitbox_idx_in, anchor=west] (wN1a) at (SN2) {11\ldots};
\node[bitbox_idx_in, right=0mm of wN1a] (wN1b) {\ldots10};
\node[bitbox_in_out, right=0.3mm of wN1b] (wN2) {1\ldots0};
\node[bitbox_in_out, right=0mm of wN2] (wN3) {0\ldots1};
\node[right=0mm of wN3, font=\fontfamily{phv}\selectfont] (wNdotsA) {$\cdots$};
\node[bitbox_in_out, right=0mm of wNdotsA] (wN4) {1\ldots1};
\node[bitbox_red_in, right=0mm of wN4] (wN5r) {0\ldots0};
\node[right=0mm of wN5r, font=\fontfamily{phv}\selectfont] (wNdotsR) {$\cdots$};
\node[bitbox_red_in, right=0mm of wNdotsR] (wNLr) {1\ldots0};

\draw [decorate, decoration={brace, amplitude=5pt}] 
        ($(w15r.north west)+(0,0.05)$) -- ($(w1Lr.north west)+(-0.1,0.05)$) 
        node [midway, xshift=-0.5mm, above=3.5pt, font=\fontfamily{phv}\selectfont\scriptsize] {$c_{\text{in}}$ guess-pars};

    \draw [decorate, decoration={brace, amplitude=5pt}] 
        ($(w1Lr.north west)+(0,0.05)$) -- ($(w1Lr.north east)+(0,0.05)$) 
        node [midway, xshift=1mm, above=3.5pt, font=\fontfamily{phv}\selectfont\scriptsize] {check-par};
    \node[] at (0,0) (x_ref) {};
   \draw[draw=gray!90, densely dashed, thick, rounded corners=3pt] 
      (6.25,1.22) rectangle (8.35,-4.2);

   \draw[draw=gray!90, densely dashed, thick, rounded corners=3pt] 
      (8.4,1.22) rectangle (15.1,-4.2);

      \node[above] at (7.3,1.15) {\fontfamily{phv}\selectfont\scriptsize\color{gray} (3) Indexing};
\node[above] at (11.75,1.15) {\fontfamily{phv}\selectfont\scriptsize\color{gray} (4.1) Inner MGC+ Code (intra-sequence binary redundancy)};

\draw[<->, black]
  ([yshift=5.5mm]w11a.north west) -- ([yshift=5.5mm]w11b.north east)
  node[midway, above, font=\fontfamily{phv}\selectfont\footnotesize] {$\ell_{\text{out}}$};
\draw[<->, black]
  ([yshift=1mm]w11a.north west) -- ([yshift=1mm]w11a.north east)
  node[midway, above, font=\fontfamily{phv}\selectfont\footnotesize] {$\ell_{\text{in}}$};
\draw[<->, black]
  ([yshift=1mm]w11b.north west) -- ([yshift=1mm]w11b.north east)
  node[midway, above, font=\fontfamily{phv}\selectfont\footnotesize] {$\ell_{\text{in}}$};
\draw[<->, black]
  ([yshift=1mm]w12.north west) -- ([yshift=1mm]w12.north east)
  node[midway, above, font=\fontfamily{phv}\selectfont\footnotesize] {$\ell_{\text{in}}$};

\draw [arrow, densely dotted, black, thin] ($(w11a.south west)+(0,0.08)$ ) -- ($(w1Lr.south east)+(0.25,0.08)$);
\draw [arrow, densely dotted, black, thin] ($(wN1a.south west)+(0,0.08)$ ) -- ($(wNLr.south east)+(0.25,0.08)$);



\coordinate (Q1)  at ([xshift=12mm]w1Lr.east);
\coordinate (Q2)  at ([xshift=12mm]w2Lr.east);
\coordinate (QK2) at ([xshift=12mm]wKLr.east);
\coordinate (QKp2) at ([xshift=12mm]wKpLr.east);
\coordinate (QN3) at ([xshift=12mm]wNLr.east);

\tikzset{
  dna_orange/.style={bitbox_idx, inner xsep=2pt},
  dna_gray/.style={bitbox_in},
  dna_blue/.style={bitbox_outer, inner xsep=2pt},
  dna_red/.style={bitbox_red, inner xsep=2pt},
  dna_marker/.style={bitbox_red, inner xsep=2pt},
  dna_barcode/.style={
    bitbox_red,
    inner xsep=2pt,   
    fill=red!15,
    path picture={
      \begin{scope}
        \clip (path picture bounding box.south west) rectangle (path picture bounding box.north east);
        \begin{scope} 
          \fill[pattern=vertical lines, pattern color=gray!70]
            (path picture bounding box.south west) rectangle (path picture bounding box.north east);
        \end{scope}
      \end{scope}
    }
  }
}

\node[dna_orange, anchor=west] (z11) at (Q1) {T\ldots C};
\node[dna_marker, right=0mm of z11] (mz11) {AC};
\node[dna_gray, right=0mm of mz11] (z12) {A\ldots G};
\node[dna_marker, right=0mm of z12] (mz12) {AC};
\node[right=2mm of mz12, font=\fontfamily{phv}\selectfont] (zdots1a) {$\cdots$};
\node[dna_marker, right=2mm of zdots1a] (mz13) {AC};
\node[dna_gray, right=0mm of mz13] (z13) {T\ldots A};
\node[dna_red, right=0mm of z13] (z14r) {C\ldots G};
\node[dna_marker, right=0mm of z14r] (mz14) {AC};
\node[right=0mm of mz14, font=\fontfamily{phv}\selectfont] (zdots1r) {$\cdots$};
\node[dna_marker, right=0mm of zdots1r] (mz15) {AC};
\node[dna_barcode, right=0mm of mz15] (z1Lr) {TC\,\ldots \,GA};

\node[dna_orange, anchor=west] (z21) at (Q2) {G\ldots A};
\node[dna_marker, right=0mm of z21] (mz21) {AC};
\node[dna_gray, right=0mm of mz21] (z22) {G\ldots C};
\node[dna_marker, right=0mm of z22] (mz22) {AC};
\node[right=2mm of mz22, font=\fontfamily{phv}\selectfont] (zdots2a) {$\cdots$};
\node[dna_marker, right=2mm of zdots2a] (mz23) {AC};
\node[dna_gray, right=0mm of mz23] (z23) {A\ldots T};
\node[dna_red, right=0mm of z23] (z24r) {T\ldots G};
\node[dna_marker, right=0mm of z24r] (mz24) {AC};
\node[right=0mm of mz24, font=\fontfamily{phv}\selectfont] (zdots2r) {$\cdots$};
\node[dna_marker, right=0mm of zdots2r] (mz25) {AC};
\node[dna_barcode, right=0mm of mz25] (z2Lr) {CG\,\ldots \,AT};

\node[below=0mm of z21, font=\fontfamily{phv}\selectfont] (anc4) {$\vdots$};
\node[below=0mm of z22, font=\fontfamily{phv}\selectfont] {$\vdots$};
\node[below=4mm of zdots2a, font=\fontfamily{phv}\selectfont] {$\cdots$};
\node[below=0mm of z23, font=\fontfamily{phv}\selectfont] {$\vdots$};
\node[below=0mm of z24r, font=\fontfamily{phv}\selectfont] {$\vdots$};
\node[below=4mm of zdots2r, font=\fontfamily{phv}\selectfont] {$\cdots$};
\node[below=0mm of z2Lr, font=\fontfamily{phv}\selectfont] {$\vdots$};

\node[dna_orange, anchor=west] (zK1) at (QK2) {C\ldots G};
\node[dna_marker, right=0mm of zK1] (mzK1) {AC};
\node[dna_gray, right=0mm of mzK1] (zK2) {G\ldots T};
\node[dna_marker, right=0mm of zK2] (mzK2) {AC};
\node[right=2mm of mzK2, font=\fontfamily{phv}\selectfont] (zdotsKa) {$\cdots$};
\node[dna_marker, right=2mm of zdotsKa] (mzK3) {AC};
\node[dna_gray, right=0mm of mzK3] (zK3) {T\ldots G};
\node[dna_red, right=0mm of zK3] (zK4r) {T\ldots A};
\node[dna_marker, right=0mm of zK4r] (mzK4) {AC};
\node[right=0mm of mzK4, font=\fontfamily{phv}\selectfont] (zdotsKr) {$\cdots$};
\node[dna_marker, right=0mm of zdotsKr] (mzK5) {AC};
\node[dna_barcode, right=0mm of mzK5] (zKLr2) {GA\,\ldots \,TC};

\node[dna_orange, anchor=west] (zKp1) at (QKp2) {T\ldots T};
\node[dna_marker, right=0mm of zKp1] (mzKp1) {AC};
\node[dna_blue, right=0mm of mzKp1] (zKp2) {G\ldots A};
\node[dna_marker, right=0mm of zKp2] (mzKp2) {AC};
\node[right=2mm of mzKp2, font=\fontfamily{phv}\selectfont] (zdotsKpa) {$\cdots$};
\node[dna_marker, right=2mm of zdotsKpa] (mzKp3) {AC};
\node[dna_blue, right=0mm of mzKp3] (zKp3) {A\ldots T};
\node[dna_red, right=0mm of zKp3] (zKp4r) {G\ldots C};
\node[dna_marker, right=0mm of zKp4r] (mzKp4) {AC};
\node[right=0mm of mzKp4, font=\fontfamily{phv}\selectfont] (zdotsKpr) {$\cdots$};
\node[dna_marker, right=0mm of zdotsKpr] (mzKp5) {AC};
\node[dna_barcode, right=0mm of mzKp5] (zKpLr2) {AT\,\ldots \,CG};

\node[below=-2mm of zKp1, font=\fontfamily{phv}\selectfont] {$\vdots$};
\node[below=-2mm of zKp2, font=\fontfamily{phv}\selectfont] {$\vdots$};
\node[below=2mm of zdotsKpa, font=\fontfamily{phv}\selectfont] {$\cdots$};
\node[below=-2mm of zKp3, font=\fontfamily{phv}\selectfont] {$\vdots$};
\node[below=-2mm of zKp4r, font=\fontfamily{phv}\selectfont] {$\vdots$};
\node[below=2mm of zdotsKpr, font=\fontfamily{phv}\selectfont] {$\cdots$};
\node[below=-2mm of zKpLr2, font=\fontfamily{phv}\selectfont] {$\vdots$};

\node[dna_orange, anchor=west] (zN1) at (QN3) {G\ldots C};
\node[dna_marker, right=0mm of zN1] (mzN1) {AC};
\node[dna_blue, right=0mm of mzN1] (zN2d) {G\ldots T};
\node[dna_marker, right=0mm of zN2d] (mzN2) {AC};
\node[right=2mm of mzN2, font=\fontfamily{phv}\selectfont] (zdotsNa) {$\cdots$};
\node[dna_marker, right=2mm of zdotsNa] (mzN3) {AC};
\node[dna_blue, right=0mm of mzN3] (zN3d) {T\ldots T};
\node[dna_red, right=0mm of zN3d] (zN4r) {A\ldots G};
\node[dna_marker, right=0mm of zN4r] (mzN4) {AC};
\node[right=0mm of mzN4, font=\fontfamily{phv}\selectfont] (zdotsNr) {$\cdots$};
\node[dna_marker, right=0mm of zdotsNr] (mzN5) {AC};
\node[dna_barcode, right=0mm of mzN5] (zNLr3) {AC\,\ldots\,GC};

   \draw[draw=gray!90, densely dashed, thick, rounded corners=3pt] 
      (15.85,1.22) rectangle (25.82,-4.2);

      \node[above] at (20.86,1.15) {\fontfamily{phv}\selectfont\scriptsize\color{gray} (4.2) Inner MGC+ Code (DNA mapping \& intra-sequence quaternary redundancy)};

\draw [decorate, decoration={brace, amplitude=5pt}] 
        ($(mz11.north west)+(0,0.05)$) -- ($(mz11.north east)+(0,0.05)$) 
        node [midway, above=3.5pt, font=\fontfamily{phv}\selectfont\scriptsize] {marker};

        \draw [decorate, decoration={brace, amplitude=5pt}] 
        ($(mz12.north west)+(0,0.05)$) -- ($(mz12.north east)+(0,0.05)$) 
        node [midway, above=3.5pt, font=\fontfamily{phv}\selectfont\scriptsize] {marker};

\draw [decorate, decoration={brace, amplitude=5pt}] 
        ($(z14r.north west)+(0,0.05)$) -- ($(z14r.north east)+(0,0.05)$) 
        node [midway, above=3.5pt, font=\fontfamily{phv}\selectfont\scriptsize] {guess-par};

\draw [decorate, decoration={brace, amplitude=5pt}] 
        ($(z1Lr.north west)+(0,0.05)$) -- ($(z1Lr.north east)+(0,0.05)$) 
        node [midway, above=3.5pt, font=\fontfamily{phv}\selectfont\scriptsize] {\shortstack{barcoded \\ check-par}};

\draw[<->, black]
  ([yshift=1.0mm]z11.north west) -- ([yshift=1.0mm]z11.north east)
  node[midway, above, font=\fontfamily{phv}\selectfont\footnotesize] {$\ell_{\text{in}}$};

\draw[<->, black]
  ([yshift=1.0mm]z12.north west) -- ([yshift=1.0mm]z12.north east)
  node[midway, above, font=\fontfamily{phv}\selectfont\footnotesize] {$\ell_{\text{in}}$};

\draw[arrow, gray, very thick] ($(anc1)!0.5!(anc2)$) -- ++(0.5,0);
\draw[arrow, gray, very thick] ($(anc3)!0.5!(anc4)+(-0.15,0)$) -- ++(0.5,0);


\coordinate (LegendCenter) at ($(v11.west)!0.5!(z1Lr.east)$);
\coordinate (LegendBase)   at ($(LegendCenter)+(-0.7,2.4)$);

\def\LegendWidth{14cm}

\coordinate (L1) at ($(LegendBase)+(-9.5,0)$);
\coordinate (L2) at ($(LegendBase)+(-6,0)$);
\coordinate (L3) at ($(LegendBase)+( 0.9,0)$);
\coordinate (L4) at ($(LegendBase)+( 8,0)$);

\node[bitbox, anchor=west] (leg1) at (L1) {\phantom{0\ldots0}};
\node[anchor=west, font=\fontfamily{phv}\selectfont\scriptsize] at ([xshift=1mm]leg1.east)
  {payload};

\node[bitbox_outer, anchor=west] (leg2) at (L2) {\phantom{0\ldots0}};
\node[anchor=west, font=\fontfamily{phv}\selectfont\scriptsize] at ([xshift=1mm]leg2.east)
  {inter-sequence redundancy (outer code)};

\node[bitbox_red, anchor=west] (leg3) at (L3) {\phantom{0\ldots0}};
\node[anchor=west, font=\fontfamily{phv}\selectfont\scriptsize] at ([xshift=1mm]leg3.east)
  {intra-sequence redundancy (inner code)};

\node[bitbox_idx, anchor=west] (leg4) at (L4) {\phantom{0\ldots0}};
\node[anchor=west, font=\fontfamily{phv}\selectfont\scriptsize] at ([xshift=1mm]leg4.east)
  {index};
\end{tikzpicture}
    }
\caption{\textbf{Schematic illustration of the DNA-MGC+ encoding process.} (1)~The input data is partitioned into $K$ fragments, each of length $k$ bits. (2)~An outer Reed-Solomon code is applied across fragments to introduce inter-sequence redundancy, where each code symbol consists of $\ell_{\text{out}}$ bits, producing $c_{\text{out}}$ additional sequences. (3)~Each sequence is prepended with a unique binary index of length $\ell_{\text{out}}$ bits. (4.1)~In the first stage of the inner MGC+ code, the indexed sequences are encoded to introduce binary intra-sequence redundancy, using symbols of $\ell_{\text{in}}$ bits and generating $c_{\text{in}}$ \emph{guess} parities in addition to a single \emph{check} parity. (4.2)~In the second stage of the inner MGC+ code, the resulting binary sequences are mapped to quaternary DNA sequences, followed by the insertion of periodic ``$\mathsf{AC}$'' markers and further barcoding of the check parity.}
\label{fig:encoding_pipeline}
\end{figure}

\noindent The encoder therefore outputs a total of $N=K+c_{\text{out}}$ reference DNA sequences, where the length of each sequence is given by
\begin{equation} \label{eqL}
L_{\text{ref}}(k)=\frac{1}{2}\bigg(\overbrace{k}^{\text{ \shortstack{payload\\fragment}}}+\overbrace{\ell_{\text{out}}}^{\text{index}}+\overbrace{c_{\text{in}}\ell_{\text{in}}}^{\text{guess-pars}}\bigg)+2\bigg(\overbrace{\frac{k+c_{\text{in}}}{2\ell_{\text{in}}}+1}^{\text{num. of markers}}\bigg)+\overbrace{\beta}^{\text{\shortstack{check-par\\barcode}}}.
\end{equation}

In summary, the DNA-MGC+ encoder is fully specified by the parameter set ($k,\, \ell_{\text{in}},\, \ell_{\text{out}},\, c_{\text{in}},\, \rho_{\text{out}},\, d_{\text{min}},\, \beta, \, \text{marker mode}$), where \(k\) denotes the payload length of each fragment (in bits); $\ell_{\text{in}}$ and $\ell_{\text{out}}$  denote the symbol sizes (in bits)  of the inner and outer codes;  \(c_{\text{in}}\) is the number of guess parity symbols generated by the inner code; $\rho_{\text{out}}=K/(K+c_{\text{out}})$ is the rate of the outer code; $d_{\text{min}}$ and $\beta$ denote the minimum barcode edit distance and the corresponding barcode length used for encoding the check parity; and the marker mode specifies whether periodic DNA markers are inserted or not. 

The fragment payload length \(k\) is a design parameter that is 
selected so that the reference DNA sequence length $L_{\text{ref}}(k)$~(see equation~\eqref{eqL}) matches a target design length \(L_{\text{target}}\) as closely as possible. In general, the target length cannot always be matched exactly due to structural constraints imposed by the codec. For instance, since the inner and outer codes operate over symbols of length $\ell_{\text{in}}$ and $\ell_{\text{out}}$ bits, respectively, $k$ must be divisible by both $\ell_{\text{in}}$ and $\ell_{\text{out}}$. Moreover, the index length $\ell_{\text{out}}$ must also be divisible by $\ell_{\text{in}}$ since it is included in the inner code. In addition, when markers are used, the total number of inner code symbols~(excluding the check parity) must be even to ensure periodicity. The underlying RS codes also impose the standard field size constraints of $2^{\ell_\mathrm{in}}\geq (k+\ell_\mathrm{out})/\ell_{\text{in}} + c_{\text{in}} + 1$ and $2^{\ell_\mathrm{out}}\geq K + c_{\text{out}}$. The latter constraint is equivalent to $N \leq 2^{\ell_\mathrm{out}}$, thereby limiting the total number of encoded DNA sequences to at most $2^{\ell_\text{out}}$ for a given file. Consequently, either $\ell_\text{out}$ must be chosen sufficiently large to enable the generation of enough sequences within a single encoding instance, or the file must be partitioned into smaller independently encoded blocks, resulting in a multi-packet representation as discussed later in this section.

Taking the aforementioned constraints into account, the fragment length $k$ is selected as follows. Given a target design length \(L_{\text{target}}\) and fixed parameters ($\ell_{\text{in}},\, \ell_{\text{out}},\, c_{\text{in}},\, \rho_{\text{out}},\, d_{\text{min}},\, \beta, \, \text{marker mode}$), the encoder determines \(k^{\star}\) that minimizes \mbox{\(|L_{\text{ref}}(k) - L_{\text{target}}|\)} subject to \(L_{\text{ref}}(k) \le L_{\text{target}}\). Thus, $L_{\text{target}}$ serves as an upper bound on $L_{\text{ref}}$, and \(k^{\star}\) minimizes the gap between the two. If an exact match is required, the remaining gap $L_{\text{target}}-L_{\text{ref}}(k^{\star})$ is filled with random DNA symbols. The encoding steps illustrated in~Fig.~\ref{fig:encoding_pipeline} are applied once \(k^{\star}\) is determined. If the input file size $B$ is not divisible by $k^{\star}$, random bits are added so that the final payload fragment has length exactly $k^{\star}$. These padding bits are removed after decoding.

Throughout all results presented in this work, the following code parameters were fixed: \(\ell_\text{in}=8\), \(\ell_\text{out}=16\), $d_{\text{min}}=5$, and $\beta=12$. The different codec configurations of DNA-MGC+ were obtained by varying 
\(k\), \(c_{\text{in}}\), \(\rho_{\text{out}}\), and the marker mode, with the full parameter specifications for each configuration provided in Supplementary Table S1. The code rate (in bits/nt) associated with each configuration follows from the ratio $B/(N\,L_{\text{ref}})$, where $B$ is the input file size in bits, $N=K+c_{\text{out}}$ is the total number of encoded DNA sequences, and $L_{\text{ref}}$ is the length in nucleotides of each sequence. 

\paragraph{Filtering.} The DNA-MGC+ codec supports filtering of the encoded DNA sequences according to arbitrary content-specific constraints prior to synthesis and storage. These constraints can be specified independently of the coding structure and may include limits on homopolymer length, bounds on GC content, exclusion of specific sequence motifs, or other biochemical and thermodynamic requirements. This capability is enabled by the outer RS code, which allows the encoder to generate an excess pool of candidate sequences from which any subset can be selected to represent the file. In particular, by setting the outer redundancy parameter $c_{\text{out}}$ to its maximal value, the encoder can generate up to $N_{\text{max}}=2^{\ell_{\text{out}}}$ candidate sequences. These candidate sequences are then screened, and a subset of $N\leq N_{\text{max}}$ sequences satisfying the desired constraints is retained for synthesis and storage, while the remaining sequences are discarded. 


From a coding-theoretic perspective, the filtering step corresponds to a customized puncturing of the outer RS code, since certain codeword coordinates are deliberately omitted. In the unfiltered case, outer RS decoding is performed using the standard Berlekamp-Massey~\cite{1054260} algorithm. When filtering is applied, decoding is carried out using the Welch–Berlekamp~\cite{welch1986error} algorithm, which is more suitable for punctured RS codes. Filtering does not affect the code rate provided that enough admissible sequences are available among the $N_{\text{max}}$ candidates. The main cost of filtering lies in the increased decoding complexity, which is $O(N^{3})$ under Welch-Berlekamp, compared to $O(N^{2})$ under Berlekamp-Massey. 


\hl{Throughout this paper, all \emph{in silico} results are obtained using unfiltered encoding, whereas the \emph{in vitro} results reported in Fig.}~\ref{fig:wetlab} \hl{evaluate both unfiltered and filtered variants of DNA-MGC+}. 

\paragraph{Decoding.} 
The DNA-MGC+ decoder takes as input either noisy sequencing reads or, when applicable, consensus sequences obtained after clustering, and attempts to reconstruct the original binary file. A high-level description of the sequential decoding stages is provided below. 

\begin{enumerate}[label=(\arabic*), leftmargin=*]

\item {\em Inner MGC+ Code: } 
The inner MGC+ decoder processes each DNA sequence independently with the objective of correcting IDS errors and recovering the indexed binary fragments prior to outer decoding. Insertions and deletions induce symbol offsets, causing misalignment of the inner code symbol boundaries. The decoder therefore seeks to estimate an offset pattern, denoted by \(\boldsymbol{\delta}\), reflecting the boundary shifts along the sequence. Given an estimate of this pattern, symbols affected by insertions or deletions are treated as erasures, while synchronized symbols are retained. In this manner, correction of IDS errors is reduced to erasure-and-substitution decoding over the underlying RS code.

\begin{enumerate}[label=(\arabic{enumi}.\arabic*)]

\item {\em DNA-domain processing: } 
The barcoded check parity is first decoded via minimum-distance decoding applied to the last $\beta$ DNA symbols. If markers were introduced during encoding, an estimate of the offset pattern is then derived by solving a marker-centric maximum a posteriori (MAP) problem of the form $\hat{\boldsymbol{\delta}} = \arg\max_{\boldsymbol{\delta}} P(\boldsymbol{\delta}\mid \text{marker observations})$,
as detailed in Supplementary Note S1. The DNA symbols at the estimated marker locations are subsequently removed, and the remaining sequence is mapped to binary according to $\mathsf{A}\mapsto 00$, $\mathsf{T}\mapsto 01$, $\mathsf{C}\mapsto 10$, and $\mathsf{G}\mapsto 11$.

\item {\em Binary-domain processing: } 
If an estimate $\hat{\boldsymbol{\delta}}$ is available, the binary sequence is parsed accordingly. Symbols with nonzero estimated offsets are marked as erasures, while the remaining symbols are retained with their recovered values. All retained symbols, excluding the check parity, are then passed to the underlying RS decoder. The decoded fragment is subsequently validated using the check parity. If validation fails, a refined set of candidate offset patterns is generated and tested via a neighborhood search around \(\hat{\boldsymbol{\delta}}\), as described in Ref.~\cite{MGCP}. If none of the candidates yields a valid result, the sequence is discarded. In the absence of markers, no initial estimate \(\hat{\boldsymbol{\delta}}\) is available; candidate offset patterns are therefore generated by exhaustive guessing and tested in turn, resulting in increased decoding complexity, as discussed in Refs.~\cite{hanna2024GC,GCP}.

\end{enumerate}

\item {\em Outer RS decoding: }
Indices for which no valid fragment is recovered at the inner decoding stage result in erasures at the outer-code level, while incorrectly decoded fragments induce residual substitution errors that remain undetected. All recovered fragments are reordered according to their prepended indices, and outer RS decoding is then applied. Let \(E\) and \(S\) denote the numbers of erasures and substitution errors, respectively, among the $N$ outer-code symbols. Successful data retrieval with an exact match is guaranteed whenever $E + 2S \leq N-K$, which is equivalent to requiring that the number of correctly decoded fragments exceeds the number of incorrectly decoded fragments by at least \(K\).

\item {\em Defragmentation: } 
The recovered payload fragments are concatenated to reconstruct the original binary file.

\end{enumerate}

\paragraph{Multi-packet coding for large files.}
For large files requiring more than $2^{\ell_{\text{out}}}$ encoded sequences, and when increasing $\ell_{\text{out}}$ is not practical, DNA-MGC+ uses a multi-packet representation. The input file is first partitioned into smaller parts, each of which is encoded independently using the same encoding procedure described above. The encoding workflow therefore remains the same as in Fig.~\ref{fig:encoding_pipeline} at the level of each packet, except that each sequence is prepended with a packet index in addition to its within-packet sequence index. The packet index is added before inner encoding and is thus protected by the inner code.

If the total number of encoded sequences required for the file is $N$, the packet size can be limited to at most $N_P\leq2^{\ell_{\text{out}}}$ sequences, resulting in $P\approx N/N_P$ packets, where $N_P$ is a user-defined codec parameter. The packet-index length is chosen to be sufficient to uniquely identify all $P$ packets, requiring at least $\log_2P$ bits, while satisfying the previously described divisibility constraints imposed by the inner code. The additional packet-index bits are included in the encoded sequence length and are therefore accounted for in the computation of the fragment length $k$ and the effective code rate.

At the decoder, the recovered packet index is first used to assign each fragment to its corresponding packet. Outer RS decoding is then applied independently within each packet using the same procedure described previously. Consequently, the condition $E+2S\leq N_P-K_P$ must be satisfied independently for each packet, where $K_P$ denotes the number of information fragments in the corresponding packet. The successfully decoded packets are subsequently reordered and concatenated to reconstruct the original file.

\hl{The multi-packet representation also provides additional flexibility for constraint-based filtering}. A limitation of the filtering mechanism in the single-packet case is that, when the constraints are strict, the number of admissible sequences among the $N_{\text{max}}=2^{\ell_{\text{out}}}$ candidates, denoted by $N_{\text{adm}}$, may be smaller than the required number of encoded sequences. One possible workaround is to increase $\ell_{\text{out}}$, thereby increasing $N_{\text{max}}$. However, this also increases the encoded sequence length (see Fig.~\ref{fig:encoding_pipeline}), which may make some constraints harder to satisfy. A more effective solution is to leverage the multi-packet coding architecture and limit the size of each packet such that $N_P \leq N_{\text{adm}}$, with filtering performed independently within each packet. Since the number of packets scales as $P\approx N/N_P$, reducing $N_P$ increases the number of packets and the corresponding packet-indexing overhead. Nevertheless, this additional redundancy is typically modest, since the required packet-index length grows logarithmically with the number of packets.

\hl{A further motivation for multi-packet coding is to control the complexity of outer decoding for large files. When an entire file is represented by a single packet, the $O(N^2)$ complexity of unfiltered outer decoding, and especially the $O(N^3)$ complexity of filtered outer decoding, can become prohibitive as the number of encoded sequences $N$ increases. For a file represented by $P$ packets of size $N_P$, with $P\approx N/N_P$, the total outer-decoding complexity instead scales as $O(P\,N_P^2)$ in the unfiltered case and $O(P\,N_P^3)$ in the filtered case. If $N_P$ is capped at a manageable value for which the per-packet outer-decoding complexity remains practical, it can then be kept fixed as the file size increases through $P$. Consequently, the overall outer-decoding complexity grows linearly with file size, owing to the linear dependence on $P$ in $O(P\,N_P^2)$ and $O(P\,N_P^3)$, at the expense of an additional packet-indexing overhead as described above}.

In the scalability experiments of Fig.~\ref{silico11}, the packet-index length is set to $\ell_{\text{out}}=16$ bits and the packet size is capped at $N_P=2^{12}$ sequences, while all other results reported in this work use a single packet. \hl{The near-linear increase in decoding time observed in Fig.}~\ref{silico11}c \hl{is consistent with the scaling described above, since fixing $N_P$ makes the overall outer-decoding complexity grow linearly in the number of packets, and therefore in file size}.

\subsection*{Computational environment and software}

All simulations and decoding experiments were conducted on a machine equipped with an AMD Ryzen™ Threadripper™ PRO 7985WX CPU and 256\,GB RAM. The minimum required coverage (or sequencing) depth was determined via binary search over the interval [1, 32] with a resolution of 0.25, under the reliability constraint of 50-out-of-50 successful decodings with an exact file match. Reported decoding times correspond to the wall-clock time measured at the minimum coverage depth. Clustering, alignment, and consensus calling were excluded from decoding time measurements. All reported decoding times correspond to using a single CPU core, except for Fig.~\ref{fig:wetlab}b, where results using 8 parallel CPU cores are also included.

Read clustering was performed using CD-HIT~\cite{cd-hit} (v4.8.1) or CBR~\cite{rashtchian2017clustering}, and multiple sequence alignment was carried out using Kalign~\cite{kalign} (v3.4.0) or MUSCLE~\cite{muscle5} (v5.1), as specified for each experiment. For CD-HIT, the identity threshold and word size were set to 0.80 and 5, respectively, with all other parameters left at their default values. For CBR, we used the implementation provided as part of the DNAStorageToolkit~\cite{dna_storage_toolkit}, modifying only the clustering diameter parameter to 50. To simulate the different codec configurations of DNA-Aeon, HEDGES, DNA-Fountain, and DNA-RS, we used the dt4dds-benchmark~\cite{gimpel2025dt4ddsbenchmark, gimpel2025comparison} (v1.0.0) benchmarking suite. For DNA-Stairloop~\cite{yan2025dna} and the LDPC-based scheme~\cite{chandak2019improved}, we used the respective GitHub repositories referenced in those works. For DNA-MGC+, we used our Python implementation, referenced in the Code Availability statement, requiring Python 3.9 or later and relying only on numpy (v1.21), scipy (v1.7), reedsolo (v1.5), and galois (v0.0.3). 

The channel model used in the first {\em in silico} setting was implemented as part of this work and is included in the provided code. The second {\em in silico} setting is based on the DT4DDS~\cite{gimpel2023digital} digital twin framework and corresponds to the low-fidelity workflow described in Ref.~\cite{gimpel2025comparison}. We used the implementation of this workflow provided in dt4dds-benchmark~\cite{gimpel2025dt4ddsbenchmark} (v1.0.0) as a black-box simulator to generate sequencing reads under specified physical redundancy and sequencing depth. For the wet-lab experiment, the primer sequences were generated using the DSP Tools software~\cite{DSPTOOLS}. The folding Gibbs free energy ($\Delta G$) was computed using the NUPACK Python API\cite{Nupack,NupackAPI} (v4.0.1.8) with default parameters at $55^{\circ}\mathrm{C}$. Nanopore basecalling was performed with Dorado~\cite{dorado2025} (v1.3.0, simplex mode) using the R10.4.1 E8.2 400 bps {\em fast}, {\em hac}, and {\em sup} models (v4.3.0), with a minimum Q-score threshold set to 7. Paired-end Illumina reads were merged using FLASH~\cite{flash}~(v1.2.11). Primer trimming and payload extraction from raw sequencing reads were performed using Cutadapt~\cite{cutadapt}~(v5.2), with the maximum mismatch rate set to 15\%. The base-level error rates reported for the {\em in vitro} results were computed by aligning the reads to their corresponding reference sequences using minimap2~\cite{li2018minimap2} (v2.30).


\subsection*{In vitro experimental workflow}

\paragraph{Stored file and codec configurations.}
A compressed 24~KB text file (zlib-compressed Universal Declaration of Human Rights in four languages) was encoded using four configurations of DNA-MGC+ and one configuration each of DNA-Aeon and HEDGES. For DNA-MGC+, two parameter sets were considered, referred to as Design~$\mathrm{A}$ and Design~$\mathrm{B}$. For each design, both unfiltered and filtered variants were evaluated. The corresponding number of reference sequences, payload length, and code rate are given in Table~\ref{tab:encoding_summary}. Full parameter values for all evaluated codecs are provided in Supplementary Tables S1--3.

\begin{table}[h!]
\centering
\caption{\textbf{Encoding characteristics for the considered codecs when applied to the 24-KB compressed text file.}}
\label{tab:encoding_summary}
\begin{tabular}{l c c c c}
\hline
\textbf{Codec} & \textbf{ Number of sequences} & \textbf{Payload length (nts)} & \textbf{Code rate (bits/nt)} & \textbf{Tag} \\
\hline
DNA-MGC+ (Design~$\mathrm{A}$, filtered) & 1532 & 124 & 1.03 & $\mathsf{GGAT}$ \\
DNA-MGC+ (Design~$\mathrm{A}$, unfiltered) & 1532 & 124 & 1.03 & $\mathsf{AGTG}$\\
DNA-MGC+ (Design~$\mathrm{B}$, filtered) & 2277 & 122 & 0.71 & $\mathsf{CAAG}$\\
DNA-MGC+ (Design~$\mathrm{B}$, unfiltered) & 2277 & 122 & 0.71 & $\mathsf{GACA}$\\
DNA-Aeon                       & 1619 & 120 & 1.00 & $\mathsf{CTGT}$ \\
HEDGES                         & 2550 & 126 & 0.61 & $\mathsf{TAGC}$\\
\hline
\end{tabular}
\end{table}

\paragraph{Oligonucleotide design and synthesis.}
For each codec configuration, the encoded payload sequences were right-padded with random nucleotides to obtain a common length of 126 nts. A configuration-specific 4-nt tag was then appended to the right of each sequence (see Table~\ref{tab:encoding_summary}), followed by the addition of a 20-nt forward primer $\mathsf{AGCGTGCGTTACTTAGATAC}$ and a 20-nt reverse primer $\mathsf{TCACCGTATTGCGTAGTATG}$. The resulting sequences had a total length of 170 nt, matching the maximum length supported by the synthesis provider GenScript. A single oligo pool containing all six configurations was synthesized.

\paragraph{Constraint-based filtering.} For the filtered DNA-MGC+ variants, an excess pool of candidate sequences was first generated using the codec. The encoded payloads were then formatted to match the 170-nt oligonucleotide structure described above, including the configuration-specific tags and primers. The sequences were subsequently screened and retained only if they satisfied the following constraints: maximum homopolymer length $\leq 4$, GC content between 45\% and 55\%, and absence of di- and trinucleotide repeat motifs with four or more repetitions. For the chosen value of $\ell_{\text{out}}=16$, the DNA-MGC+ codec allows the generation of up to $2^{\ell_{\text{out}}}=65536$ candidate sequences, of which $35815$ ($54.65\%$) and $42349$ ($64.62\%$) satisfied all three constraints for designs A and B, respectively. For the retained sequences, the folding Gibbs free energy ($\Delta G$, in kcal/mol) was computed at~$55^{\circ}\mathrm{C}$, corresponding to the primer annealing temperature ($T_a$). The $1532$ and $2277$ sequences for designs A and B, respectively, whose $\Delta G$ values were closest to zero were then preferentially selected. Histograms of the maximum homopolymer length, GC content, and $\Delta G$ for the encoded sequences of each of the six codec configurations are shown in Supplementary Fig. S7.

\paragraph{PCR, library preparation, and sequencing.} 
The oligo pool from GenScript, received dry, was resuspended to a concentration of 4.5~ng~$\mu$L$^{-1}$ using nuclease-free water. For PCR amplification, 2~$\mu$L of the resuspended oligo pool (4.5~ng~$\mu$L$^{-1}$) was mixed with 25~$\mu$L KAPA HiFi HotStart ReadyMix~2$\times$ (Roche), 1~$\mu$L of 10~$\mu$M forward primer, 1~$\mu$L of 10~$\mu$M reverse primer, and 21~$\mu$L nuclease-free water, yielding a final reaction volume of 50~$\mu$L. The reaction was prepared in three separate tubes. Thermocycling followed established protocols, consisting of an initial denaturation at $95^{\circ}$C for 5~min, followed by 30 cycles of denaturation at $95^{\circ}$C for 20~s, annealing at $55^{\circ}$C for 30~s, and extension at $72^{\circ}$C for 20~s. PCR products were subsequently purified using SPRI magnetic beads (Beckman Coulter) and analyzed on a BioAnalyzer system using the High Sensitivity DNA kit and reagents (Agilent Technologies).

Libraries for Oxford Nanopore sequencing were prepared using the SQK-LSK114 ligation sequencing kit (Oxford Nanopore Technologies). Nanopore sequencing was performed on a PromethION platform using R10.4.1 PromethION flow cells. Libraries for Illumina sequencing were prepared using the NEBNext UltraExpress\textsuperscript{\textregistered} DNA Library Prep Kit (New England Biolabs). Illumina sequencing was performed on a NextSeq 2000 system (Illumina) using a P2 flow cell and a 600-cycle kit to generate paired-end $2\times300$~bp reads. Coverage distributions derived from the Illumina and Nanopore sequencing data for each of the six codec configurations are shown in Supplementary Fig.~S8.

\paragraph{Read processing and decoding.}
Raw sequencing reads were first processed to locate and trim the primer sequences and to extract the payload in between. For Illumina data, paired-end reads were merged prior to primer trimming, whereas Nanopore reads were processed directly. Reads with an extracted payload length within 10 nucleotides of the design length of 130 nts, that is between 120 and 140 nts, were retained for further analysis. The retained payload sequences were then demultiplexed into six separate files according to their configuration-specific 4-nt tags listed in Table~\ref{tab:encoding_summary}. Specifically, a read was assigned to a codec configuration if the last four nucleotides of its extracted payload exactly matched the corresponding tag. \hl{Of the raw reads, 52.65\% were retained after merging, 43.40\% after primer trimming, 43.31\% after length filtering, and 42.11\% after demultiplexing for Illumina data, while for Nanopore data (dorado-sup) 41.28\%, 36.09\%, and 33.21\% of raw reads were retained after primer trimming, length filtering, and demultiplexing, respectively}.

For each demultiplexed set, clustering, alignment, and consensus calling were performed. The resulting consensus sequences were directly provided to the corresponding decoder of each codec without additional processing. To determine the minimum sequencing depth required for reliable decoding, progressive downsampling was applied to the demultiplexed reads prior to clustering. For each codec configuration and each tested depth value, a number of reads equal to the product of the depth and the corresponding number of reference sequences was randomly sampled. Clustering, alignment, consensus calling, and decoding were then repeated at each depth value until the minimum depth enabling reliable decoding was identified.





\section*{Data availability}
The reference sequences and the corresponding FASTQ sequencing data generated in the \emph{in vitro} experiments are available on Zenodo~\cite{khabbaz2026dnamgcdata} at \url{https://doi.org/10.5281/zenodo.18848901}.


\section*{Code availability}
The source code of DNA-MGC+ is publicly available on GitHub at \url{https://github.com/ramy-khabbaz/MGCP}. An archived version of the code corresponding to this study is available on Zenodo~\cite{khabbaz2026mgcpcode} at \url{https://doi.org/10.5281/zenodo.22169494}.


\bibliography{Refs} 

@article{welzel2023dna,
  title={{DNA-Aeon} provides flexible arithmetic coding for constraint adherence and error correction in {DNA} storage},
  author={Welzel, Marius and Schwarz, Peter Michael and L{\"o}chel, Hannah F and Kabdullayeva, Tolganay and Clemens, Sandra and Becker, Anke and Freisleben, Bernd and Heider, Dominik},
  journal={Nature Communications},
  volume={14},
  number={1},
  pages={628},
  year={2023},
  publisher={Nature Publishing Group UK London}
}

@article{press2020hedges,
  title={HEDGES error-correcting code for {DNA} storage corrects indels and allows sequence constraints},
  author={Press, William H and Hawkins, John A and Jones Jr, Stephen K and Schaub, Jeffrey M and Finkelstein, Ilya J},
  journal={Proceedings of the National Academy of Sciences},
  volume={117},
  number={31},
  pages={18489--18496},
  year={2020},
  publisher={National Academy of Sciences}
}

@article{erlich2017dna,
  title={{DNA} Fountain enables a robust and efficient storage architecture},
  author={Erlich, Yaniv and Zielinski, Dina},
  journal={science},
  volume={355},
  number={6328},
  pages={950--954},
  year={2017},
  publisher={American Association for the Advancement of Science}
}

@article{antkowiak2020low,
  title={Low cost {DNA} data storage using photolithographic synthesis and advanced information reconstruction and error correction},
  author={Antkowiak, Philipp L and Lietard, Jory and Darestani, Mohammad Zalbagi and Somoza, Mark M and Stark, Wendelin J and Heckel, Reinhard and Grass, Robert N},
  journal={Nature communications},
  volume={11},
  number={1},
  pages={5345},
  year={2020},
  publisher={Nature Publishing Group UK London}
}

@article{meiser2020reading,
  title={Reading and writing digital data in {DNA}},
  author={Meiser, Linda C and Antkowiak, Philipp L and Koch, Julian and Chen, Weida D and Kohll, A Xavier and Stark, Wendelin J and Heckel, Reinhard and Grass, Robert N},
  journal={Nature protocols},
  volume={15},
  number={1},
  pages={86--101},
  year={2020},
  publisher={Nature Publishing Group UK London}
}

@article{yan2025dna,
  title={{DNA} StairLoop: enabling high-fidelity data recovery and robust error correction in {DNA}-based data storage},
  author={Yan, Zihui and Qu, Guanjin and Chen, Xin and Zheng, Gang and Wu, Huaming},
  journal={Nature Communications},
  volume={16},
  number={1},
  pages={9191},
  year={2025},
  publisher={Nature Publishing Group UK London}
}

@inproceedings{chandak2019improved,
  title={Improved read/write cost tradeoff in {DNA}-based data storage using {LDPC} codes},
  author={Chandak, Shubham and Tatwawadi, Kedar and Lau, Billy and Mardia, Jay and Kubit, Matthew and Neu, Joachim and Griffin, Peter and Wootters, Mary and Weissman, Tsachy and Ji, Hanlee},
  booktitle={2019 57th Annual Allerton Conference on Communication, Control, and Computing (Allerton)},
  pages={147--156},
  year={2019},
  organization={IEEE}
}

@article{reed1960polynomial,
  title={Polynomial codes over certain finite fields},
  author={Reed, Irving S and Solomon, Gustave},
  journal={Journal of the society for industrial and applied mathematics},
  volume={8},
  number={2},
  pages={300--304},
  year={1960},
  publisher={SIAM}
}

@INPROCEEDINGS{MGCP,
  author={Khabbaz, Ramy and Antonini, Marc and {Kas Hanna}, Serge},
  booktitle={2025 13th International Symposium on Topics in Coding (ISTC)}, 
  title={{Marker Guess \& Check Plus} ({MGC+}): An Efficient Short Blocklength Code for Random Edit Errors}, 
  year={2025},
  volume={},
  number={},
  pages={1-5},
  doi={10.1109/ISTC65386.2025.11154528}}

@article{cd-hit,
  title={{CD-HIT}: accelerated for clustering the next-generation sequencing data},
  author={Fu, Limin and Niu, Beifang and Zhu, Zhengwei and Wu, Sitao and Li, Weizhong},
  journal={Bioinformatics},
  volume={28},
  number={23},
  pages={3150--3152},
  year={2012},
  publisher={Oxford University Press}
}

@article{kalign,
  title={Kalign--an accurate and fast multiple sequence alignment algorithm},
  author={Lassmann, Timo and Sonnhammer, Erik LL},
  journal={BMC bioinformatics},
  volume={6},
  number={1},
  pages={298},
  year={2005},
  publisher={Springer}
}

@article{gimpel2023digital,
  title={A digital twin for {DNA} data storage based on comprehensive quantification of errors and biases},
  author={Gimpel, Andreas L and Stark, Wendelin J and Heckel, Reinhard and Grass, Robert N},
  journal={Nature Communications},
  volume={14},
  number={1},
  pages={6026},
  year={2023},
  publisher={Nature Publishing Group UK London}
}

@article{chen2020quantifying,
  title={Quantifying molecular bias in {DNA} data storage},
  author={Chen, Yuan-Jyue and Takahashi, Christopher N and Organick, Lee and Bee, Callista and Ang, Siena Dumas and Weiss, Patrick and Peck, Bill and Seelig, Georg and Ceze, Luis and Strauss, Karin},
  journal={Nature communications},
  volume={11},
  number={1},
  pages={3264},
  year={2020},
  publisher={Nature Publishing Group UK London}
}

@article{rashtchian2017clustering,
  title={Clustering billions of reads for {DNA} data storage},
  author={Rashtchian, Cyrus and Makarychev, Konstantin and Racz, Miklos and Ang, Siena and Jevdjic, Djordje and Yekhanin, Sergey and Ceze, Luis and Strauss, Karin},
  journal={Advances in Neural Information Processing Systems},
  volume={30},
  year={2017}
}

@inproceedings{hanna2025reliability,
  author = {Kas Hanna, Serge},
  title = {On the Reliability of Information Retrieval From {MDS} Coded Data in {DNA} Storage},
  booktitle    = {2025 IEEE International Symposium on Information Theory (ISIT)},
  year         = {2025},
  note         = {Extended version available at: \url{https://arxiv.org/abs/2502.06618}}
}

@ARTICLE{bar-lev25,
  author={Bar-Lev, Daniella and Sabary, Omer and Gabrys, Ryan and Yaakobi, Eitan},
  journal={IEEE Transactions on Information Theory}, 
  title={Cover Your Bases: How to Minimize the Sequencing Coverage in {DNA} Storage Systems}, 
  year={2025},
  volume={71},
  number={1},
  pages={192-218},
  doi={10.1109/TIT.2024.3496587}}

@ARTICLE{11164904,
  author={Weindel, Franziska and Gimpel, Andreas L. and Grass, Robert N. and Heckel, Reinhard},
  journal={IEEE Transactions on Molecular, Biological, and Multi-Scale Communications}, 
  title={Embracing Errors Can Be More Efficient Than Avoiding Them Through Constrained Coding for {DNA} Data Storage}, 
  year={2026},
  volume={12},
  number={},
  pages={146-156},
  doi={10.1109/TMBMC.2025.3610330}}

@software{gimpel2025dt4ddsbenchmark,
  title        = {dt4dds-benchmark: Benchmarking suite for {DNA} data storage},
  author       = {Gimpel, Andreas L. and others},
   howpublished = {{\em GitHub:} \url{https://github.com/fml-ethz/dt4dds-benchmark}},
  year         = {2025}
}

@inproceedings{dna_storage_toolkit,
  title={{DNA} {Storage} {Toolkit}: A Modular End-to-End DNA Data Storage Codec and Simulator},
  author={Sharma, Puru and Goh Yipeng, Gary and Gao, Bin and Ou, Longshen and Lin, Dehui and Sharma, Deepak and Jevdjic, Djordje},
  booktitle={2024 IEEE International Symposium on Performance Analysis of Systems and Software (ISPASS)},
  year={2024},
  abbr={ISPASS'24},
  organization={IEEE},
note = {{\em GitHub:} \url{https://github.com/prongs1996/DNAStorageToolkit}},
}

@article{muscle5,
  title={Muscle5: High-accuracy alignment ensembles enable unbiased assessments of sequence homology and phylogeny},
  author={Edgar, Robert C},
  journal={Nature communications},
  volume={13},
  number={1},
  pages={6968},
  year={2022},
  publisher={Nature Publishing Group UK London}
}

@article{cutadapt,
  title={Cutadapt removes adapter sequences from high-throughput sequencing reads},
  author={Martin, Marcel},
  journal={EMBnet. journal},
  volume={17},
  number={1},
  pages={10--12},
  year={2011}
}

@ARTICLE{1054260,
  author={Massey, J.},
  journal={IEEE Transactions on Information Theory}, 
  title={Shift-register synthesis and BCH decoding}, 
  year={1969},
  volume={15},
  number={1},
  pages={122-127},
  doi={10.1109/TIT.1969.1054260}}

@patent{welch1986error,
  author       = {Lloyd R. Welch and Elwyn R. Berlekamp},
  title        = {Error Correction for Algebraic Block Codes},
  number       = {US4633470A},
  year         = {1986},
  month        = dec,
  day          = {30},
  note         = {U.S. Patent 4,633,470}
}

@article{li2018minimap2,
  title={Minimap2: pairwise alignment for nucleotide sequences},
  author={Li, Heng},
  journal={Bioinformatics},
  volume={34},
  number={18},
  pages={3094--3100},
  year={2018},
  publisher={Oxford University Press}
}

@misc{dorado2025,
  author       = {{Oxford Nanopore Technologies}},
  title        = {Dorado basecalling software for Oxford Nanopore sequencing data},
  year         = {2025},
  howpublished = {{\em GitHub:} \url{https://github.com/nanoporetech/dorado}}
}

@article{flash,
  title={FLASH: fast length adjustment of short reads to improve genome assemblies},
  author={Mago{\v{c}}, Tanja and Salzberg, Steven L},
  journal={Bioinformatics},
  volume={27},
  number={21},
  pages={2957--2963},
  year={2011},
  publisher={Oxford University Press}
}

@inproceedings{DSPTOOLS,
  TITLE = {{Primer design for DNA storage random access}},
  AUTHOR = {Mateos, Jeremy and Lavenier, Dominique and Dimopoulou, Melpomeni and Genot, Anthony and Antonini, Marc},
  URL = {https://hal.science/hal-04739300},
  BOOKTITLE = {{CORESA  2024 - 23{\`e}me conf{\'e}rence sur COmpression et REpr{\'e}sentation des Signaux Audiovisuels}},
  ADDRESS = {Rennes, France},
  PAGES = {1-3},
  YEAR = {2024},
  MONTH = Nov,
  HAL_ID = {hal-04739300},
  HAL_VERSION = {v1},
}

@article{rydning2022worldwide,
  title={Worldwide IDC Global datasphere forecast, 2022--2026: enterprise organizations driving most of the data growth},
  author={Rydning, John},
  journal={International Data Corporation (IDC)},
  year={2022}
}

@article{church2012,
  title={Next-generation digital information storage in {DNA}},
  author={Church, George M and Gao, Yuan and Kosuri, Sriram},
  journal={Science},
  volume={337},
  number={6102},
  pages={1628--1628},
  year={2012},
  publisher={American Association for the Advancement of Science}
}

@article{grass2015robust,
  title={Robust chemical preservation of digital information on DNA in silica with error-correcting codes},
  author={Grass, Robert N and Heckel, Reinhard and Puddu, Michela and Paunescu, Daniela and Stark, Wendelin J},
  journal={Angewandte Chemie International Edition},
  volume={54},
  number={8},
  pages={2552--2555},
  year={2015},
  publisher={Wiley Online Library}
}

@article{heckel2019characterization,
  title={A characterization of the DNA data storage channel},
  author={Heckel, Reinhard and Mikutis, Gediminas and Grass, Robert N},
  journal={Scientific reports},
  volume={9},
  number={1},
  pages={9663},
  year={2019},
  publisher={Nature Publishing Group UK London}
}

@article{milenkovic2024dna,
  title={{DNA}-based data storage systems: A review of implementations and code constructions},
  author={Milenkovic, Olgica and Pan, Chao},
  journal={IEEE Transactions on Communications},
  volume={72},
  number={7},
  pages={3803--3828},
  year={2024},
  publisher={IEEE}
}

@article{sabary2024survey_coding,
  title={Survey for a decade of coding for {DNA} storage},
  author={Sabary, Omer and Kiah, Han Mao and Siegel, Paul H and Yaakobi, Eitan},
  journal={IEEE Transactions on Molecular, Biological, and Multi-Scale Communications},
  volume={10},
  number={2},
  pages={253--271},
  year={2024},
  publisher={IEEE}
}

@article{shomorony2022information,
  title={Information-theoretic foundations of {DNA} data storage},
  author={Shomorony, Ilan and Heckel, Reinhard and others},
  journal={Foundations and Trends in Communications and Information Theory},
  volume={19},
  number={1},
  pages={1--106},
  year={2022},
  publisher={Now Publishers, Inc.}
}

@article{goldman2013towards,
  title={Towards practical, high-capacity, low-maintenance information storage in synthesized {DNA}},
  author={Goldman, Nick and Bertone, Paul and Chen, Siyuan and Dessimoz, Christophe and LeProust, Emily M and Sipos, Botond and Birney, Ewan},
  journal={nature},
  volume={494},
  number={7435},
  pages={77--80},
  year={2013},
  publisher={Nature Publishing Group UK London}
}

@inproceedings{bornholt2016dna,
  title={A {DNA}-based archival storage system},
  author={Bornholt, James and Lopez, Randolph and Carmean, Douglas M and Ceze, Luis and Seelig, Georg and Strauss, Karin},
  booktitle={Proceedings of the twenty-first international conference on architectural support for programming languages and operating systems},
  pages={637--649},
  year={2016}
}

@article{blawat2016forward,
  title={Forward error correction for {DNA} data storage},
  author={Blawat, Meinolf and Gaedke, Klaus and Huetter, Ingo and Chen, Xiao-Ming and Turczyk, Brian and Inverso, Samuel and Pruitt, Benjamin W and Church, George M},
  journal={Procedia Computer Science},
  volume={80},
  pages={1011--1022},
  year={2016},
  publisher={Elsevier}
}

@article{organick2018random,
  title={Random access in large-scale {DNA} data storage},
  author={Organick, Lee and Ang, Siena Dumas and Chen, Yuan-Jyue and Lopez, Randolph and Yekhanin, Sergey and Makarychev, Konstantin and Racz, Miklos Z and Kamath, Govinda and Gopalan, Parikshit and Nguyen, Bichlien and others},
  journal={Nature biotechnology},
  volume={36},
  number={3},
  pages={242--248},
  year={2018},
  publisher={Nature Publishing Group}
}

@inproceedings{appuswamy2019oligoarchive,
  title={OligoArchive: Using {DNA} in the {DBMS} storage hierarchy},
  author={Appuswamy, Raja and Lebrigand, Kevin and Barbry, Pascal and Antonini, Marc and Madderson, Oliver and Freemont, Paul and MacDonald, James and Heinis, Thomas},
  booktitle={Biennal Conference on Innovative Data Systems Research (CIDR 2019)},
  pages={p98},
  year={2019}
}

@inproceedings{jpegdnaSPIE2025,
    address = {San Diego, United States},
    title = {Overview of {JPEG DNA} coding system for image storage on synthetic {DNA}},
    booktitle = {Applications of {Digital} {Image} {Processing} {XLVIII}},
    publisher = {SPIE},
    author = {Lazzarotto, Davi and Testolina, Michela and Kas Hanna, Serge and Watanabe, Osamu and Gil San Antonio, Eva and Pic, Xavier and Dimopoulou, Melpomeni and Antonini, Marc and Ebrahimi, Touradj},
    year = {2025}
}

@article{song2022robust,
  title={Robust data storage in {DNA} by de Bruijn graph-based de novo strand assembly},
  author={Song, Lifu and Geng, Feng and Gong, Zi-Yi and Chen, Xin and Tang, Jijun and Gong, Chunye and Zhou, Libang and Xia, Rui and Han, Ming-Zhe and Xu, Jing-Yi and others},
  journal={Nature communications},
  volume={13},
  number={1},
  pages={5361},
  year={2022},
  publisher={Nature Publishing Group UK London}
}

@article{tabatabaei2015rewritable,
  title={A rewritable, random-access {DNA}-based storage system},
  author={Tabatabaei Yazdi, SM Hossein and Yuan, Yongbo and Ma, Jian and Zhao, Huimin and Milenkovic, Olgica},
  journal={Scientific reports},
  volume={5},
  number={1},
  pages={14138},
  year={2015},
  publisher={Nature Publishing Group UK London}
}

@article{yazdi2017portable,
  title={Portable and error-free {DNA}-based data storage},
  author={Yazdi, SM Hossein Tabatabaei and Gabrys, Ryan and Milenkovic, Olgica},
  journal={Scientific reports},
  volume={7},
  number={1},
  pages={5011},
  year={2017},
  publisher={Nature Publishing Group UK London}
}

@article{ping2022towards,
  title={Towards practical and robust {DNA}-based data archiving using the yin--yang codec system},
  author={Ping, Zhi and Chen, Shihong and Zhou, Guangyu and Huang, Xiaoluo and Zhu, Sha Joe and Zhang, Haoling and Lee, Henry H and Lan, Zhaojun and Cui, Jie and Chen, Tai and others},
  journal={Nature Computational Science},
  volume={2},
  number={4},
  pages={234--242},
  year={2022},
  publisher={Nature Publishing Group US New York}
}

@article{anavy2019data,
  title={Data storage in {DNA} with fewer synthesis cycles using composite DNA letters},
  author={Anavy, Leon and Vaknin, Inbal and Atar, Orna and Amit, Roee and Yakhini, Zohar},
  journal={Nature biotechnology},
  volume={37},
  number={10},
  pages={1229--1236},
  year={2019},
  publisher={Nature Publishing Group US New York}
}

@article{choi2019high,
  title={High information capacity {DNA}-based data storage with augmented encoding characters using degenerate bases},
  author={Choi, Yeongjae and Ryu, Taehoon and Lee, Amos C and Choi, Hansol and Lee, Hansaem and Park, Jaejun and Song, Suk-Heung and Kim, Seojoo and Kim, Hyeli and Park, Wook and others},
  journal={Scientific reports},
  volume={9},
  number={1},
  pages={6582},
  year={2019},
  publisher={Nature Publishing Group UK London}
}

@article{zhao2024composite,
  title={Composite Hedges Nanopores codec system for rapid and portable {DNA} data readout with high INDEL-Correction},
  author={Zhao, Xuyang and Li, Junyao and Fan, Qingyuan and Dai, Jing and Long, Yanping and Liu, Ronghui and Zhai, Jixian and Pan, Qing and Li, Yi},
  journal={Nature Communications},
  volume={15},
  number={1},
  pages={9395},
  year={2024},
  publisher={Nature Publishing Group UK London}
}

@article{NupackAPI,
  title = {A Unified Dynamic Programming Framework for the Analysis of Interacting Nucleic Acid Strands: Enhanced Models, Scalability, and Speed},
  author = {Fornace, Michael E. and Porubsky, Nicholas J. and Pierce, Niles A.},
  journal = {ACS Synthetic Biology},
  year = {2020},
  volume = {9},
  number = {10},
  pages = {2665--2678},
  doi = {10.1021/acssynbio.9b00523}
}

@article{Nupack,
  title = {Thermodynamic Analysis of Interacting Nucleic Acid Strands},
  author = {Dirks, Robert M. and Bois, Justin S. and Schaeffer, Joseph M. and Winfree, Erik and Pierce, Niles A.},
  journal = {SIAM Review},
  year = {2007},
  volume = {49},
  number = {1},
  pages = {65--88},
  doi = {10.1137/060651100}
}

@INPROCEEDINGS{hanna2024GC,
  author={Kas Hanna, Serge},
  booktitle={2024 IEEE International Symposium on Information Theory (ISIT)}, 
  title={Short Systematic Codes for Correcting Random Edit Errors in {DNA} Storage}, 
  year={2024},
  volume={},
  number={},
  pages={663-668},
  doi={10.1109/ISIT57864.2024.10619614},
}

@article{GCP,
title={{GC+} Code: A Systematic Short Blocklength Code for Correcting Random Edit Errors in {DNA} Storage}, 
  author={Serge {Kas Hanna}},
  journal = {arXiv preprint, arXiv:2402.01244},
  year    = {2025}
}

@article{gimpel2025comparison,
  title={Comparison of state-of-the-art error-correction coding for sequence-based {DNA} data storage},
  author={Gimpel, Andreas L and Remschak, Alex and Stark, Wendelin J and Heckel, Reinhard and Grass, Robert N},
  journal={bioRxiv},
  pages={2025--07},
  year={2025},
  publisher={Cold Spring Harbor Laboratory}
}

@article{zhang2026gungnir,
  title={Gungnir codec enabling high error-tolerance and low-redundancy DNA storage through substantial computing power},
  author={Zhang, Jingcheng and Chen, Lei and Sun, Jinlin and Li, Shumin and Zhou, Yekai and Wu, Zhenqin and Li, Can and Zheng, Zhenxian and Luo, Ruibang},
  journal={Nature Communications},
  year={2026},
  publisher={Nature Publishing Group UK London}
}

@article{bar2025scalable,
  title={Scalable and robust DNA-based storage via coding theory and deep learning},
  author={Bar-Lev, Daniella and Orr, Itai and Sabary, Omer and Etzion, Tuvi and Yaakobi, Eitan},
  journal={Nature Machine Intelligence},
  volume={7},
  number={4},
  pages={639--649},
  year={2025},
  publisher={Nature Publishing Group UK London}
}

@article{aharoni2025neural,
  title={Neural Polar Decoders for DNA Data Storage},
  author={Aharoni, Ziv and Pfister, Henry D},
  journal={IEEE Journal on Selected Areas in Information Theory},
  year={2025},
  publisher={IEEE}
}

@article{welter2026end,
  title={An end-to-end coding scheme for DNA-based data storage with nanopore-sequenced reads},
  author={Welter, Lorenz and Sokolovskii, Roman and Heinis, Thomas and Wachter-Zeh, Antonia and Rosnes, Eirik and i Amat, Alexandre Graell},
  journal={IEEE Journal on Selected Areas in Information Theory},
  year={2026},
  publisher={IEEE}
}

@article{lietard2021chemical,
  title={Chemical and photochemical error rates in light-directed synthesis of complex DNA libraries},
  author={Lietard, Jory and Leger, Adrien and Erlich, Yaniv and Sadowski, Norah and Timp, Winston and Somoza, Mark M},
  journal={Nucleic acids research},
  volume={49},
  number={12},
  pages={6687--6701},
  year={2021},
  publisher={Oxford University Press}
}

@article{gaspar2018ngmerge,
  title={NGmerge: merging paired-end reads via novel empirically-derived models of sequencing errors},
  author={Gaspar, John M},
  journal={BMC bioinformatics},
  volume={19},
  number={1},
  pages={536},
  year={2018},
  publisher={Springer}
}

@inproceedings{shao2025hsfc,
  title={{DNA} Sequence Clustering in High Error Rates via Hash Sketches Fuzzy Clustering for Efficient Stored Data Reconstruction},
  author={Shao, Qi and Zheng, Yanfen and Cao, Ben and Liu, Zhenlu and Wang, Bin and Zhou, Shihua and Zheng, Pan},
  booktitle={Pacific-Asia Conference on Knowledge Discovery and Data Mining},
  pages={142--154},
  year={2025},
  organization={Springer}
}

@dataset{khabbaz2026dnamgcdata,
  author       = {Khabbaz, Ramy and Mateos, Jérémy and Antonini, Marc and Kas Hanna, Serge},
  title        = {Data of the {DNA-MGC+} In Vitro Experiment},
  year         = {2026},
  publisher    = {Zenodo},
  doi          = {10.5281/zenodo.18848901},
  url          = {https://doi.org/10.5281/zenodo.18848901}
}

@misc{khabbaz2026mgcpcode,
  author       = {Khabbaz, Ramy and Kas Hanna, Serge},
  title        = {{DNA-MGC+} Source Code},
  year         = {2026},
  publisher    = {Zenodo},
  version      = {1.2},
  doi          = {10.5281/zenodo.22169494},
  url          = {https://doi.org/10.5281/zenodo.22169494}
}

\section*{Acknowledgements}
The authors thank the Institute of Molecular and Cellular Pharmacology (IPMC), Sophia Antipolis, France, and its director Pascal Barbry for providing access to laboratory facilities and technical support that made the experimental work possible.

\section*{Author contributions}
R.K. and S.K. designed the DNA-MGC+ codec and performed simulations and data analysis. R.K. developed the implementation code and prepared the figures and tables. S.K. wrote the manuscript, with contributions from R.K. and J.M. All authors participated in the design of the wet-lab experiment. J.M. performed the experiments and devised the thermodynamic sequence filtering procedure. M.A. and S.K. supervised the study. All authors read, revised, and approved the final manuscript.

\section*{Funding statement}
This work was supported by the France 2030 investment plan, managed by the National Research Agency (ANR), under references ANR-22-PEXM-003 and ANR-22-CPJ2-0054-01.

\section*{Competing interests}
The authors declare no competing interests.

\clearpage
\onecolumn
\beginsupplement
\supplementcaptions

\suptitle{Supplementary Information for: \\ DNA-MGC+: A versatile codec for reliable and resource-efficient data storage on synthetic DNA}
\supauthors{Ramy Khabbaz, Jérémy Mateos, Marc Antonini, and Serge Kas Hanna}

\section*{Supplementary Notes}

\noindent

\subsection*{Supplementary Note S1: Detailed marker-based offset estimation for inner MGC+ code}
\renewcommand{\thefigure}{A}
\captionsetup[figure]{name=Figure}  
\begin{figure}[htbp]
    \centering
    \includegraphics[width=0.4\textwidth]{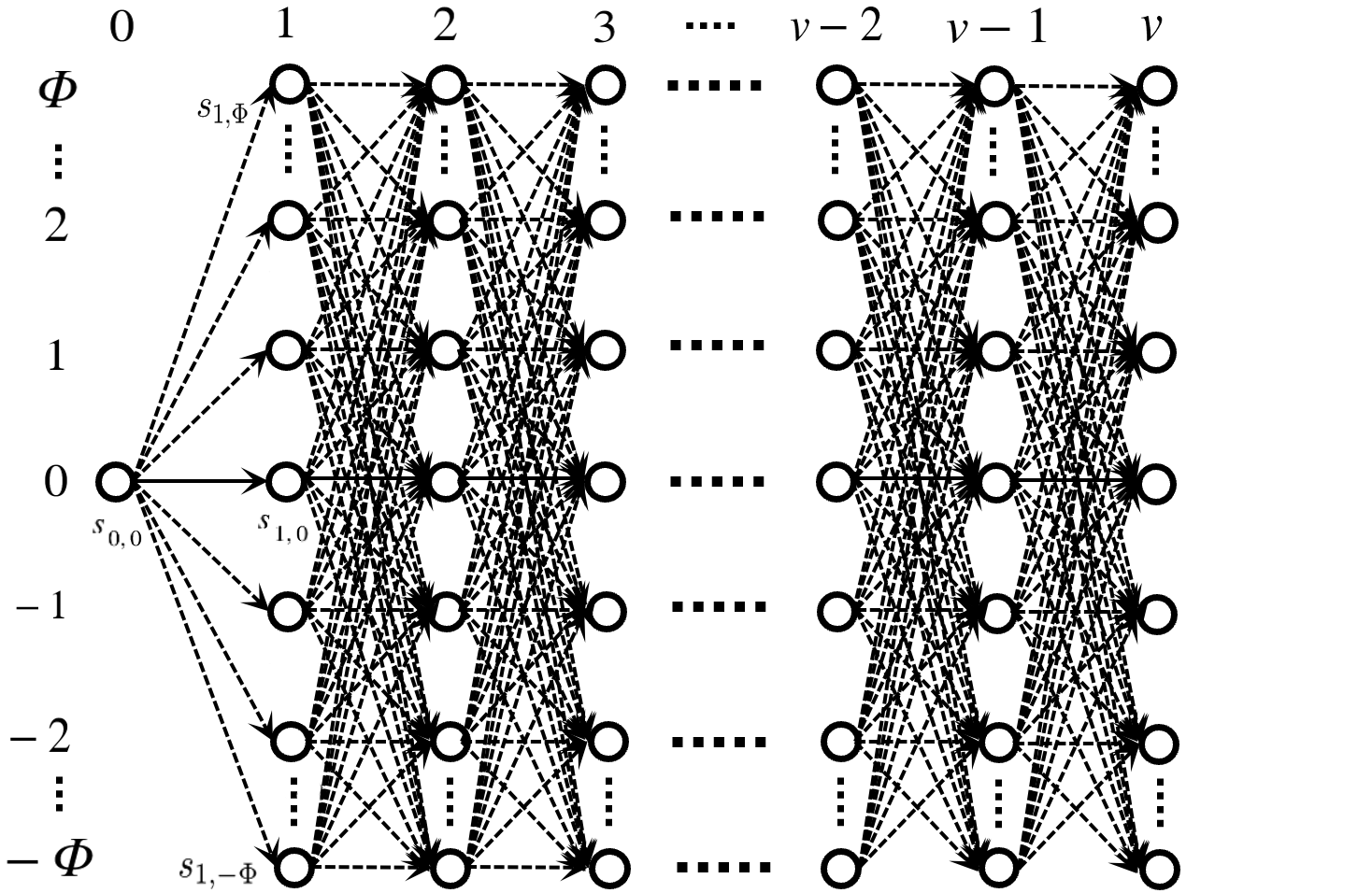} 
    \caption{Trellis representation of the drift sequence $\mathbf{z} = (z_0, z_1, \dots, z_v)$}
    \label{fig:trellis}
    \vspace{-0.5cm}
\end{figure}
\addtocounter{figure}{-1}
\renewcommand{\thefigure}{S\arabic{figure}}
\captionsetup[figure]{name=Supplementary Figure}  

Supplementary Note S1 extends the offset estimation framework originally introduced in Ref.~\cite{MGCP}, where the MGC+ code (the inner code of the DNA-MGC+ codec) was developed and analyzed in the {\em binary} domain. While the overall marker-based synchronization and trellis-based decoding principles remain conceptually the same, the analysis below reformulates the system in the {\em DNA} domain. The key differences arise from (i) base-level insertion, deletion, and substitution (IDS) errors in the underlying channel model, which modify the transition probability structure, and (ii)~the explicit derivation of received DNA marker transition probabilities under quaternary alphabet statistics (Table~\ref{tab:dna_marker_transition_probabilities}).

\paragraph*{Drift and Trellis Formulation}

For marker-based encoding, a fixed DNA marker ($\mathsf{AC}$) is inserted periodically after every $\ell_{\text{in}}$ DNA symbols. At the decoder, synchronization is recovered by constructing a trellis whose states represent the cumulative synchronization drift after each marked block. Let the trellis state space be defined as $\Omega = \{-\Phi, -\Phi+1, \dots, \Phi\},$ where $\Phi$ is an upper bound on the cumulative drift induced by insertion and deletion events. Let $\mathbf{z} = (z_0, z_1, \dots, z_v)$ denote the drift sequence, where $z_i \in \Omega$ is the cumulative offset after the $i$-th block and $z_0 = 0$. The drift evolves according to $z_i = z_{i-1} + d_i,$ where $d_i$ is the offset increment associated with block~$i$. The received DNA sequence $\mathbf{y}$ is segmented into blocks according to the hypothesized drift trajectory as
\begin{equation}
\tilde{\mathbf{b}}^{(i)} \triangleq
\mathbf{y}_{[(i-1)\ell' + z_{i-1} + 1,\; i\ell' + z_i]},
\quad i \in [v],
\label{eq:block_segmentation}
\end{equation}
where $\ell'=\ell+2$. The offset increment is equivalently expressed as $d_i = z_i - z_{i-1}.$ The last $\mu = 2$ nucleotides of each block correspond to the received marker sequence, denoted by $\tilde{\mathbf{m}}^{(i)}$.

\paragraph*{Marker-centric MAP Drift Estimation}

Assuming independent and identically distributed IDS errors across nucleotide positions, the joint probability of observing the received sequence and a particular drift trajectory factorizes as
\begin{equation}
\Pr(\tilde{\mathbf{y}}, \mathbf{z})
=
\prod_{i=1}^{v}
\Pr(\tilde{\mathbf{b}}^{(i)}, z_i \mid z_{i-1}),
\label{eq:joint_probability}
\end{equation}
with $z_0 = 0$. To reduce decoding complexity, drift estimation is performed using only the marker observations. This leads to the following marker-centric maximum a posteriori (MAP) objective:
\begin{equation}
\hat{\mathbf{z}}
=
\arg\max_{\mathbf{z} \in \mathcal{Z}}
\prod_{i=1}^{v}
\Pr(\tilde{\mathbf{m}}^{(i)}, z_i \mid z_{i-1}),
\label{eq:map_estimate}
\end{equation}
where $\mathcal{Z}$ denotes the set of all admissible trellis paths constrained to the state space $\Omega$. A graphical illustration of the resulting trellis and its admissible transitions is provided in Fig.~\ref{fig:trellis}.

\renewcommand{\thetable}{A}
\captionsetup[table]{name=Table}
\begin{table}[h!]
\centering
\caption{Marker transition probabilities $\boldsymbol{\Pr(\tilde{\boldsymbol{m}},\,\rv{D}_{\mathrm{mar}} = t)}$ for the DNA marker $\mathsf{AC}$.}
\label{tab:dna_marker_transition_probabilities}
\medskip
\begin{tabular}{c c l}
\toprule
\textbf{$\tilde{\boldsymbol{m}}$}
& \textbf{$t$}
& $\boldsymbol{\Pr(\tilde{\boldsymbol{m}},\,\rv{D}_{\mathrm{mar}} = t)}$ \\
\midrule
\multirow{5}{*}{$\mathsf{AC}$}
& $0$  & $P_0 P_i P_d + P_r^2$ \\
& $+1$ & $P_r P_i + (1 - P_d - P_i) P_i P_0$ \\
& $+2$ & $P_i^2 P_0$ \\
& $-1$ & $P_0 P_r P_d + P_0^2 P_d P_s$ \\
& $-2$ & $P_0^2 P_d^2$ \\
\midrule
\multirow{4}{*}{$\mathsf{AT}$, $\mathsf{AG}$}
& $0$  & $P_s P_0 P_r$ \\
& $+1$ & $P_i P_0 P_s$ \\
& $-1$ & $2 P_d P_0^2 P_s$ \\
& $-2$ & $P_d^2 P_0^2$ \\
\midrule
\multirow{4}{*}{$\mathsf{AA}$}
& $0$  & $P_r P_s P_0 + P_i P_d P_0$ \\
& $+1$ & $P_i P_s P_0$ \\
& $-1$ & $P_d P_0 P_r + P_0^2 P_d P_s$ \\
& $-2$ & $P_d^2 P_0^2$ \\
\midrule
\multirow{5}{*}{$\mathsf{CC}$, $\mathsf{TC}$, $\mathsf{GC}$}
& $0$  & $P_0 P_s P_r + P_d P_i P_0$ \\
& $+1$ & $(1 - P_d - P_i) P_i P_0$ \\
& $+2$ & $P_i^2 P_0$ \\
& $-1$ & $P_0 P_r P_d + P_0^2 P_d P_s$ \\
& $-2$ & $P_0^2 P_d^2$ \\
\midrule
\multirow{3}{*}{
\begin{tabular}{c}
$\mathsf{CT}$, $\mathsf{CG}$, $\mathsf{TT}$ \\
$\mathsf{GT}$, $\mathsf{GG}$, $\mathsf{TG}$
\end{tabular}
}
& $0$  & $P_s^2 P_0^2$ \\
& $-1$ & $2 P_0^2 P_s P_d$ \\
& $-2$ & $P_0^2 P_d^2$ \\
\midrule
\multirow{3}{*}{$\mathsf{CA}$, $\mathsf{TA}$, $\mathsf{GA}$}
& $0$  & $P_0 P_i P_d + P_s^2 P_0^2$ \\
& $-1$ & $P_0 P_r P_d + P_0^2 P_s P_d$ \\
& $-2$ & $P_0^2 P_d^2$ \\
\bottomrule
\end{tabular}
\end{table}
\addtocounter{table}{-1}
\renewcommand{\thetable}{S\arabic{table}}
\captionsetup[table]{name=Supplementary Table}  

\paragraph*{Transition Probabilities}

A trellis transition from state $z_{i-1}$ to $z_i = z_{i-1} + d$ is weighted by the probability of observing a received DNA marker $\tilde{\mathbf{m}}$ together with a net offset $d$ over the corresponding block. Since each block consists of a data segment followed by a fixed DNA marker, this probability decomposes as
\begin{equation}
\Pr(\tilde{\mathbf{m}}, d)
=
\sum_{t=-\mu}^{\mu}
\Pr(\rv{D}_{\mathrm{data}} = d - t)\,
\Pr(\tilde{\mathbf{m}}, \rv{D}_{\mathrm{mar}} = t),
\label{eq:transition_probability}
\end{equation}
where $\rv{D}_{\mathrm{data}}$ and $\rv{D}_{\mathrm{mar}}$ are random variables representing the cumulative offsets introduced by insertion and deletion events in the data portion and the marker portion of the block, respectively.

\smallskip

\noindent
The data portion of each block consists of $\ell_{\text{in}}$ DNA nucleotides. Under the standard base-level IDS channel with deletion probability $P_d$, insertion probability $P_i$, substitution probability $P_s$, and $P_r = 1 - P_d - P_i - P_s$, the offset distribution
$\Pr(\rv{D}_{\mathrm{data}} = t)$
depends only on the block length and channel parameters. Mathematically, it is given by
\begin{equation}
\Pr(\rv{D}_{\mathrm{data}} = t)
=
\sum_{j=\max\{0,-t\}}^{\left\lfloor\frac{\ell_{\text{in}} - t}{2}\right\rfloor}
\binom{\ell_{\text{in}}}{j,\, j+t,\, \ell_{\text{in}} - 2j - t}
\times
P_d^{\,j}\, P_i^{\,j+t}\, P_r^{\,\ell_{\text{in}} - 2j - t}.
\label{eq:data_offset_pmf}
\end{equation}

\smallskip

\noindent
The marker contribution $\Pr(\tilde{\mathbf{m}}, \rv{D}_{\mathrm{mar}} = t)$ depends explicitly on the known transmitted DNA marker $\mathbf{m} = \mathsf{AC},$ the channel parameters, and the received marker sequence
$\tilde{\mathbf{m}} \in \{\mathsf{A},\mathsf{C},\mathsf{G},\mathsf{T}\}^{\ast}$.
Since the marker length is short ($\mu = 2$), this probability can be evaluated exactly by enumerating all possible insertion, deletion, and substitution events affecting the marker nucleotides. For completeness, Table~\ref{tab:dna_marker_transition_probabilities} reports $\Pr(\tilde{\mathbf{m}}, \rv{D}_{\mathrm{mar}} = t)$ for all possible received marker sequences $\tilde{\mathbf{m}}$ and offsets $|t| \leq 2$, expressed in closed form as functions of $P_d$, $P_i$, $P_s$, $P_r$ and $P_0$ where $P_0 = 0.25$.

\paragraph*{Dynamic Programming Solution}

The MAP problem in \eqref{eq:map_estimate} is solved efficiently using dynamic programming. Let $\alpha_i(z_i)$ denote the maximum path metric ending at drift state $z_i$ after processing $i$ blocks. The recursion is
\begin{equation}
\alpha_i(z_i)
=
\max_{z_{i-1} \in \Omega}
\alpha_{i-1}(z_{i-1})\,
\Pr(\tilde{\mathbf{m}}^{(i)}, z_i \mid z_{i-1}),
\label{eq:dp_metric}
\end{equation}
with initialization $\alpha_0(0) = 1$ and $\alpha_0(z) = 0$ for $z \neq 0$. The maximizing predecessor state is stored as
\begin{equation}
\mathrm{pred}_i(z_i)
=
\arg\max_{z_{i-1} \in \Omega}
\alpha_{i-1}(z_{i-1})\,
\Pr(\tilde{\mathbf{m}}^{(i)}, z_i \mid z_{i-1}).
\label{eq:dp_predecessor}
\end{equation}

After processing all $v$ blocks, the estimated drift sequence $\hat{\mathbf{z}}$ is recovered by traceback starting from
\begin{equation}
\hat{z}_v = \arg\max_{z_v \in \Omega} \alpha_v(z_v),
\label{eq:final_drift}
\end{equation}
and the corresponding offset pattern is obtained as $\hat{d}_i = \hat{z}_i - \hat{z}_{i-1},$ which is subsequently used for inner decoding.

\subsection*{Supplementary Note S2: Additional details and comments about some of the comparison codecs}

\paragraph*{LDPC Codec} The LDPC-based scheme~\cite{chandak2019improved} uses a conceptually different design from the inner--outer architectures adopted by most existing DNA storage codecs. The entire data stream is encoded using a single long binary LDPC code, whose encoded bits are then segmented into short fragments and mapped to DNA symbols. Each fragment carries a BCH-protected index and a marker to assist in handling insertions and deletions during decoding. 

In our \emph{in silico} evaluations, we used the implementation provided at \url{https://github.com/shubhamchandak94/LDPC_DNA_Storage}, keeping its default parameters and setting the \texttt{oligo\_length} parameter to $150$. To obtain different code rates, we replaced the underlying parity-check matrices using three matrices from \url{https://github.com/shubhamchandak94/LDPC_DNA_storage_data/tree/master/matrices}, resulting in three codec configurations with code rates of $0.58$, $0.67$, and $0.80$ bits/nt. Among these, the configuration with rate $0.58$ achieved the best overall performance under the synthetic channel model, as shown in the subsequent Supplementary Tables. However, none of the three configurations was able to decode successfully under the conditions used in the second \emph{in silico} setting.

\paragraph*{DNA-Stairloop} 
The DNA-StairLoop codec~\cite{yan2025dna} employs a concatenated coding structure consisting of a row code and a column code, but differs from the classical inner--outer architecture through the use of a staircase interleaver that couples the two component codes across adjacent blocks. Decoding follows an iterative soft-information exchange between the row and column decoders. The decoder is designed to operate directly on unclustered sequencing reads and to exploit the presence of multiple reads within its own probabilistic decoding algorithm, without relying on consensus calling. To remain consistent with the decoding procedure proposed by the authors, clustering and alignment were therefore not performed for this codec in our evaluations.

In our \emph{in silico} experiments, we used the reference implementation available at \url{https://github.com/Guanjinqu/StairLoop}. The parameter \texttt{msg\_length} was fixed to $19$ to obtain encoded DNA sequences of length $147$, close to our desired target length of $150$. Two values of the parameter \texttt{block\_num}, $34$ and $68$, were used to vary the total number of encoded sequences, resulting in two codec configurations with code rates of $0.98$ and $0.49$ bits/nt, respectively. Our results indicate that this codec is highly sensitive to sequence dropouts under the considered parameters, leading to degraded performance under strong bias conditions. The configuration with rate 0.98 bits/nt outperformed the configuration with rate 0.49 bits/nt, suggesting that increasing the number of encoded sequences while keeping the sequence length fixed does not necessarily improve the performance of this codec. This observation is also implicit in Ref.~\cite{yan2025dna}, where the authors evaluate their codec using sequence lengths closer to 200, whereas comparison codecs are evaluated at lengths closer to~150, suggesting that DNA-Stairloop may benefit from allocating more redundancy within individual sequences, and hence operating at longer sequence lengths.

\paragraph*{DNA-Aeon}
As reflected in the results in Supplementary Tables S8--S11, the decoding time of DNA-Aeon is significantly higher than that of the other codecs evaluated in this work. Decoding also becomes substantially slower at lower code rates and higher error rates. When the medium-rate configuration of DNA-Aeon was tested at an error rate of 15\%, decoding ran for more than two hours without producing a result and was therefore terminated; consequently, no results are reported for this setting. The same occurred for the low-rate configuration at error rates of 10\% and 15\%.

\subsection*{Supplementary Note S3: Discussion about the effect of $\Delta G$ on the coverage distribution}
As noted in the Discussion section of the main text, the filtered variants of DNA-MGC+ provide modest but consistent gains in terms of sequencing depth (and read cost) compared to their unfiltered counterparts. This is reflected in the coverage distributions shown in Supplementary Fig.~S8, where the unfiltered variants consistently exhibit a substantially larger fraction of underrepresented reference sequences than the filtered variants. For example, under Illumina sequencing for Design~B of DNA-MGC+, approximately 200 reference sequences in the unfiltered variant receive between 0 and 10\% of the mean coverage, whereas this number decreases to roughly 50 sequences in the filtered variant. The same qualitative behavior is observed across both designs and under both sequencing platforms, demonstrating that filtering systematically reduces the number of poorly covered sequences. These strongly underrepresented sequences contribute to increased sequence dropouts at low sequencing depths, which explains the observed difference in the minimum sequencing depth required for reliable decoding between the filtered and unfiltered variants.

Our analysis of the per-sequence coverage shows that the decoding performance gains of the filtered variants are primarily attributable to the filtering criterion based on the Gibbs free energy $\Delta G$. We support this claim through three complementary statistical analyses of the per-sequence coverage distribution obtained for the unfiltered DNA-MGC+ variants. The numerical results provided below correspond to Nanopore sequencing; qualitatively similar trends were observed under Illumina sequencing.

\paragraph*{Characteristics of underrepresented sequences.}
We first identify the subset of reference sequences whose read count falls below 10\% of the mean pool coverage. Across the complete unfiltered pool, the mean~$\pm$~standard deviation~(SD) of the maximum homopolymer length, GC content, and $\Delta G$ are $4.2 \pm 1.1$, $48.7\% \pm 3.3\%$, and $-4.1 \pm 2.3$~kcal/mol, respectively. Within the underrepresented subset, the corresponding means are $3.9$, $49.9\%$, and $-6.8$~kcal/mol, yielding standardized differences from the pool-wide means of $-0.27$~SD, $+0.36$~SD, and $-1.17$~SD. The mean $\Delta G$ of the underrepresented sequences is thus shifted by approximately $1.2$ pool-wide standard deviations toward more negative values, which is \hl{significant} on its own and also substantially larger than the shifts observed for GC content or homopolymer length. Notably, the homopolymer-length shift is small and \emph{negative}, whereas a larger and \emph{positive} shift would be expected if long homopolymers were a major contributor to lower coverage.

\paragraph*{Marginal Spearman correlations.}
Spearman correlation analysis further supports the dominant role of $\Delta G$. Across the full unfiltered pool, $\Delta G$ shows the strongest marginal association with per-sequence read count ($\rho = +0.315$, $p$-value $< 10^{-50}$), compared with GC content ($\rho = -0.205$) and maximum homopolymer length ($\rho = +0.082$). The positive sign of the $\Delta G$--coverage correlation confirms that sequences with $\Delta G$ closer to zero (less negative) tend to receive more reads. The weak and positive marginal correlation between maximum homopolymer length and coverage is inconsistent with the hypothesis that long homopolymers cause sequence underrepresentation.

\paragraph*{Partial Spearman correlations.}
To verify that the association between $\Delta G$ and coverage is not merely a consequence of its relationship with GC content or homopolymer length, we compute partial Spearman correlations that isolate the $\Delta G$--coverage association after mathematically removing the contribution explained by each of the other variables. The partial correlation between $\Delta G$ and coverage remains largely unchanged when controlling for GC content ($\rho = +0.231$) or for homopolymer length ($\rho = +0.295$), indicating that the association of $\Delta G$ with coverage is largely independent of these two variables. By contrast, the marginal association between GC content and coverage ($\rho = -0.205$) decreases markedly after controlling for $\Delta G$ ($\rho = -0.062$), suggesting that a substantial portion of the apparent GC--coverage association can be attributed to the correlation between GC content and $\Delta G$ rather than to an independent effect of GC content on coverage.

\paragraph*{Summary.}
Taken together, these three statistical results indicate that $\Delta G$ is the sequence-level characteristic most strongly associated with coverage variability among those examined here, while GC content and homopolymer length show little to no independent association with sequence underrepresentation in our experiments.

Taken together, these three statistical results indicate that $\Delta G$ is the \hl{dominant sequence-level feature associated with coverage variability among those examined here, while GC content and homopolymer length show little to no independent association with sequence underrepresentation in our experiments}. A plausible mechanistic explanation is that sequences with more negative $\Delta G$ tend to form stable secondary structures, which may interfere with PCR amplification efficiency and thereby reduce their representation in the sequenced pool. This is also consistent with the observation that the fraction of underrepresented sequences is more pronounced under Illumina sequencing, which involves additional PCR steps during library preparation and cluster generation compared with Nanopore sequencing (see Supplementary Fig.~S8).

\subsection*{Supplementary Note S4: Comparison with the Gungnir codec}
We provide here a comparison between DNA-MGC+ and the recently introduced Gungnir codec~\cite{zhang2026gungnir} under the synthetic error and bias models considered in the main text. Gungnir provides strong IDS error-correction capability by using a complex decoding procedure, in which candidate reconstructions are exhaustively tested until a hash signature associated with the encoded fragment is satisfied. This mechanism gives Gungnir high error tolerance; however, the codec also has two important limitations.

First, as explicitly acknowledged by the authors in Ref.~\cite{zhang2026gungnir}, Gungnir does not provide robustness to sequence dropouts.As a result, successful file retrieval requires that each encoded sequence be read at least once.  \hl{This limitation is architectural and independent of the inner-code IDS error-correction capability, and can therefore substantially affect the required coverage depth and read cost even under ideal inner decoding.} Second, Gungnir relies on an exhaustive hypothesis-search procedure whose computational cost becomes substantial at high error rates. For example, the authors report that decoding a 19.4-KB file required more than 37 hours at high error rate on a dual 32-core Xeon Platinum 8369C server. Therefore, to keep the comparison computationally tractable, we evaluate Gungnir on a small, randomly generated 5-KB file at a total error rate of 5\%, for the same bias regimes (no bias, moderate bias, strong bias) considered in Fig.~1 of the main text. Unlike the experiments reported in Fig.~1, which restrict each codec to a single CPU core, the Gungnir decoder is allowed to use the full 64-core CPU of our workstation, equipped with an AMD Ryzen Threadripper PRO 7985WX processor. The same computing resources are provided to DNA-MGC+ for a fair comparison. 

\hl{For reference, we also attempted the no-bias case at a 10\% error rate for a 5-KB file, for which Gungnir decoding ran for more than 8 hours without converging while using the full 64-core CPU, exceeding the 2-hour decoding limit adopted throughout our evaluation. In comparison, for the exact same scenario (5-KB file, 10\% error rate, no bias), DNA-MGC+ incurs a decoding time of 0.95s at a coverage depth of 16$\times$, and 1.26s at the minimum required coverage depth of 3.75$\times$.}

Following the evaluation framework of Fig.~1, we consider three code-rate classes. For Gungnir, the high- and medium-rate configurations are obtained using the Gungnir-Trit mode of the public implementation (\url{https://github.com/HKU-BAL/Gungnir}), with rates of 1.50 and 1.00 bits/nt, respectively. The low-rate configuration is obtained using the standard Gungnir mode, with a rate of 0.50 bits/nt. In all three cases, the encoded sequence length is set to 150 nts. For each configuration and bias regime, we determine the minimum coverage depth and associated read cost required for successful file retrieval at the 5\% error rate. To reduce the computational burden, the reliability requirement is relaxed relative to Fig.~1: instead of requiring successful retrieval in 50 out of 50 independent trials, we require exact file recovery in a single trial at each evaluated coverage depth. The resulting coverage depth, read cost, and decoding time are reported in Fig.~B below.

\renewcommand{\thefigure}{B}
\captionsetup[figure]{name=Figure}  
\begin{figure}[h!]
    \centering
    \centering
    \includegraphics[width=0.9\textwidth]{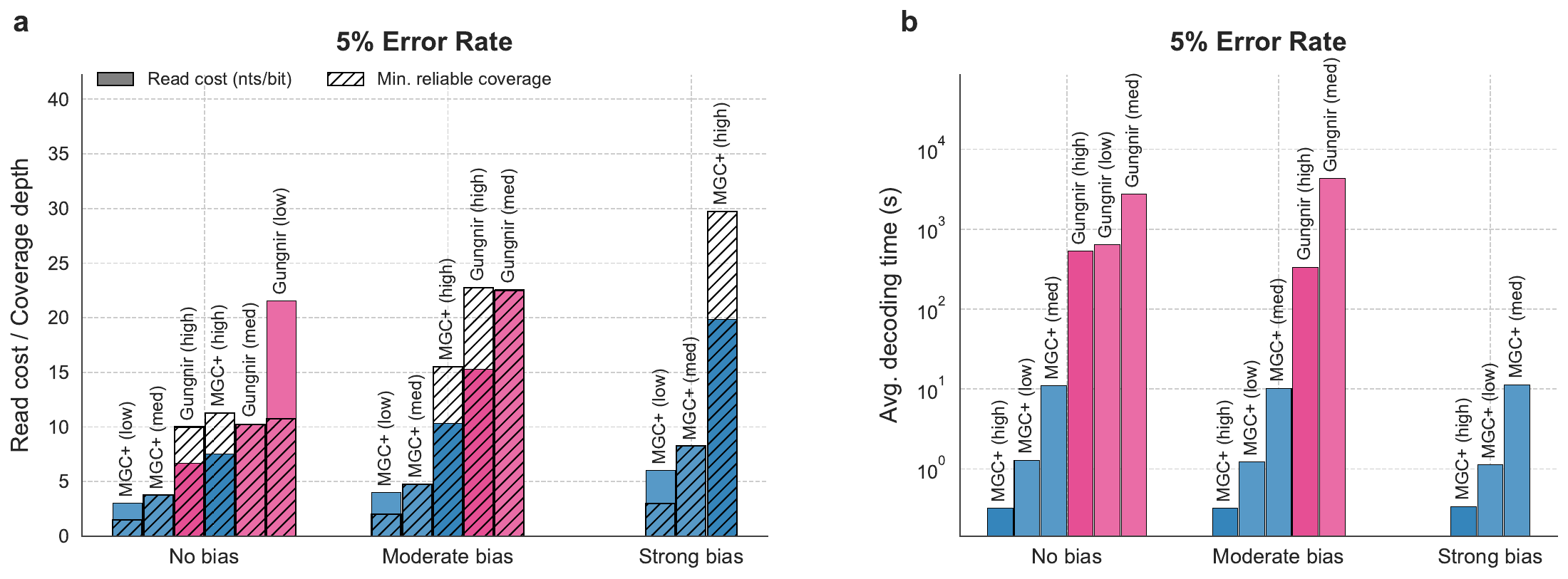} 
    \caption{\textbf{In silico performance comparison of DNA-MGC+ and Gungnir.} (\textbf{a}) Minimum sequencing depth and corresponding read cost required for reliable decoding at a 5\% error rate under three different bias regimes. (\textbf{b}) Average decoding time measured at the minimum coverage depth required for reliable decoding.}
    \label{fig:gungnir}
\end{figure}

The results show that DNA-MGC+ requires lower coverage depth and lower read cost than Gungnir across the evaluated bias regimes and code-rate classes, while also achieving orders-of-magnitude faster decoding. In the strong-bias case, Gungnir does not achieve successful retrieval within the evaluated coverage-depth range of [1, 32], which is expected since the codec is not robust to dropouts. \hl{This dropout limitation is architectural and remains regardless of the IDS error rate, while extending the comparison to higher error rates is computationally prohibitive, as previously noted for the 10\% error-rate case.} 

\addtocounter{figure}{-1}
\renewcommand{\thefigure}{S\arabic{figure}}
\captionsetup[figure]{name=Supplementary Figure}  

\subsection*{\hl{Supplementary Note S5: Additional DNA-MGC+ results for larger file sizes and higher error rates}}

\renewcommand{\thefigure}{C}
\captionsetup[figure]{name=Figure}  
\begin{figure}[h!]
    \centering
    \includegraphics[width=0.9\textwidth]{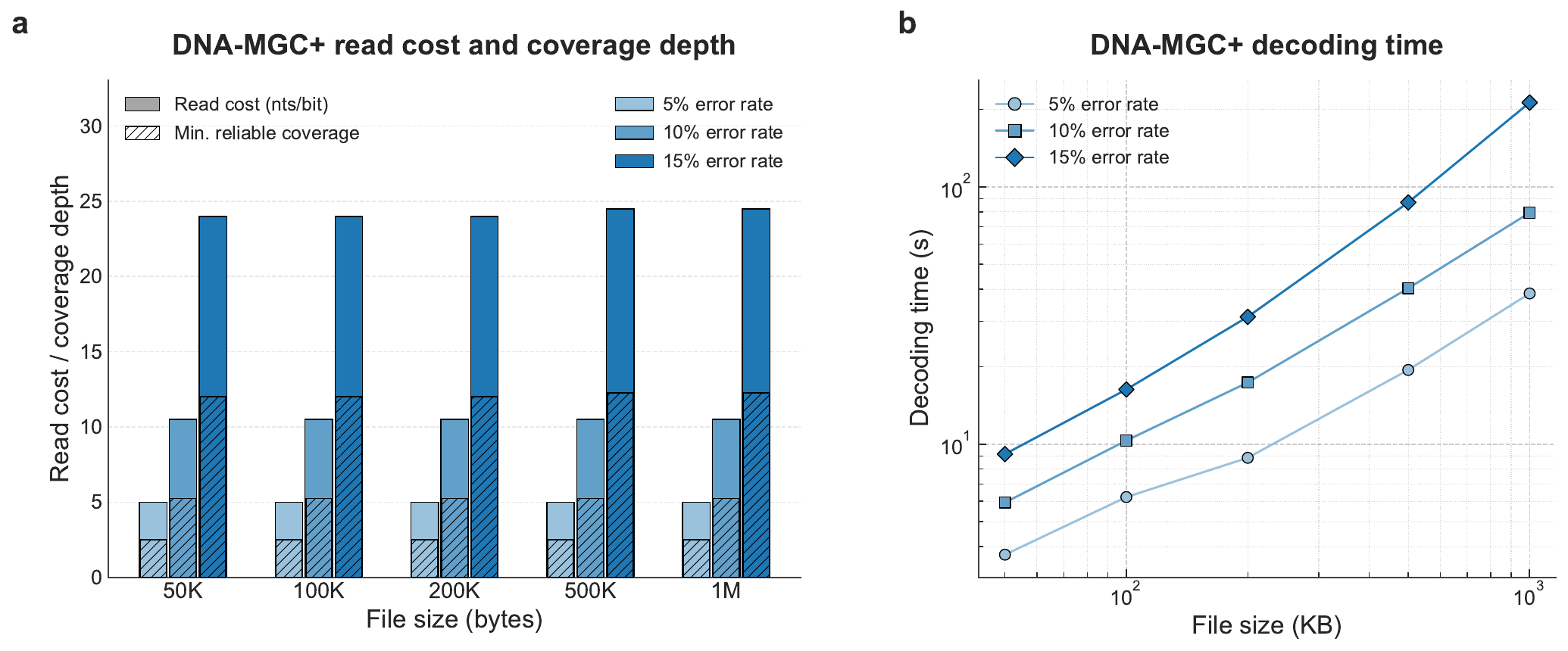} 
    \caption{\textbf{In silico DNA-MGC+ performance across varying file sizes and error rates under the moderate-bias scenario of the synthetic channel model.} (\textbf{a})~Minimum coverage depth required for reliable decoding and corresponding read cost (nts/bit) as a function of file size and error rate. (\textbf{b})~Average decoding time measured at the minimum coverage depth required for reliable decoding.}
    \label{fig:MGCPP}
\end{figure}
\addtocounter{figure}{-1}
\renewcommand{\thefigure}{S\arabic{figure}}
\captionsetup[figure]{name=Supplementary Figure}  

In Fig.~2 of the main text, we assessed the system-level scalability of DNA-MGC+ and several comparison codecs by varying the input file size from a few kilobytes to hundreds of megabytes, while fixing the error rate at 5\% and considering the moderate-bias setting of the synthetic channel model. Here, we complement these results by instead fixing the file size at several representative values (50~KB, 100~KB, 200~KB, 500~KB, and 1~MB) and varying the error rate among 5\%, 10\%, and 15\%, under the same moderate-bias setting. In Fig.~\ref{fig:MGCPP}, for each file size--error rate combination, we report the minimum coverage depth required for reliable decoding together with the associated read cost, as well as the average decoding time measured at the minimum coverage depth (Fig.~\ref{fig:MGCPP}~b), following the same 50-out-of-50 reliability criterion and the same preprocessing pipeline (CBR clustering and Kalign alignment) used in the evaluation corresponding to Fig.~2.

The results show that, for a fixed error rate, the minimum coverage depth and read cost of DNA-MGC+ remain essentially stable as the file size increases from 50~KB to 1~MB, consistent with the trend already reported in Fig.~2 at 5\% error rate. As expected, coverage depth, read cost, and decoding time all increase with the error rate; however, this increase is consistent across file sizes, indicating a favorable scalability of DNA-MGC+ over the investigated scenarios. We note that the error rates considered here, in particular 10\% and 15\%, are substantially higher than those typically observed in the conventional wet-lab experimental workflows and are instead intended to stress-test the scalability of the codec.

\clearpage
\section*{Supplementary Tables}

\begin{table}[h!]
\centering
\caption{Selected parameters for the DNA-MGC+ codec in the {\em in silico} and {\em in vitro} evaluations.}
\renewcommand{\arraystretch}{1.3}

\newcolumntype{E}{>{\centering\arraybackslash}m{1.8cm}}

\begin{tabular}{>{\centering\arraybackslash}m{4.5cm} c c c c c E E}
\toprule
\multirow{2}{*}{\textbf{Parameter}} &
\multicolumn{4}{c}{\textbf{In-silico studies}} &
\multicolumn{2}{c}{\textbf{In-vitro experiment}} \\
\cmidrule(lr){2-5} \cmidrule(lr){6-7}
 & Low & Low (large file) & Medium & High & Opt & Design A & Design B \\
\midrule
Inner Redundancy ($c_{\text{in}}$)     & 12  & 10 & 6   & 0   &  6 &6    & 6    \\
Outer code rate (($K/(K+c_{\text{out}})$)     & 0.679 & 0.679 & 0.712 & 0.796 &  0.356 & 0.80  & 0.77  \\
Inner symbol length ($\ell_{\text{in}}$, bits)        & 8 & 8  & 8   & 8  &  8 & 8    & 8    \\
Outer symbol length ($\ell_{\text{out}}$, bits)      & 16 & 16 & 16  & 16  & 16 & 16   & 16   \\
Sequence index length (bits)      & 16 & 16 & 16  & 16  & 16 & 16   & 16   \\
Packet index length (bits)      & -- & 16 & --  & --  & -- & --   & --   \\
Periodic markers              &  Yes   & Yes &   No  &   No  & No & No   & Yes  \\
Code rate (bits/nt)   & 0.5 & 0.5 & 1   & 1.5 &  0.5 &1.03 & 0.71 \\
Sequence length ($L_{\text{ref}}$, nts) & 152 & 152 & 148 & 152 & 148  &124  & 122  \\
\bottomrule
\end{tabular}

\end{table}

\begin{table}[h!]
\centering
\caption{Selected parameters for the DNA-Aeon codec used in the {\em in silico} and {\em in vitro} evaluations. These parameter settings are taken from Refs.~\cite{gimpel2025comparison,gimpel2025dt4ddsbenchmark}.}
\renewcommand{\arraystretch}{1.3}

\begin{tabular}{>{\centering\arraybackslash}m{3.3cm} c c c c}
\toprule
\multirow{2}{*}{\textbf{Parameter}} &
\multicolumn{3}{c}{\textbf{In-silico studies}} &
\multirow{2}{*}{\textbf{In-vitro experiment}} \\
\cmidrule(lr){2-4}
 & Low & Medium & High & \\
\midrule
Homopolymer & 4 & 4 & 4 & 4\\
GC-content & 0.0--1.0 & 0.0--1.0 & 0.0--1.0 & 0.0--1.0\\
Package redundancy & 1.68 & 0.34 & 0.031 & 0.32\\
Chunk size & 25 & 25 & 28 & 20\\
Sync value & 4 & 4 & 8 &  4\\
Error correction & CRC & CRC & CRC &  CRC \\
Codeword length & 10 & 10 & 10 &  10\\
CRC threshold & 3 & 3 & 3 & 3 \\
Loop & 1 & 1 & 1 & 1 \\
Finish & 0 & 0 & 0 & 0 \\
Penalty (CRC) & 0.1 & 0.1 & 0.1 & 0.1 \\
Penalty (No-Hit) & 8 & 8 & 8 & 8 \\
Code rate (bits/nt)   & 0.5 & 1.01   & 1.5 & 1 \\
Sequence length (nts) & 149 & 149 & 144 & 120 \\
\bottomrule
\end{tabular}
\end{table}

\begin{table}[h!]
\centering
\caption{Selected parameters for the Hedges codec used in the {\em in silico} and {\em in vitro} evaluations. The {\em in silico} parameter settings are taken from Refs.~\cite{gimpel2025comparison,gimpel2025dt4ddsbenchmark}.}
\renewcommand{\arraystretch}{1.3}

\begin{tabular}{>{\centering\arraybackslash}m{3.3cm} c c c}
\toprule
\multirow{2}{*}{\textbf{Parameter}} &
\multicolumn{2}{c}{\textbf{In-silico studies}} &
\multirow{2}{*}{\textbf{In-vitro experiment}} \\
\cmidrule(lr){2-3}
 & Low & Medium &  \\
\midrule
Code rate index & 3 & 1 & 3 \\
Homopolymer & 4 & 4 &  4\\
GC window size & 12 & 12 & 12 \\
Max. GC (in window) & 8 & 8 & 8 \\
Code rate (bits/nt) & 0.65  & 1.09 & 0.61 \\
Sequence length (nts) & 148 & 147  &  126 \\
\bottomrule
\end{tabular}

\end{table}

\begin{table}[h!]
\centering
\caption{Selected parameters for the RS codec used in the {\em in silico} evaluations. These parameter settings are taken from Refs.~\cite{gimpel2025comparison,gimpel2025dt4ddsbenchmark}.}
\renewcommand{\arraystretch}{1.3}

\begin{tabular}{>{\centering\arraybackslash}m{4cm} c c c}
\toprule
\multirow{2}{*}{\textbf{Parameter}} & \multicolumn{3}{c}{\textbf{In-silico studies}} \\
\cmidrule(lr){2-4}
 & Low & Medium & High \\
\midrule
Inner RS symbol length            & 6   & 6   & 6  \\
Outer RS symbol length          & 14     & 14     & 14     \\
Index length        & 24      & 24      & 24     \\
Inner Red. symbols & 4    & 2    & 2    \\
Number of sequences & 1602    & 834    & 556    \\
Code rate (bits/nt) & 0.5  & 1 & 1.5 \\
Sequence length (nts) & 150 & 144  &  144 \\
\bottomrule
\end{tabular}
\end{table}

\begin{table}[h!]
\centering
\caption{Selected parameters for the DNA-Fountain codec used in the {\em in silico} evaluations. These parameter settings are taken from Refs.~\cite{gimpel2025comparison,gimpel2025dt4ddsbenchmark}.}
\renewcommand{\arraystretch}{1.3}

\begin{tabular}{>{\centering\arraybackslash}m{4cm} c c c}
\toprule
\multirow{2}{*}{\textbf{Parameter}} & \multicolumn{3}{c}{\textbf{In-silico studies}} \\
\cmidrule(lr){2-4}
 & Low & Medium & High \\
\midrule
Alpha            & 2.35   & 0.68   & 0.19   \\
Payload          & 32     & 32     & 34     \\
RS length        & 2      & 2      & 0      \\
Hamming distance & 100    & 100    & 100    \\
GC-content       & 0.0--1.0 & 0.0--1.0 & 0.0--1.0 \\
Homopolymer      & 4      & 4      & 4      \\
Delta            & 0.05   & 0.1    & 0.1    \\
C-Dist           & 0.1    & 0.025  & 0.025  \\
Header size      & 4      & 4      & 4      \\
Code rate (bits/nt) & 0.5  & 1 & 1.5 \\
Sequence length (nts) & 152 & 152  &  152 \\
\bottomrule
\end{tabular}
\end{table}

\begin{table*}[ht]
\centering
\caption{Minimum reliable coverage depth for different codec configurations, error rates, and bias conditions, under the synthetic channel model (in silico). Entries are reported as unequal | equal; equal denotes $P_d=P_s=P_i$, whereas unequal denotes $P_s=0.572P_e$, $P_d=0.447P_e$, and $P_i=0.026P_e$.}
\renewcommand{\arraystretch}{0.85}
\setlength{\tabcolsep}{6pt}
\begin{tabular}{c l c c c c c}
\toprule
\raisebox{1.2ex}{\textbf{Codec}} & 
\raisebox{1.2ex}{\textbf{Class}} & 
\raisebox{1.2ex}{\textbf{Code rate}} & 
\raisebox{1.2ex}{\textbf{Error rate}} &
\shortstack{\textbf{No bias}\\Unequal | Equal} & \shortstack{\textbf{Moderate bias}\\Unequal | Equal} & \shortstack{\textbf{Strong bias}\\Unequal | Equal} \\
\midrule
\multirow{8}{*}{\textbf{Aeon}} & \multirow{2}{*}{Low} & \multirow{2}{*}{0.50} & 0.01 & 1.50 | 1.75 & 1.75 | 1.25 & 2.25 | 1.25 \\
 & & & 0.05 & 3.75 | 4.50 & 4 | 5.50 & 5.25 | 8.75 \\
\cmidrule(lr){2-7}
 & \multirow{3}{*}{Medium} & \multirow{3}{*}{1} & 0.01 & 3 | 3.25 & 2.75 | 3.25 & 4.25 | 5 \\
 & & & 0.05 & 5 | 6.25 & 5.50 | 6.25 & 10.25 | 13.25 \\
 & & & 0.10 & 10 | 11.50 & 11.50 | 12 & 20.75 | 22.50 \\
\cmidrule(lr){2-7}
 & \multirow{3}{*}{High} & \multirow{3}{*}{1.50} & 0.01 & 6.25 | 7.25 & 9.25 | 12 & 29.50 | - \\
 & & & 0.05 & 10.75 | 11.25 & 21.25 | 19.50 & - \\
 & & & 0.10 & 22.25 | - & - & - \\
\midrule
\multirow{3}{*}{\textbf{Fountain}} & \multirow{2}{*}{Medium} & \multirow{2}{*}{1} & 0.01 & 12.25 | 11.25 & 23.50 | 26 & - \\
 & & & 0.05 & 25.75 | - & - & - \\
\cmidrule(lr){2-7}
 & \multirow{1}{*}{High} & \multirow{1}{*}{1.50} & 0.01 & 16 | - & - & - \\
\midrule
\multirow{7}{*}{\textbf{HEDGES}} & \multirow{4}{*}{Low} & \multirow{4}{*}{0.65} & 0.01 & 3.25 | 3.50 & 4.75 | 4.75 & 11.50 | 14.25 \\
 & & & 0.05 & 3.75 | 4 & 6.50 | 5.25 & 14 | 14.25 \\
 & & & 0.10 & 6.25 | 6.50 & 9.25 | 9.25 & 22 | 20 \\
 & & & 0.15 & 16.75 | 13 & 26.25 | 20.25 & - \\
\cmidrule(lr){2-7}
 & \multirow{3}{*}{Medium} & \multirow{3}{*}{1.09} & 0.01 & 3.25 | 3.25 & 4.50 | 4.75 & 11.25 | 11 \\
 & & & 0.05 & 6 | 6.25 & 9 | 9.50 & 23.25 | 23.75 \\
 & & & 0.10 & 10 | 10 & 16.75 | 15.50 & - \\
\midrule
\multirow{8}{*}{\textbf{LDPC}} & \multirow{3}{*}{Low} & \multirow{3}{*}{0.58} & 0.01 & 2.75 | 3.75 & 3.25 | 4.50 & 5.75 | 8 \\
 & & & 0.05 & 6.75 | 6.75 & 8.75 | 8.75 & 16.25 | 17.75 \\
 & & & 0.10 & 18.50 | 18.50 & 24.25 | 28 & - \\
\cmidrule(lr){2-7}
 & \multirow{3}{*}{Low} & \multirow{3}{*}{0.67} & 0.01 & 3.50 | 4.75 & 4.50 | 6 & 10 | 12.75 \\
 & & & 0.05 & 8.75 | 8.50 & 11.50 | 12.25 & 25.50 | 27 \\
 & & & 0.10 & 23 | - & 32 | - & - \\
\cmidrule(lr){2-7}
 & \multirow{2}{*}{Medium-Low} & \multirow{2}{*}{0.80} & 0.01 & 7 | 8.25 & 11.50 | 14.75 & - \\
 & & & 0.05 & 15.50 | 15.25 & 27.75 | 27.25 & - \\
\midrule
\multirow{9}{*}{\textbf{MGC+}} & \multirow{4}{*}{Low} & \multirow{4}{*}{0.50} & 0.01 & 1.25 | 1.25 & 1.50 | 1.50 & 2.50 | 2.50 \\
 & & & 0.05 & 1.75 | 1.75 & 2 | 1.75 & 3.25 | 3 \\
 & & & 0.10 & 3.75 | 3.75 & 4.25 | 4.50 & 7.50 | 7.50 \\
 & & & 0.15 & 12.50 | 9.75 & 16.25 | 11.50 & 30.25 | 21 \\
\cmidrule(lr){2-7}
 & \multirow{3}{*}{Medium} & \multirow{3}{*}{1} & 0.01 & 1.50 | 1.50 & 1.75 | 1.75 & 3.25 | 3 \\
 & & & 0.05 & 4 | 4 & 5 | 4.75 & 9.25 | 8.75 \\
 & & & 0.10 & 8.75 | 7 & 11.25 | 8.75 & 22.50 | 17.75 \\
\cmidrule(lr){2-7}
 & \multirow{2}{*}{High} & \multirow{2}{*}{1.50} & 0.01 & 4.25 | 4.50 & 5.25 | 5.50 & 11.25 | 11.25 \\
 & & & 0.05 & 11.50 | 9.50 & 14.75 | 12.75 & 31.25 | 28 \\
\midrule
\multirow{9}{*}{\textbf{RS}} & \multirow{3}{*}{Low} & \multirow{3}{*}{0.50} & 0.01 & 1 | 1.25 & 1 | 1.25 & 1.25 | 1.50 \\
 & & & 0.05 & 3.25 | 3.25 & 3.25 | 3.25 & 4.25 | 4 \\
 & & & 0.10 & 8.75 | 6.75 & 8.75 | 7.25 & 10.25 | 9 \\
\cmidrule(lr){2-7}
 & \multirow{3}{*}{Medium} & \multirow{3}{*}{1} & 0.01 & 2.75 | 3 & 3 | 3.25 & 4.50 | 5 \\
 & & & 0.05 & 5.50 | 5.50 & 6.50 | 6.50 & 10.25 | 10 \\
 & & & 0.10 & 13.50 | 11.50 & 16 | 14 & 25.75 | 24 \\
\cmidrule(lr){2-7}
 & \multirow{3}{*}{High} & \multirow{3}{*}{1.50} & 0.01 & 4.75 | 5.25 & 6.50 | 7.25 & 14.25 | 16.25 \\
 & & & 0.05 & 9.25 | 8.75 & 12.50 | 12.25 & 30.50 | - \\
 & & & 0.10 & 23 | 24.25 & - & - \\
\midrule
\multirow{5}{*}{\textbf{Stairloop}} & \multirow{2}{*}{Low} & \multirow{2}{*}{0.49} & 0.01 & 6.25 | 6.75 & 10.25 | 16.75 & - \\
 & & & 0.05 & 12 | 15.25 & 31 | 29 & - \\
\cmidrule(lr){2-7}
 & \multirow{3}{*}{Medium} & \multirow{3}{*}{0.98} & 0.01 & 4.75 | 6 & 12.25 | 10.75 & - \\
 & & & 0.05 & 12.25 | 12 & 24 | - & - \\
 & & & 0.10 & 30 | - & - & - \\
\bottomrule
\end{tabular}
\label{tab:min_reliable_depth}
\end{table*}

\begin{table*}[ht]
\centering
\caption{Read cost (nts/bit) for different codec configurations, error rates, and bias conditions, under the synthetic channel model (in silico). Entries are reported as unequal | equal; equal denotes $P_d=P_s=P_i$, whereas unequal denotes $P_s=0.572P_e$, $P_d=0.447P_e$, and $P_i=0.026P_e$.}
\renewcommand{\arraystretch}{0.85}
\setlength{\tabcolsep}{6pt}
\begin{tabular}{c l c c c c c}
\toprule
\raisebox{1.2ex}{\textbf{Codec}} & 
\raisebox{1.2ex}{\textbf{Class}} & 
\raisebox{1.2ex}{\textbf{Code rate}} & 
\raisebox{1.2ex}{\textbf{Error rate}} &
\shortstack{\textbf{No bias}\\Unequal | Equal} & \shortstack{\textbf{Moderate bias}\\Unequal | Equal} & \shortstack{\textbf{Strong bias}\\Unequal | Equal} \\
\midrule
\multirow{8}{*}{\textbf{Aeon}} & \multirow{2}{*}{Low} & \multirow{2}{*}{0.50} & 0.01 & 3 | 3.50 & 3.50 | 2.50 & 5.50 | 2.50 \\
 & & & 0.05 & 7.5 | 9 & 8 | 11 & 10.50 | 17.50 \\
\cmidrule(lr){2-7}
 & \multirow{3}{*}{Medium} & \multirow{3}{*}{1} & 0.01 & 3 | 3.25 & 2.75 | 3.25 & 4.25 | 5 \\
 & & & 0.05 & 5 | 6.25 & 5.50 | 6.25 & 10.25 | 13.25 \\
 & & & 0.10 & 10 | 11.50 & 11.50 | 12 & 20.75 | 22.51 \\
\cmidrule(lr){2-7}
 & \multirow{3}{*}{High} & \multirow{3}{*}{1.50} & 0.01 & 4.15 | 4.83 & 6.15 | 8 & 19.60 | - \\
 & & & 0.05 & 7.14 | 7.50 & 14.12 | 13 & - \\
 & & & 0.10 & 14.78 | - & - & - \\
\midrule
\multirow{3}{*}{\textbf{Fountain}} & \multirow{2}{*}{Medium} & \multirow{2}{*}{1} & 0.01 & 12.23 | 11.23 & 23.46 | 25.95 & - \\
 & & & 0.05 & 25.70 | - & - & - \\
\cmidrule(lr){2-7}
 & \multirow{1}{*}{High} & \multirow{1}{*}{1.50} & 0.01 & 10.65 | - & - & - \\
\midrule
\multirow{7}{*}{\textbf{HEDGES}} & \multirow{4}{*}{Low} & \multirow{4}{*}{0.65} & 0.01 & 4.99 | 5.37 & 7.29 | 7.29 & 17.66 | 21.88 \\
 & & & 0.05 & 5.76 | 6.14 & 9.98 | 8.06 & 21.50 | 21.88 \\
 & & & 0.10 & 9.60 | 9.98 & 14.20 | 14.20 & 33.78 | 30.71 \\
 & & & 0.15 & 25.72 | 19.96 & 40.31 | 31.10 & - \\
\cmidrule(lr){2-7}
 & \multirow{3}{*}{Medium} & \multirow{3}{*}{1.09} & 0.01 & 2.97 | 2.97 & 4.12 | 4.35 & 10.30 | 10.07 \\
 & & & 0.05 & 5.49 | 5.72 & 8.24 | 8.69 & 21.28 | 21.74 \\
 & & & 0.10 & 9.15 | 9.15 & 15.33 | 14.18 & - \\
\midrule
\multirow{8}{*}{\textbf{LDPC}} & \multirow{3}{*}{Low} & \multirow{3}{*}{0.58} & 0.01 & 4.71 | 6.42 & 5.57 | 7.71 & 9.85 | 13.70 \\
 & & & 0.05 & 11.56 | 11.56 & 14.99 | 14.99 & 27.83 | 30.40 \\
 & & & 0.10 & 31.68 | 31.68 & 41.53 | 47.95 & - \\
\cmidrule(lr){2-7}
 & \multirow{3}{*}{Low} & \multirow{3}{*}{0.67} & 0.01 & 5.20 | 7.05 & 6.68 | 8.91 & 14.84 | 18.93 \\
 & & & 0.05 & 12.99 | 12.62 & 17.07 | 18.18 & 37.85 | 40.08 \\
 & & & 0.10 & 34.14 | - & 47.50 | - & - \\
\cmidrule(lr){2-7}
 & \multirow{2}{*}{Medium-Low} & \multirow{2}{*}{0.80} & 0.01 & 8.79 | 10.36 & 14.45 | 18.53 & - \\
 & & & 0.05 & 19.47 | 19.16 & 34.86 | 34.23 & - \\
\midrule
\multirow{9}{*}{\textbf{MGC+}} & \multirow{4}{*}{Low} & \multirow{4}{*}{0.50} & 0.01 & 2.50 | 2.50 & 3 | 3 & 5 | 5 \\
 & & & 0.05 & 3.50 | 3.50 & 4 | 3.50 & 6.50 | 6 \\
 & & & 0.10 & 7.51 | 7.51 & 8.51 | 9.01 & 15.01 | 15.01 \\
 & & & 0.15 & 25.02 | 19.51 & 32.52 | 23.02 & 60.54 | 42.03 \\
\cmidrule(lr){2-7}
 & \multirow{3}{*}{Medium} & \multirow{3}{*}{1} & 0.01 & 1.50 | 1.50 & 1.75 | 1.75 & 3.25 | 3 \\
 & & & 0.05 & 4 | 4 & 5 | 4.75 & 9.26 | 8.76 \\
 & & & 0.10 & 8.76 | 7.01 & 11.26 | 8.76 & 22.50 | 17.77 \\
\cmidrule(lr){2-7}
 & \multirow{2}{*}{High} & \multirow{2}{*}{1.50} & 0.01 & 2.83 | 3 & 3.50 | 3.67 & 7.50 | 7.50 \\
 & & & 0.05 & 7.67 | 6.33 & 9.83 | 8.50 & 20.84 | 18.67 \\
\midrule
\multirow{9}{*}{\textbf{RS}} & \multirow{3}{*}{Low} & \multirow{3}{*}{0.50} & 0.01 & 2 | 2.50 & 2 | 2.50 & 2.50 | 3 \\
 & & & 0.05 & 6.51 | 6.51 & 6.51 | 6.51 & 8.51 | 8.01 \\
 & & & 0.10 & 17.51 | 13.51 & 17.52 | 14.51 & 20.52 | 18.02 \\
\cmidrule(lr){2-7}
 & \multirow{3}{*}{Medium} & \multirow{3}{*}{1} & 0.01 & 2.76 | 3.01 & 3.01 | 3.26 & 4.51 | 5.01 \\
 & & & 0.05 & 5.51 | 5.51 & 6.51 | 6.51 & 10.27 | 10.02 \\
 & & & 0.10 & 13.53 | 11.52 & 16.03 | 14.03 & 25.80 | 24.05 \\
\cmidrule(lr){2-7}
 & \multirow{3}{*}{High} & \multirow{3}{*}{1.50} & 0.01 & 3.17 | 3.51 & 4.34 | 4.84 & 9.52 | 10.85 \\
 & & & 0.05 & 6.18 | 5.84 & 8.35 | 8.18 & 20.37 | - \\
 & & & 0.10 & 15.36 | 16.20 & - & - \\
\midrule
\multirow{5}{*}{\textbf{Stairloop}} & \multirow{2}{*}{Low} & \multirow{2}{*}{0.49} & 0.01 & 12.71 | 13.73 & 20.85 | 34.06 & - \\
 & & & 0.05 & 24.40 | 31.01 & 63 | 58.98 & - \\
\cmidrule(lr){2-7}
 & \multirow{3}{*}{Medium} & \multirow{3}{*}{0.98} & 0.01 & 4.83 | 6.10 & 12.46 | 10.93 & - \\
 & & & 0.05 & 12.46 | 12.20 & 24.40 | - & - \\
 & & & 0.10 & 30.51 | - & - & - \\
\bottomrule
\end{tabular}
\label{tab:read_cost}
\end{table*}

\begin{table*}[ht]
\centering
\caption{Average decoding time (seconds), measured at minimum reliable coverage depth, for different codec configurations, error rates, and bias conditions, under the synthetic channel model (in silico). Entries are reported as unequal | equal; equal denotes $P_d=P_s=P_i$, whereas unequal denotes $P_s=0.572P_e$, $P_d=0.447P_e$, and $P_i=0.026P_e$.}
\renewcommand{\arraystretch}{0.85}
\setlength{\tabcolsep}{6pt}
\begin{tabular}{c l c c c c c}
\toprule
\raisebox{1.2ex}{\textbf{Codec}} & \raisebox{1.2ex}{\textbf{Class}} & \raisebox{1.2ex}{\textbf{Code rate}} & \raisebox{1.2ex}{\textbf{Error rate}} &
\shortstack{\textbf{No bias}\\Unequal | Equal} & \shortstack{\textbf{Moderate bias}\\Unequal | Equal} & \shortstack{\textbf{Strong bias}\\Unequal | Equal} \\
\midrule
\multirow{8}{*}{\textbf{Aeon}} & \multirow{2}{*}{Low} & \multirow{2}{*}{0.50} & 0.01 & 1715.46 | 4141.76 & 1721.73 | 3571.20 & 925.15 | 2626.28 \\
 & & & 0.05 & 5931.75 | 6544.29 & 6539.69 | 6784.33 & 6408.26 | 6556.18 \\
\cmidrule(lr){2-7}
 & \multirow{3}{*}{Medium} & \multirow{3}{*}{1} & 0.01 & 496.54 | 1314.22 & 587.18 | 1395.69 & 451.27 | 1119.79 \\
 & & & 0.05 & 1985.66 | 1285.73 & 2802.16 | 2727.98 & 2503 | 2130.38 \\
 & & & 0.10 & 3035.07 | 3452.06 & 4020.46 | 4669.22 & 4613.66 | 5347.59 \\
\cmidrule(lr){2-7}
 & \multirow{3}{*}{High} & \multirow{3}{*}{1.50} & 0.01 & 66.19 | 80.08 & 74.64 | 75.92 & 61.05 | - \\
 & & & 0.05 & 58.88 | 71.48 & 62.67 | 91.97 & - \\
 & & & 0.10 & 1494.41 | - & - & - \\
\midrule
\multirow{2}{*}{\textbf{Fountain}} & \multirow{2}{*}{Medium} & \multirow{2}{*}{1} & 0.01 & 0.61 | 0.62 & 0.59 | 0.61 & - \\
 & & & 0.05 & 0.62 | - & - & - \\
\midrule
\multirow{7}{*}{\textbf{HEDGES}} & \multirow{4}{*}{Low} & \multirow{4}{*}{0.65} & 0.01 & 10.02 | 6.68 & 7.56 | 6.68 & 6.71 | 6.56 \\
 & & & 0.05 & 11.33 | - & 8.73 | 10.80 & 7.80 | 8.10 \\
 & & & 0.10 & 52.70 | 52 & 55.27 | 58.26 & 64.61 | 72.21 \\
 & & & 0.15 & 2248.37 | 1510.76 & 2921.46 | 1906.64 & - \\
\cmidrule(lr){2-7}
 & \multirow{3}{*}{Medium} & \multirow{3}{*}{1.09} & 0.01 & 5.18 | 5.49 & 5.34 | 5.90 & 4.66 | 5.06 \\
 & & & 0.05 & 13.20 | 13 & 13.33 | 12.83 & 10.60 | 12.75 \\
 & & & 0.10 & 35.01 | 42.23 & 40.12 | 47.39 & - \\
\midrule
\multirow{8}{*}{\textbf{LDPC}} & \multirow{3}{*}{Low} & \multirow{3}{*}{0.58} & 0.01 & 164.53 | 151.25 & 161.02 | 174.68 & 155.83 | - \\
 & & & 0.05 & 171.66 | 183.83 & 168.34 | 177.13 & 163.43 | 169.40 \\
 & & & 0.10 & 176.77 | 192.31 & 174.87 | 192.49 & - \\
\cmidrule(lr){2-7}
 & \multirow{3}{*}{Low} & \multirow{3}{*}{0.67} & 0.01 & 138.75 | 157.88 & 147.18 | 145.56 & 141.58 | 134.17 \\
 & & & 0.05 & 147.44 | 157 & 112.61 | 154.67 & 138.27 | 122.01 \\
 & & & 0.10 & 101 | - & 129.22 | - & - \\
\cmidrule(lr){2-7}
 & \multirow{2}{*}{Medium-Low} & \multirow{2}{*}{0.80} & 0.01 & 123.31 | 135.56 & 131.28 | 132.21 & - \\
 & & & 0.05 & 125.03 | 130.78 & 124.84 | 125.98 & - \\
\midrule
\multirow{9}{*}{\textbf{MGC+}} & \multirow{4}{*}{Low} & \multirow{4}{*}{0.50} & 0.01 & 38.51 | 44.41 & 37.97 | 42.63 & 39.08 | 43.46 \\
 & & & 0.05 & 62.64 | 85.75 & 58.18 | 80.59 & 54.74 | 70.32 \\
 & & & 0.10 & 218.19 | 221.23 & 216.29 | 224.36 & 193.94 | 216.05 \\
 & & & 0.15 & 3624.93 | 2466.74 & 4142.30 | 2644.65 & 5687.49 | 3541.33 \\
\cmidrule(lr){2-7}
 & \multirow{3}{*}{Medium} & \multirow{3}{*}{1} & 0.01 & 40.88 | 139.09 & 37.56 | 116.78 & 31.34 | 90.04 \\
 & & & 0.05 & 594.15 | 1079.25 & 529.17 | 1063.09 & 657.98 | 886.38 \\
 & & & 0.10 & 1223.52 | 1774.63 & 1046 | 1764.74 & 1599.47 | 1764.34 \\
\cmidrule(lr){2-7}
 & \multirow{2}{*}{High} & \multirow{2}{*}{1.50} & 0.01 & 6.65 | 6.78 & 6.88 | 6.36 & 6.77 | 9.79 \\
 & & & 0.05 & 6.57 | 6.49 & 6.67 | 6.46 & 6.61 | 6.22 \\
\midrule
\multirow{9}{*}{\textbf{RS}} & \multirow{3}{*}{Low} & \multirow{3}{*}{0.50} & 0.01 & 2210.36 | 2625.96 & 2410.36 | 2652.63 & 2347.29 | 2603.80 \\
 & & & 0.05 & 2251.17 | 2466.59 & 2419.03 | 2601.99 & 2327.53 | - \\
 & & & 0.10 & 2224.46 | 2566.19 & 2496.38 | 2469.72 & 2644.44 | 2507.66 \\
\cmidrule(lr){2-7}
 & \multirow{3}{*}{Medium} & \multirow{3}{*}{1} & 0.01 & 169.56 | 180.53 & 189.42 | 206.43 & 197 | 200.26 \\
 & & & 0.05 & 191.69 | 191.59 & 191.16 | 190.71 & 198.28 | 207.99 \\
 & & & 0.10 & 203.49 | 209.68 & - | 211.17 & 205.81 | 211.01 \\
\cmidrule(lr){2-7}
 & \multirow{3}{*}{High} & \multirow{3}{*}{1.50} & 0.01 & 8.20 | 6.47 & 8.16 | 6.95 & 8.48 | 8.21 \\
 & & & 0.05 & 8.64 | 9.19 & 9.44 | 9.67 & 7.86 | - \\
 & & & 0.10 & 8.64 | 8.43 & - & - \\
\midrule
\multirow{5}{*}{\textbf{Stairloop}} & \multirow{2}{*}{Low} & \multirow{2}{*}{0.49} & 0.01 & 606.85 | 689.71 & 1023.71 | 1688.34 & - \\
 & & & 0.05 & 1632.23 | 1828.34 & 4598.35 | 3464.44 & - \\
\cmidrule(lr){2-7}
 & \multirow{3}{*}{Medium} & \multirow{3}{*}{0.98} & 0.01 & 247.24 | 313.80 & 590.52 | 552.66 & - \\
 & & & 0.05 & 882.21 | 731.87 & 1628.82 | - & - \\
 & & & 0.10 & 2747.86 | - & - & - \\
\bottomrule
\end{tabular}
\label{tab:avg_decode_time}
\end{table*}

\begin{table*}[ht]
\centering
\caption{Performance metrics for different codec configurations under the low-fidelity experimentally derived error and bias profiles (in silico) for a fixed physical redundancy of 100\texttimes.}
\renewcommand{\arraystretch}{1.2}
\setlength{\tabcolsep}{6pt}
\begin{tabular}{c l c c c c}
\toprule
\textbf{Codec} & \textbf{Class} & \textbf{Code rate} & \textbf{Sequencing depth} &
\textbf{Read cost (nts/bit)} & \textbf{Avg. decoding time (s)} \\
\midrule
\multirow{3}{*}{\textbf{Aeon}} & \multirow{1}{*}{Low} & \multirow{1}{*}{0.50} & 12.25 & 24.49 & 2440.82 \\
\cmidrule(lr){2-6}
 & \multirow{1}{*}{Medium} & \multirow{1}{*}{1} & 9.75 & 9.75 & 1533.13 \\
\midrule
\multirow{2}{*}{\textbf{HEDGES}} & \multirow{1}{*}{Low} & \multirow{1}{*}{0.65} & 16.75 & 25.72 & 7.25 \\
\cmidrule(lr){2-6}
 & \multirow{1}{*}{Medium} & \multirow{1}{*}{1.09} & 20.25 & 18.53 & 10.65 \\
\midrule
\multirow{3}{*}{\textbf{MGC+}} & \multirow{1}{*}{Low} & \multirow{1}{*}{0.50} & 2.50 & 5 & 41.82 \\ 
\cmidrule(lr){2-6}
 & \multirow{1}{*}{Medium} & \multirow{1}{*}{1} & 4.50 & 4.50 & 666.87 \\
 \cmidrule(lr){2-6}
 & \multirow{1}{*}{Optimized} & \multirow{1}{*}{0.5} & 1 & 2 & 1251.82 \\
\midrule
\multirow{3}{*}{\textbf{RS}} & \multirow{1}{*}{Low} & \multirow{1}{*}{0.50} & 3.25 & 6.51 & 2352.07 \\
\cmidrule(lr){2-6}
 & \multirow{1}{*}{Medium} & \multirow{1}{*}{1} & 12.75 & 12.77 & 210.18 \\
\bottomrule
\end{tabular}
\label{tab:sequencing_depth}
\end{table*}

\begin{table*}[ht]
\centering
\caption{Performance metrics for different codec configurations under the low-fidelity experimentally derived error and bias profiles (in silico) for a fixed sequencing depth of 30\texttimes.}
\renewcommand{\arraystretch}{1.2}
\setlength{\tabcolsep}{6pt}
\begin{tabular}{c l c c c c}
\toprule
\textbf{Codec} & \textbf{Class} & \textbf{Code rate} & \textbf{Physical redundancy} & 
\textbf{Write cost (nts/bit)} & \textbf{Avg. decoding time (s)} \\
\midrule
\multirow{3}{*}{\textbf{Aeon}} & \multirow{1}{*}{Low} & \multirow{1}{*}{0.50} & 13.50 & 26.99 & 3223.77 \\
\cmidrule(lr){2-6}
 & \multirow{1}{*}{Medium} & \multirow{1}{*}{1} & 14.50 & 14.51 & 1584.41 \\
\midrule
\multirow{2}{*}{\textbf{HEDGES}} & \multirow{1}{*}{Low} & \multirow{1}{*}{0.65} & 15.75 & 24.19 & 8.73 \\
\cmidrule(lr){2-6}
 & \multirow{1}{*}{Medium} & \multirow{1}{*}{1.09} & 30.75 & 28.14 & 8.12 \\
\midrule
\multirow{3}{*}{\textbf{MGC+}} & \multirow{1}{*}{Low} & \multirow{1}{*}{0.50} & 2.25 & 4.50 & 40.78  \\
\cmidrule(lr){2-6}
 & \multirow{1}{*}{Medium} & \multirow{1}{*}{1} & 4.25 & 4.25 & 642.49 \\
 \cmidrule(lr){2-6}
 & \multirow{1}{*}{Optimized} & \multirow{1}{*}{0.50} & 1 & 2 & 1190.30 \\
\midrule
\multirow{3}{*}{\textbf{RS}} & \multirow{1}{*}{Low} & \multirow{1}{*}{0.50} & 5 & 10.01 & 2492.93 \\
\cmidrule(lr){2-6}
 & \multirow{1}{*}{Medium} & \multirow{1}{*}{1} & 20.75 & 20.79 & 214.40 \\
\bottomrule
\end{tabular}
\label{tab:physical_redundancy}
\end{table*}

\newcolumntype{C}[1]{>{\centering\arraybackslash}p{#1}}

\begin{table}[t]
\centering
\caption{Minimum reliable sequencing depth, read cost, and average decoding time~(serial and parallel), under Illumina sequencing and Nanopore sequencing with different basecallers (in vitro).}
\label{tab:nanopore_supp}
\renewcommand{\arraystretch}{1.2}
\setlength{\tabcolsep}{6pt}
\begin{tabular}{C{2cm} lcccc}
\hline
\textbf{Sequencing} &
\multirow{2}{*}{\textbf{Codec}} &
\textbf{Min. reliable} &
\textbf{Code rate} &
\textbf{Read cost} &
\textbf{Avg. decoding time (s)} \\
 \textbf{method} & & \textbf{sequencing depth} & \textbf{(bits/nt)} & \textbf{(nts/bit)} & \textbf{1 core | 8 cores} \\
\hline

\multirow{6}{*}{Illumina}
& MGC+ (A, unfiltered)   & 2.75 & 1.032 & 2.66 & 33.34 | 8.08 \\
& MGC+ (A, filtered)    & 2.50 & 1.032 & 2.42 & 207.74 | 13.98 \\
& MGC+ (B, unfiltered)  & 2.75 & 0.706 & 3.90 & 58.66 | 7.51 \\
& MGC+ (B, filtered)   & 2.25 & 0.706 & 3.19 & 438.83 | 20.09 \\
& HEDGES            & 9.50 & 0.610 & 15.57 & 32.48 | -- \\
& Aeon              & 3.00 & 1.000 & 3.00 & 571.12 | 119.23 \\
\hline

\multirow{6}{*}{\begin{tabular}{c}Nanopore\\(dorado-sup)\end{tabular}}
 & MGC+ (A, unfiltered) & 4.00 & 1.032 & 3.88 & 114.65 | 14.53 \\
 & MGC+ (A, filtered)  & 3.50 & 1.032 & 3.39 & 193.58 | 22.43 \\
 & MGC+ (B, unfiltered) & 3.00 & 0.706 & 4.25 & 63.60 | 7.98 \\
 & MGC+ (B, filtered)  & 2.75 & 0.706 & 3.90 & 406.03 | 20.89 \\
 & HEDGES              & 7.50 & 0.610 & 12.30 & 31.99 | -- \\
 & Aeon                & 5.00 & 1.000 & 5.00 & 2494.72 | 305.97 \\
\hline

\multirow{6}{*}{\begin{tabular}{c}Nanopore\\(dorado-hac)\end{tabular}}
 & MGC+ (A, unfiltered) & 5.25 & 1.032 & 5.09 & 154.66 | 19.07 \\
 & MGC+ (A, filtered)  & 4.25 & 1.032 & 4.12 & 310.40 | 27.45 \\
 & MGC+ (B, unfiltered) & 3.50 & 0.706 & 4.96 & 67.05 | 8.22 \\
 & MGC+ (B, filtered)  & 3.25 & 0.706 & 4.60 & 471.67 | 21.98 \\
 & HEDGES              & 8.25 & 0.610 & 13.52 & 36.28 | -- \\
 & Aeon                & 6.50 & 1.000 & 6.50 & 3256.71 | 355.40 \\
\hline

\multirow{6}{*}{\begin{tabular}{c}Nanopore\\(dorado-fast)\end{tabular}}
 & MGC+ (A, unfiltered) & 10.75 & 1.032 & 10.42 & 389.59 | 40.42 \\
 & MGC+ (A, filtered)  & 9.25  & 1.032 & 8.96  & 502.40 | 44.36 \\
 & MGC+ (B, unfiltered) & 6.75  & 0.706 & 9.56  & 86.09 | 9.76 \\
 & MGC+ (B, filtered)  & 6.50  & 0.706 & 9.21  & 503.16 | 23.76 \\
 & HEDGES              & 11.50 & 0.610 & 18.85 & 85.37 | -- \\
 & Aeon                & 13.00 & 1.000 & 13.00 & 5071.89 | 537.22 \\
\hline
\end{tabular}
\end{table}


\section*{Supplementary Figures}

\setcounter{figure}{0}

\begin{figure}[h!]
    \centering
    \includegraphics[width=0.85\textwidth]{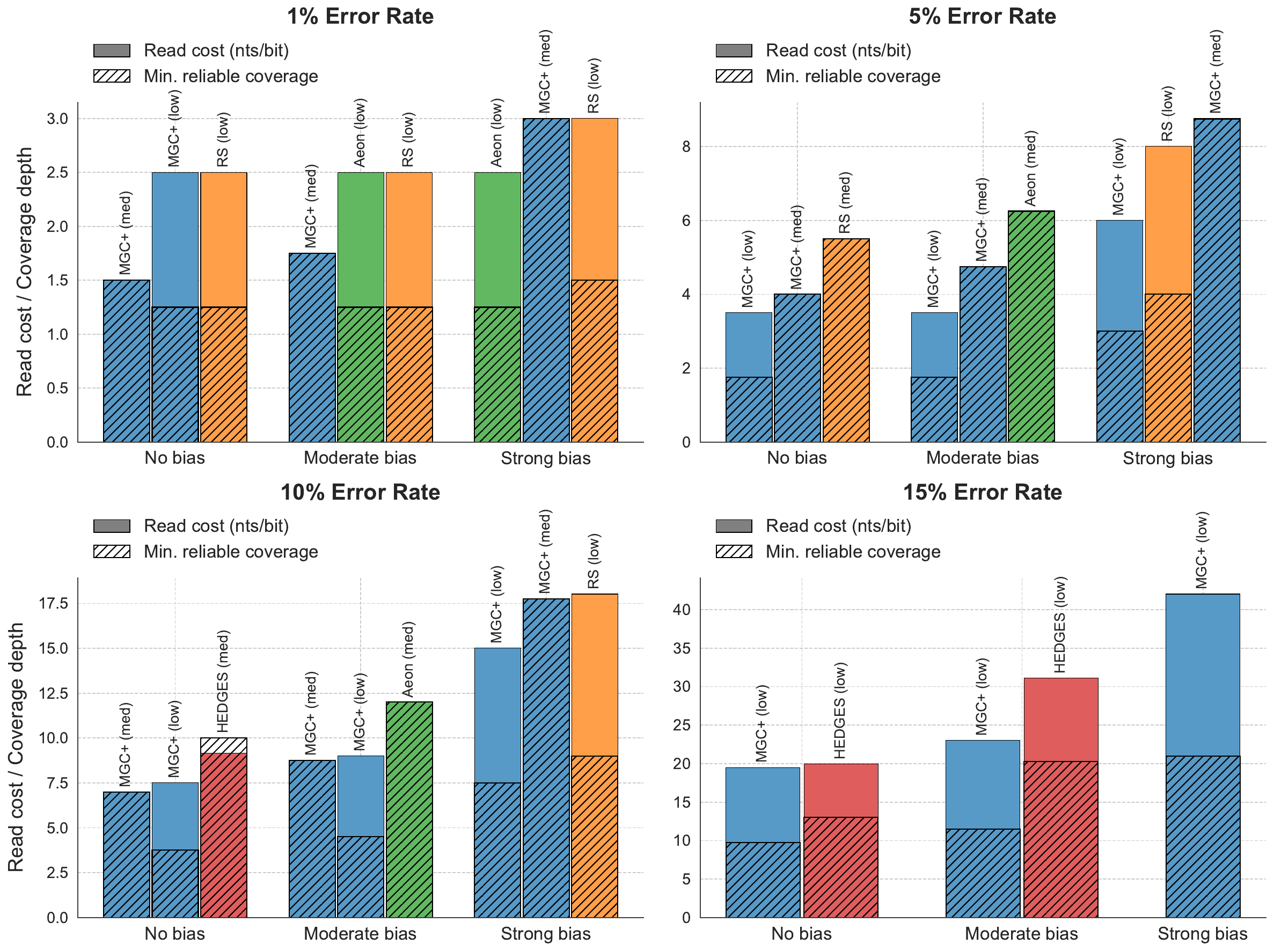} 
    \caption{Minimum coverage depth and associated read cost achieved by the three best-performing codecs across the different bias and error combinations for the case where $P_s=P_d=P_i$.}
\end{figure}

\begin{figure}[h!]
    \centering
    \includegraphics[width=0.55\textwidth]{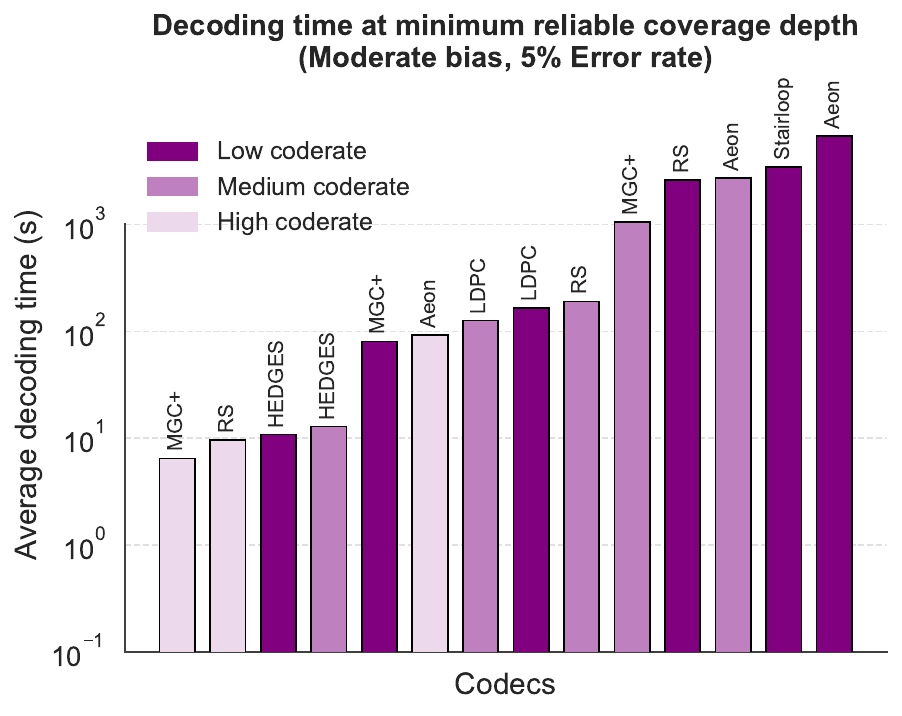} 
    \caption{Average decoding times for the moderate bias and 5\% error rate scenario, measured at the minimum coverage depth required for reliable decoding for the case where $P_s=P_d=P_i$.}
\end{figure}

\begin{figure}[h!]
    \centering
    \includegraphics[width=0.59\textwidth]{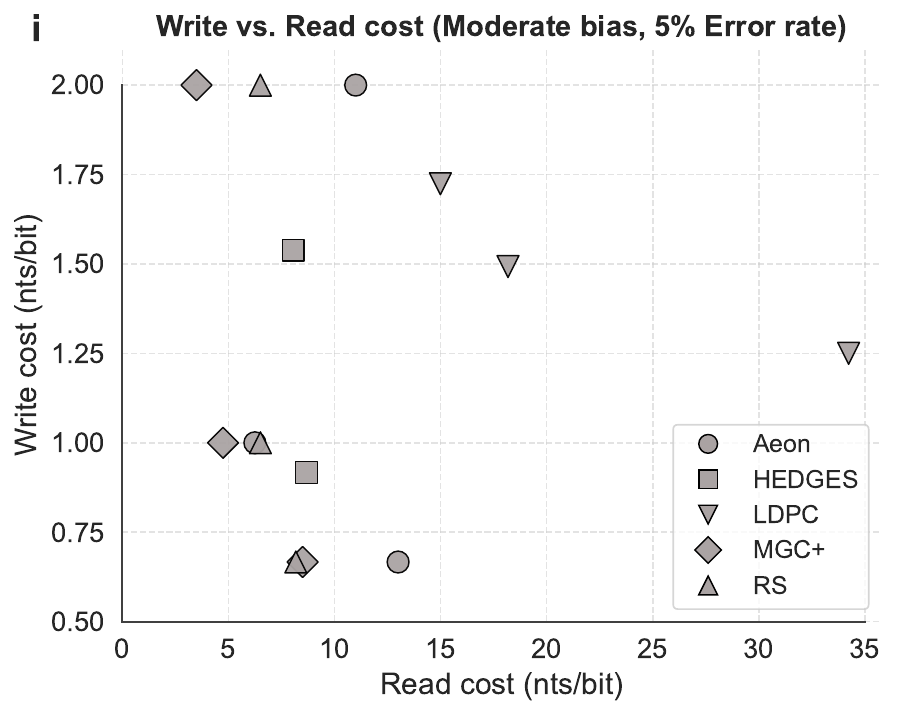} 
    \caption{Trade-off between write cost and read cost for the moderate bias and 5\% error rate scenario for the case where $P_s=P_d=P_i$.}
\end{figure}

\begin{figure}[h!]
    \centering
    \includegraphics[width=0.65\textwidth]{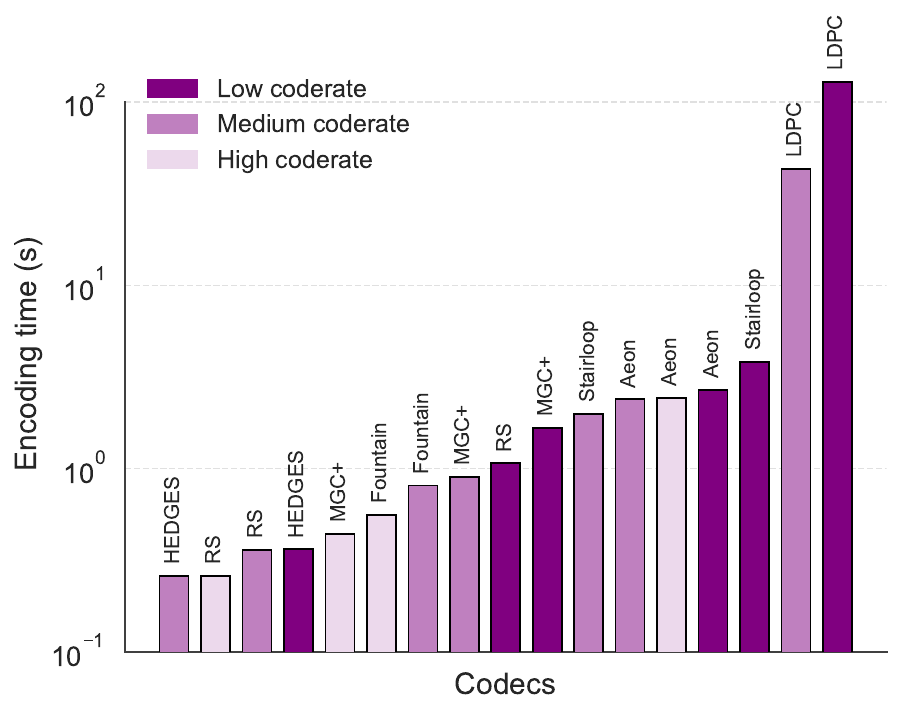} 
    \caption{Encoding times required by different codecs for a 15KB file.}
\end{figure}

\begin{figure}[h!]
    \centering
    \includegraphics[width=0.59\textwidth]{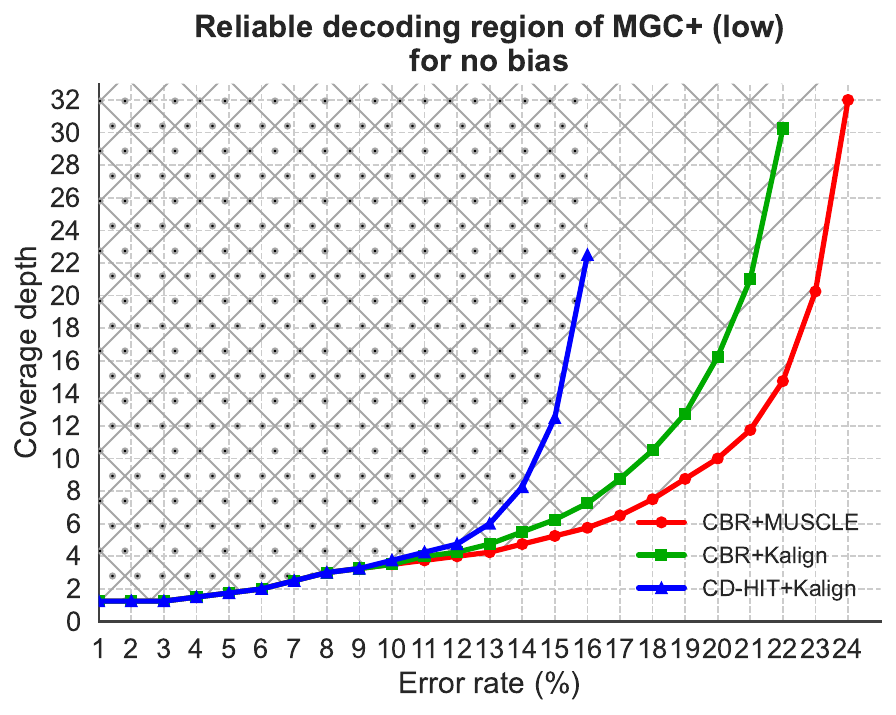} 
    \caption{Reliable decoding region of DNA-MGC+ (low-rate) for the no-bias case.}
\end{figure}

\begin{figure}[h!]
    \centering
    \includegraphics[width=0.59\textwidth]{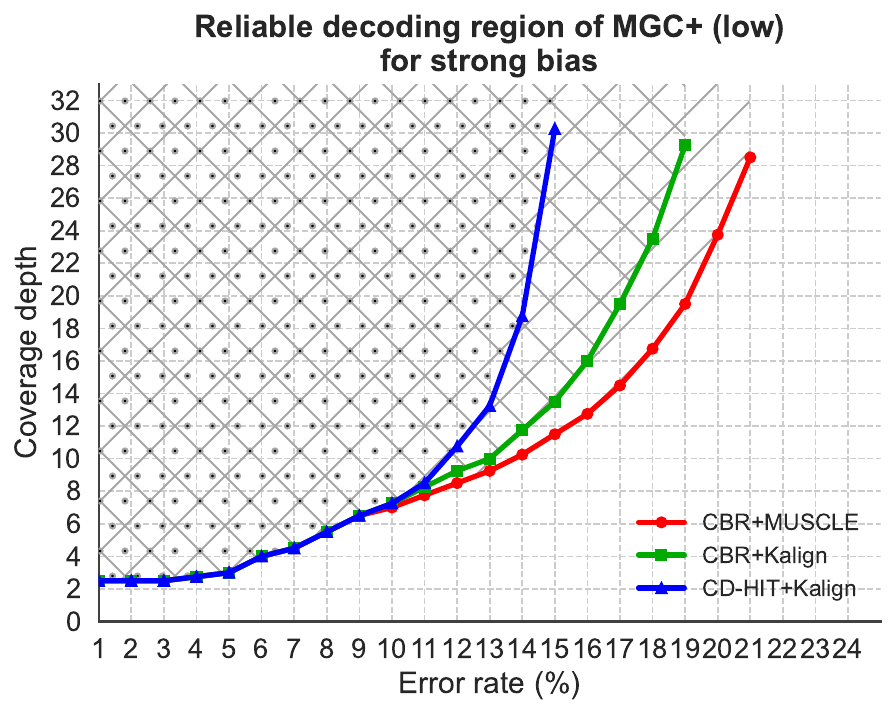}
    \caption{Reliable decoding region of DNA-MGC+ (low-rate) for the strong-bias case.}
\end{figure}

\begin{figure*}[htbp]
\centering
\begin{minipage}{0.8\textwidth}  

\begin{minipage}[c]{0.08\textwidth}
    \centering\rotatebox{90}{\small DNA-Aeon}
\end{minipage}%
\begin{minipage}[c]{0.92\textwidth}
    \includegraphics[width=\textwidth]{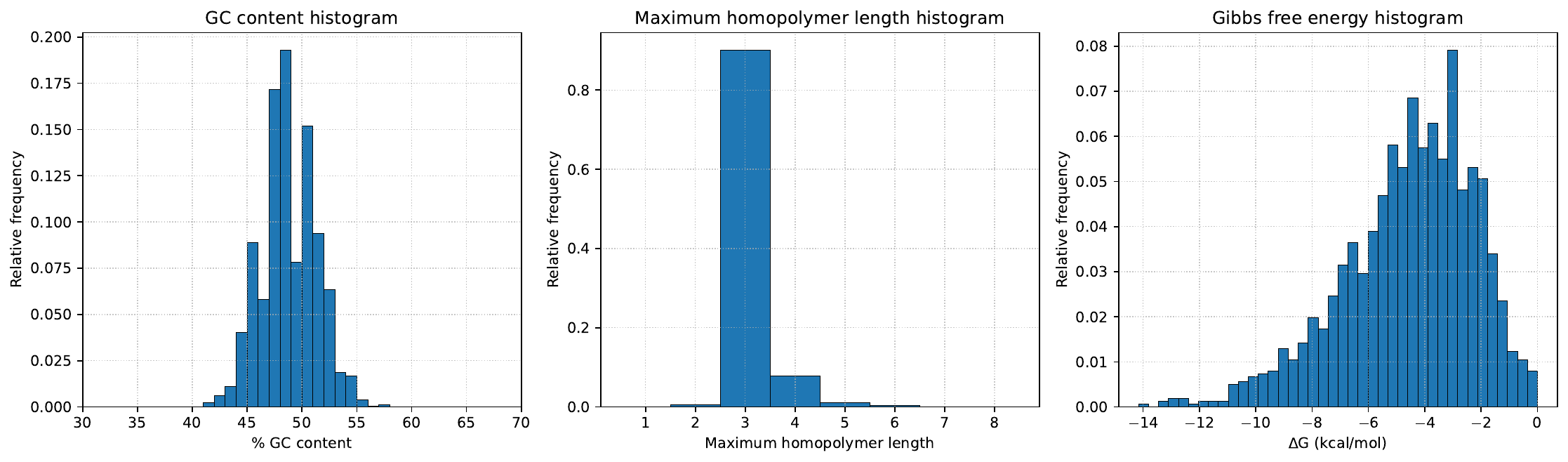}
\end{minipage}

\begin{minipage}[c]{0.08\textwidth}
    \centering\rotatebox{90}{\small Hedges}
\end{minipage}%
\begin{minipage}[c]{0.92\textwidth}
    \includegraphics[width=\textwidth]{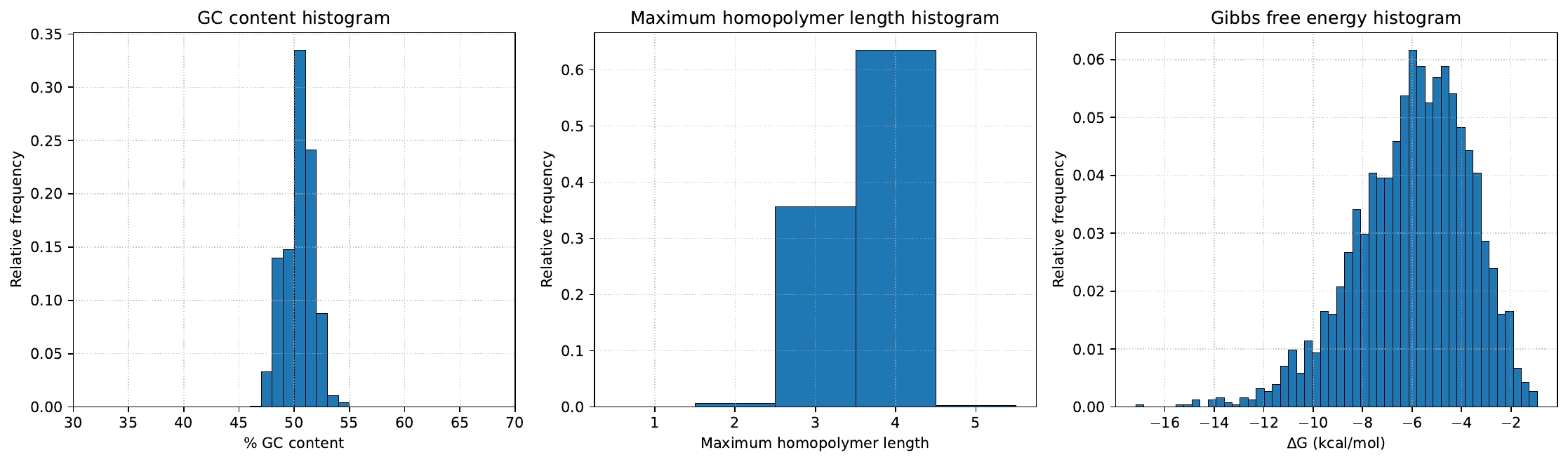}
\end{minipage}

\begin{minipage}[c]{0.08\textwidth}
    \centering\rotatebox{90}{\small MGC+ (A, filtered)}
\end{minipage}%
\begin{minipage}[c]{0.92\textwidth}
    \includegraphics[width=\textwidth]{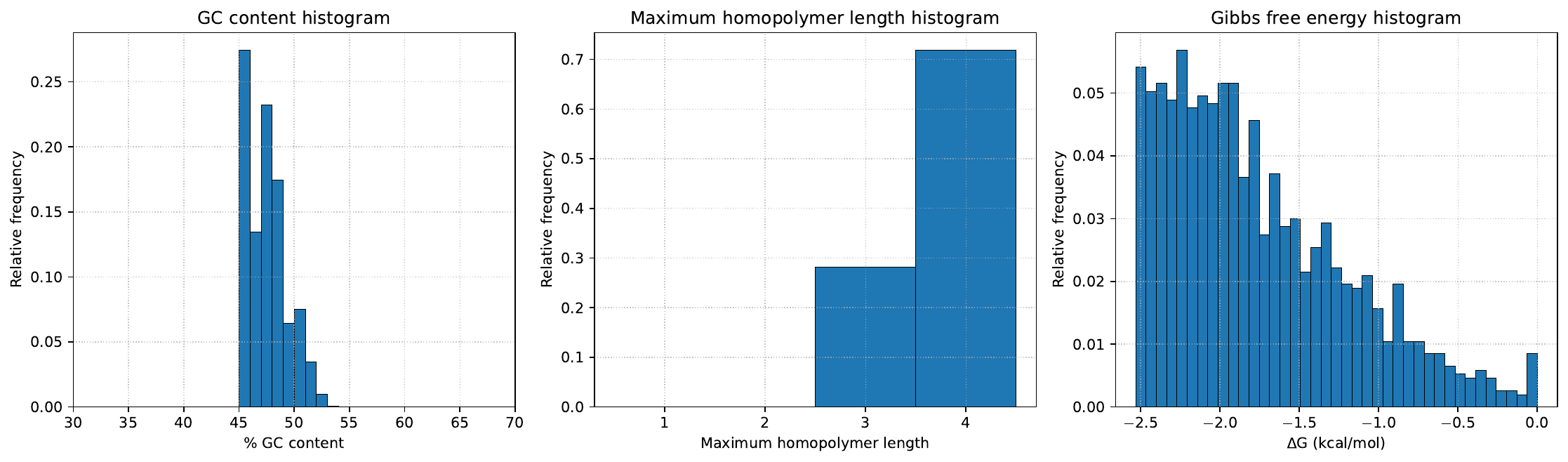}
\end{minipage}

\begin{minipage}[c]{0.08\textwidth}
    \centering\rotatebox{90}{\small MGC+ (A, unfiltered)}
\end{minipage}%
\begin{minipage}[c]{0.92\textwidth}
    \includegraphics[width=\textwidth]{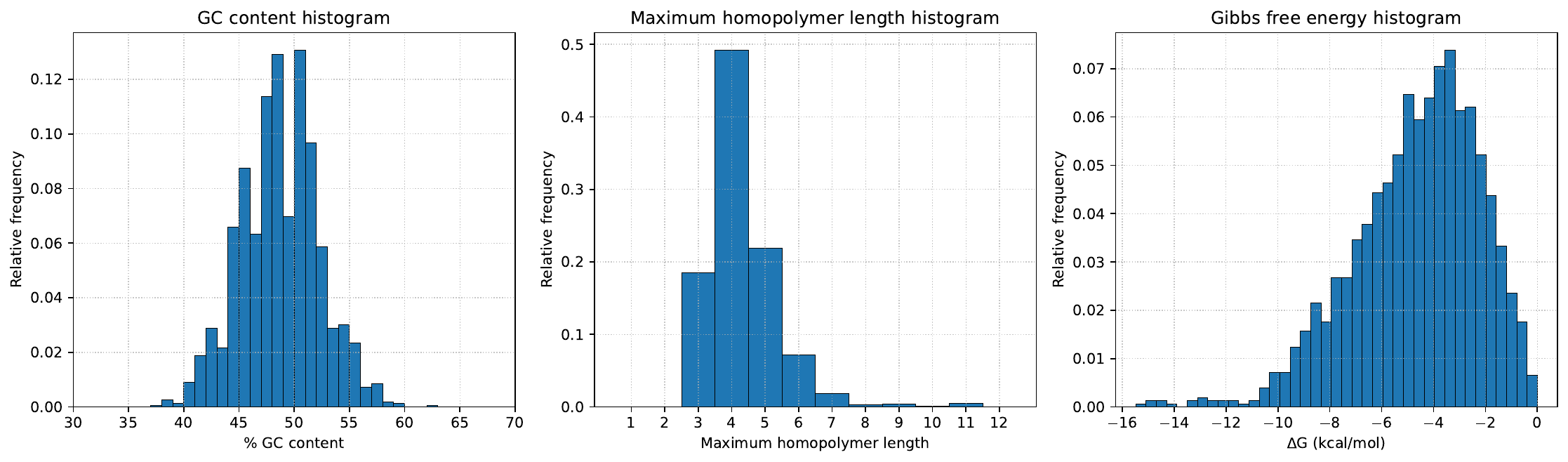}
\end{minipage}

\begin{minipage}[c]{0.08\textwidth}
    \centering\rotatebox{90}{\small MGC+ (B, filtered)}
\end{minipage}%
\begin{minipage}[c]{0.92\textwidth}
    \includegraphics[width=\textwidth]{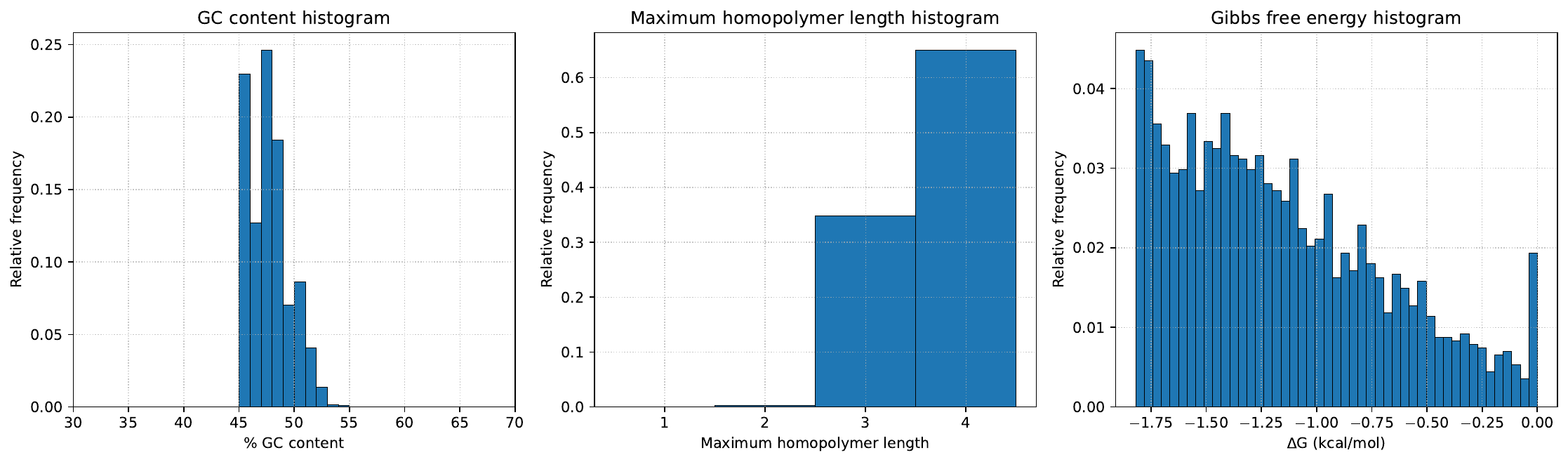}
\end{minipage}

\begin{minipage}[c]{0.08\textwidth}
    \centering\rotatebox{90}{\small MGC+ (B, unfiltered)}
\end{minipage}%
\begin{minipage}[c]{0.92\textwidth}
    \includegraphics[width=\textwidth]{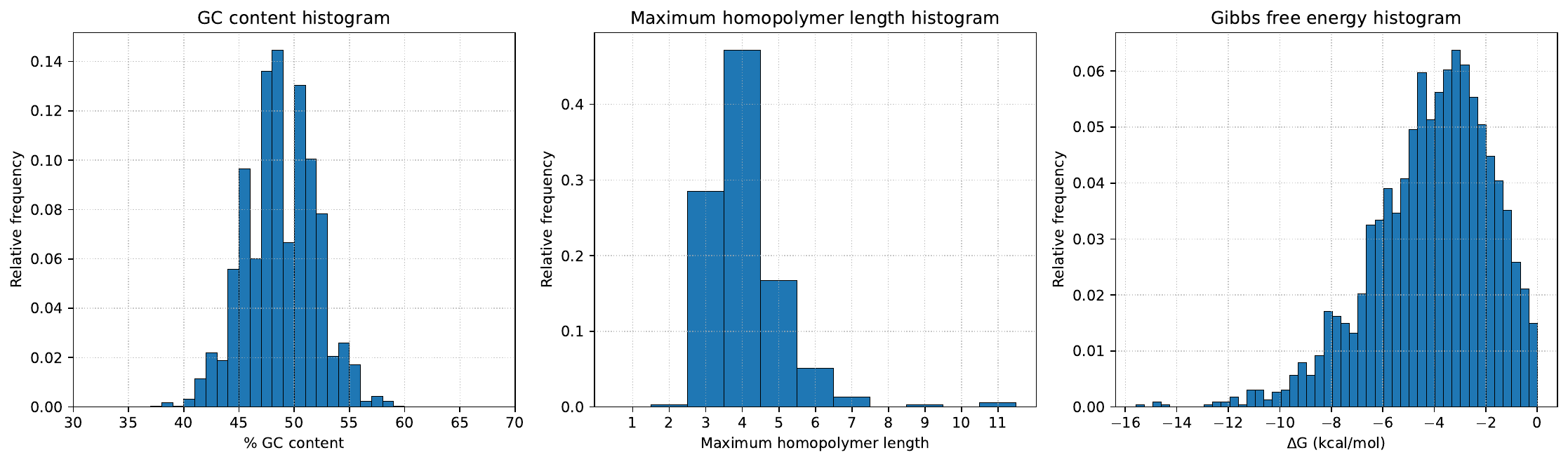}
\end{minipage}

\end{minipage}
\caption{Normalized histograms of GC content, maximum homopolymer length, and Gibbs free energy ($\Delta G$) for the encoded sequences of the different configurations included in the \emph{in vitro} experiment.
}
\end{figure*}

\begin{figure*}[htbp]
\centering


\hspace{-3em}

\begin{minipage}[t]{0.46\textwidth}
    \centering
    \textbf{Illumina}
\end{minipage}
\hspace{-4em}
\begin{minipage}[t]{0.46\textwidth}
    \centering
    \textbf{Nanopore (dorado-sup)}
\end{minipage}

\vspace{0.2em}
\hspace{4em}%
\begin{minipage}[c]{0.05\textwidth}
    \centering
    \rotatebox{90}{\small DNA-Aeon}
\end{minipage}
\begin{minipage}[c]{0.46\textwidth}
    \includegraphics[width=0.62\textwidth]{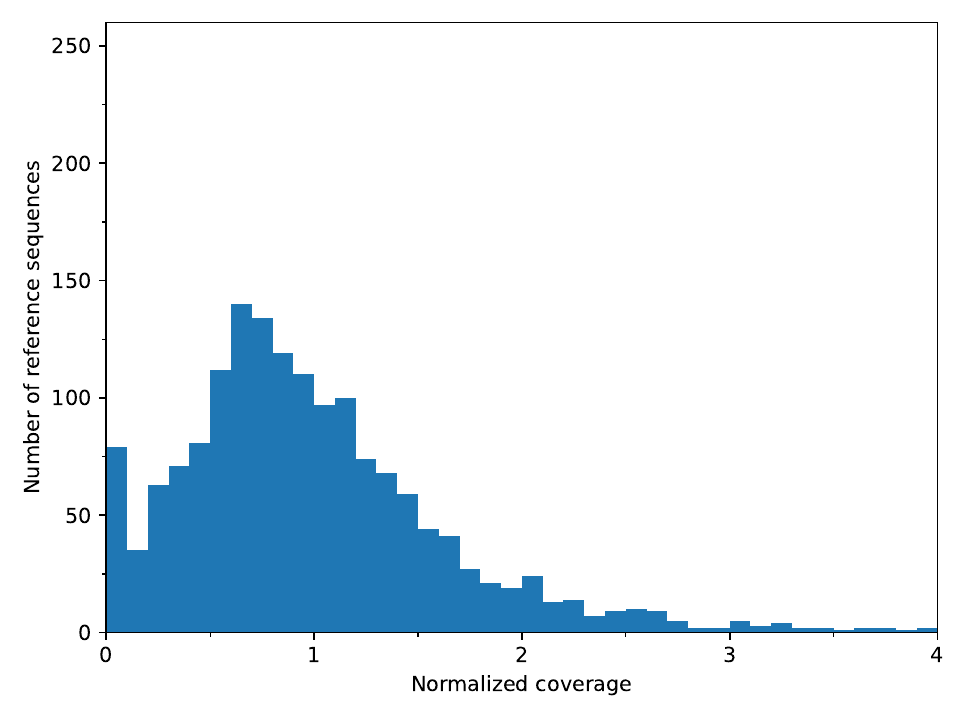}
\end{minipage}
\hspace{-4em}
\begin{minipage}[c]{0.46\textwidth}
    \includegraphics[width=0.62\textwidth]{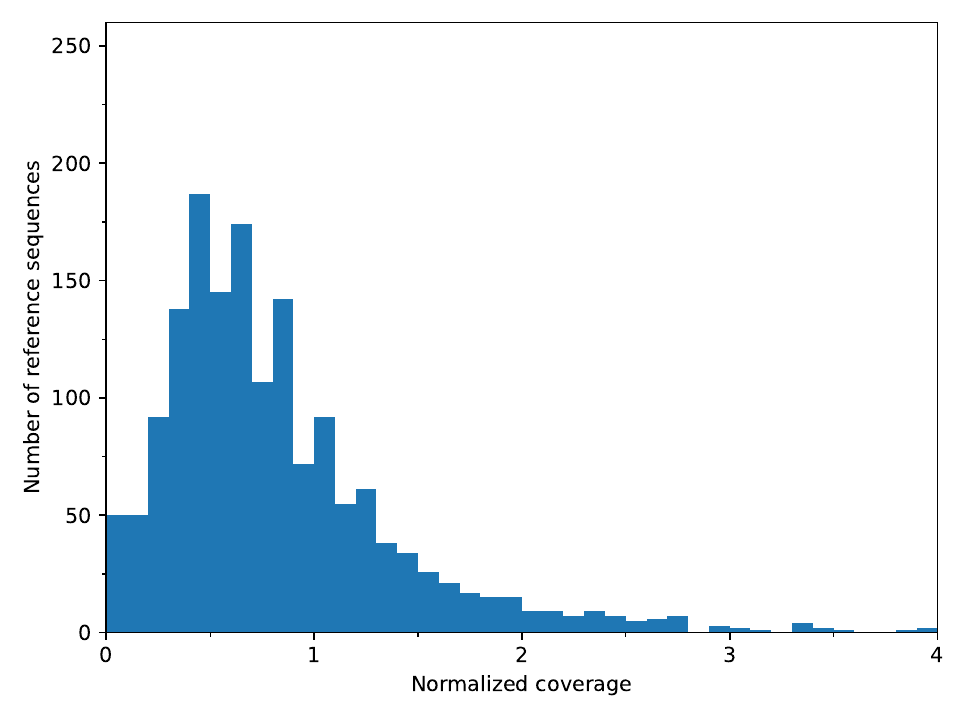}
\end{minipage}

\vspace{0.1em}
\hspace{4em}%
\begin{minipage}[c]{0.05\textwidth}
    \centering
    \rotatebox{90}{\small Hedges}
\end{minipage}
\begin{minipage}[c]{0.46\textwidth}
    \includegraphics[width=0.62\textwidth]{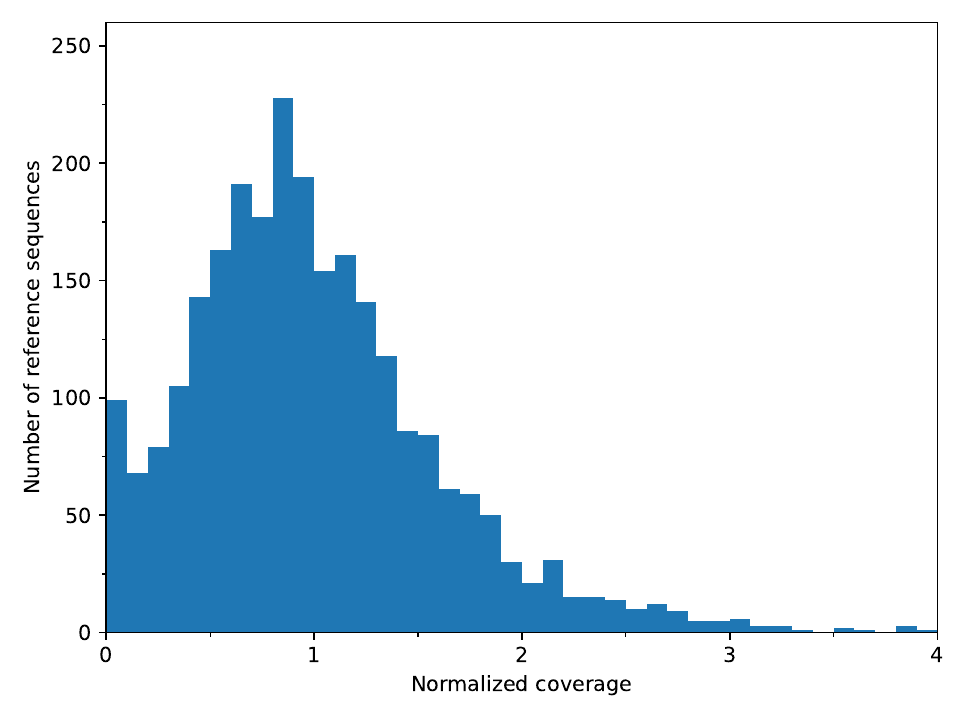}
\end{minipage}
\hspace{-4em}
\begin{minipage}[c]{0.46\textwidth}
    \includegraphics[width=0.62\textwidth]{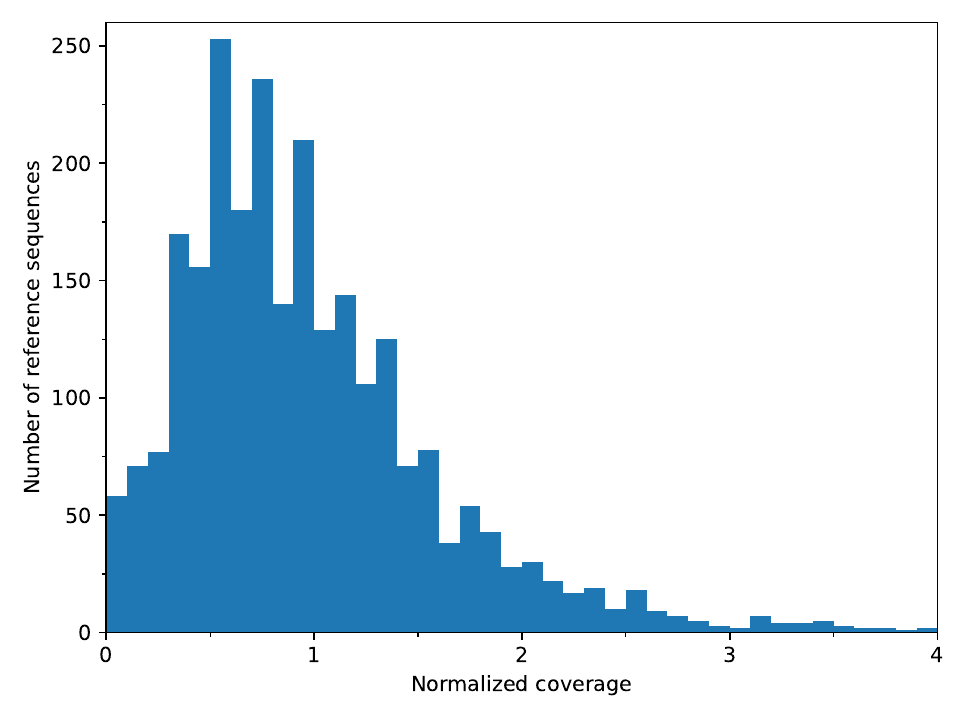}
\end{minipage}

\vspace{0.1em}
\hspace{4em}%
\begin{minipage}[c]{0.05\textwidth}
    \centering
    \rotatebox{90}{\small MGC+ (A, filtered)}
\end{minipage}
\begin{minipage}[c]{0.46\textwidth}
    \includegraphics[width=0.62\textwidth]{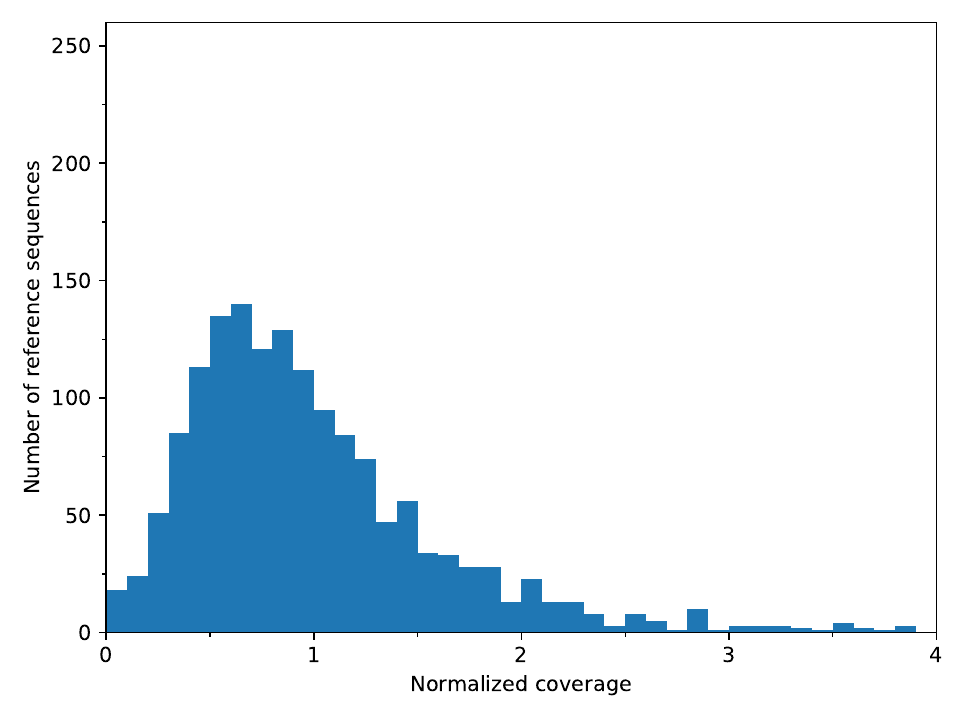}
\end{minipage}
\hspace{-4em}
\begin{minipage}[c]{0.46\textwidth}
    \includegraphics[width=0.62\textwidth]{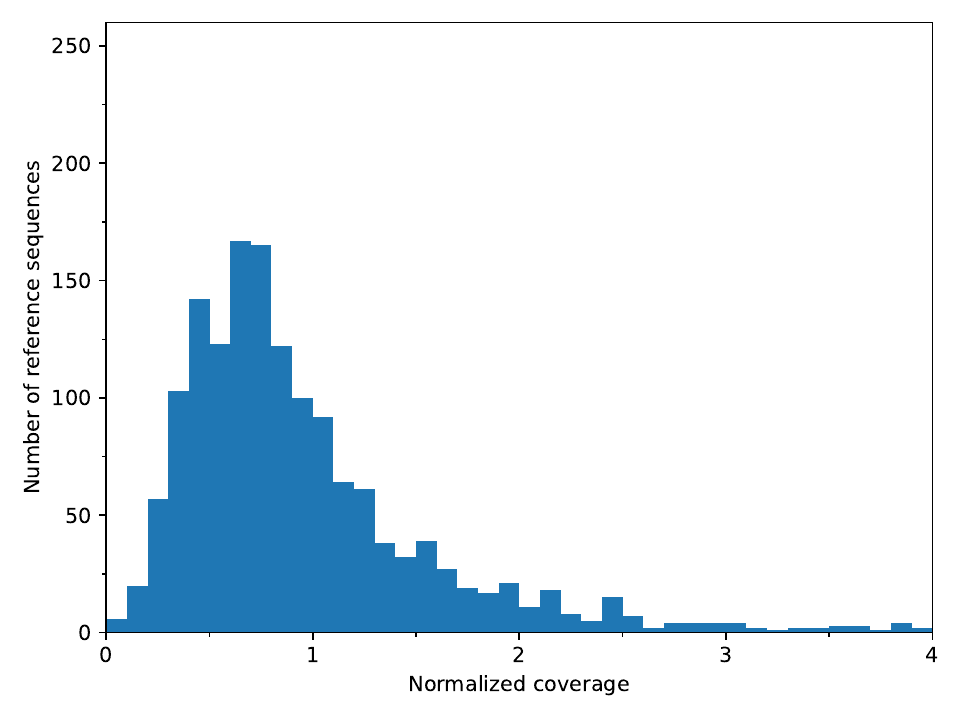}
\end{minipage}

\vspace{0.1em}
\hspace{4em}%
\begin{minipage}[c]{0.05\textwidth}
    \centering
    \rotatebox{90}{\small MGC+ (A, unfiltered)}
\end{minipage}
\begin{minipage}[c]{0.46\textwidth}
    \includegraphics[width=0.62\textwidth]{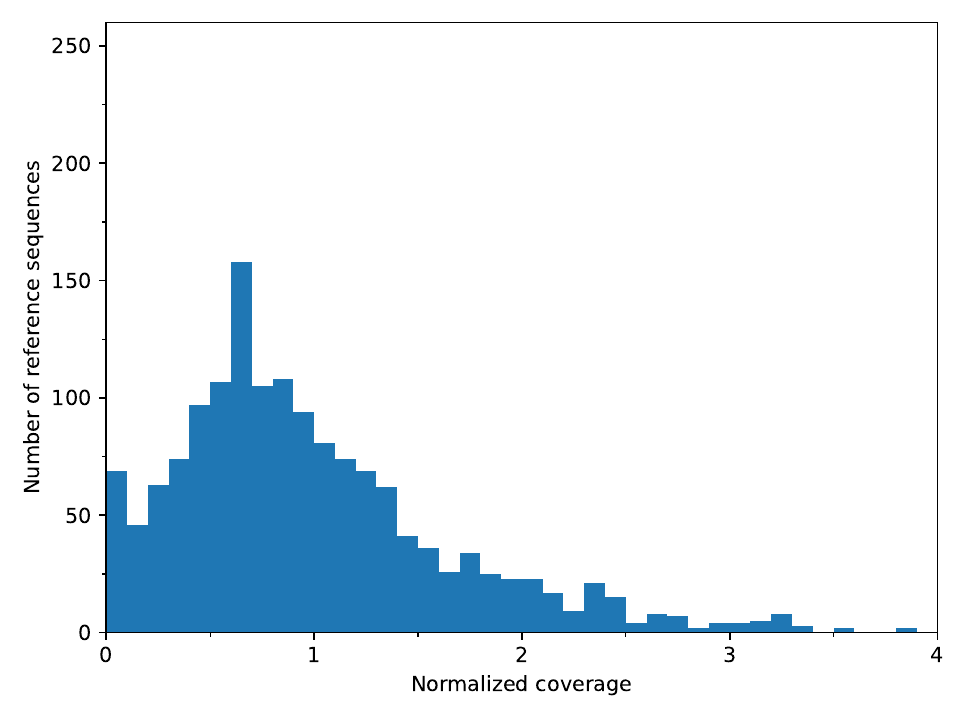}
\end{minipage}
\hspace{-4em}
\begin{minipage}[c]{0.46\textwidth}
    \includegraphics[width=0.62\textwidth]{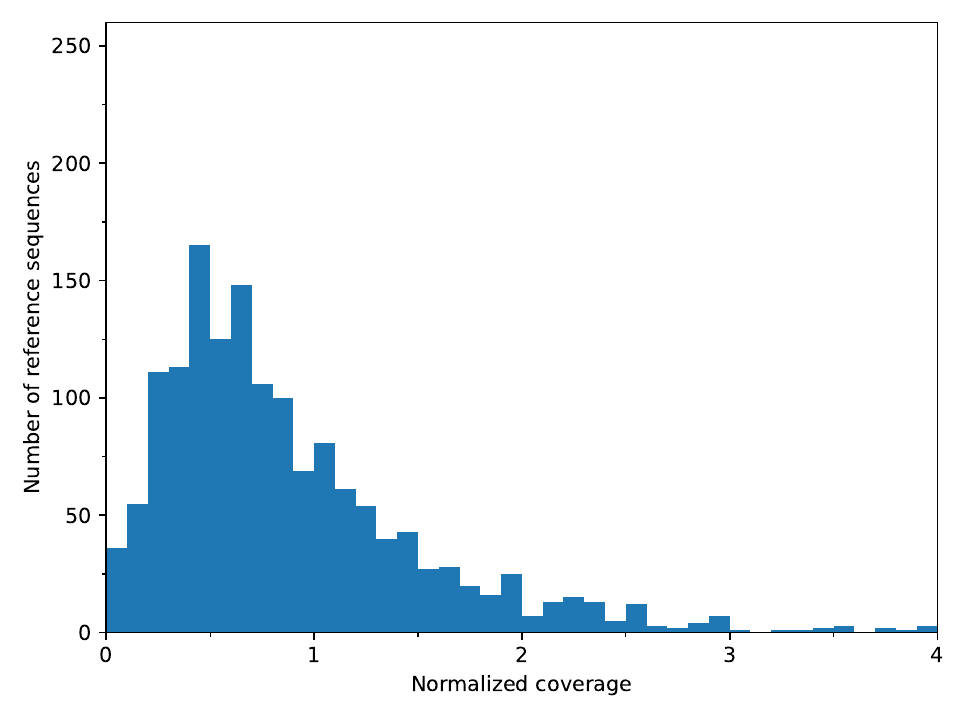}
\end{minipage}

\vspace{0.1em}
\hspace{4em}%
\begin{minipage}[c]{0.05\textwidth}
    \centering
    \rotatebox{90}{\small MGC+ (B, filtered)}
\end{minipage}
\begin{minipage}[c]{0.46\textwidth}
    \includegraphics[width=0.62\textwidth]{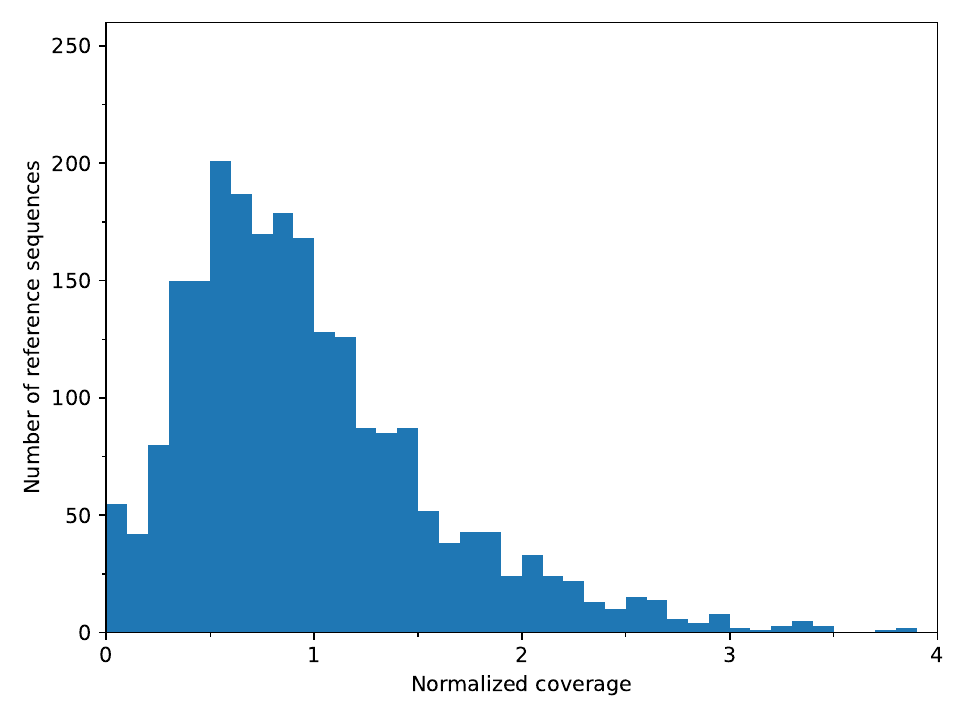}
\end{minipage}
\hspace{-4em}
\begin{minipage}[c]{0.46\textwidth}
    \includegraphics[width=0.62\textwidth]{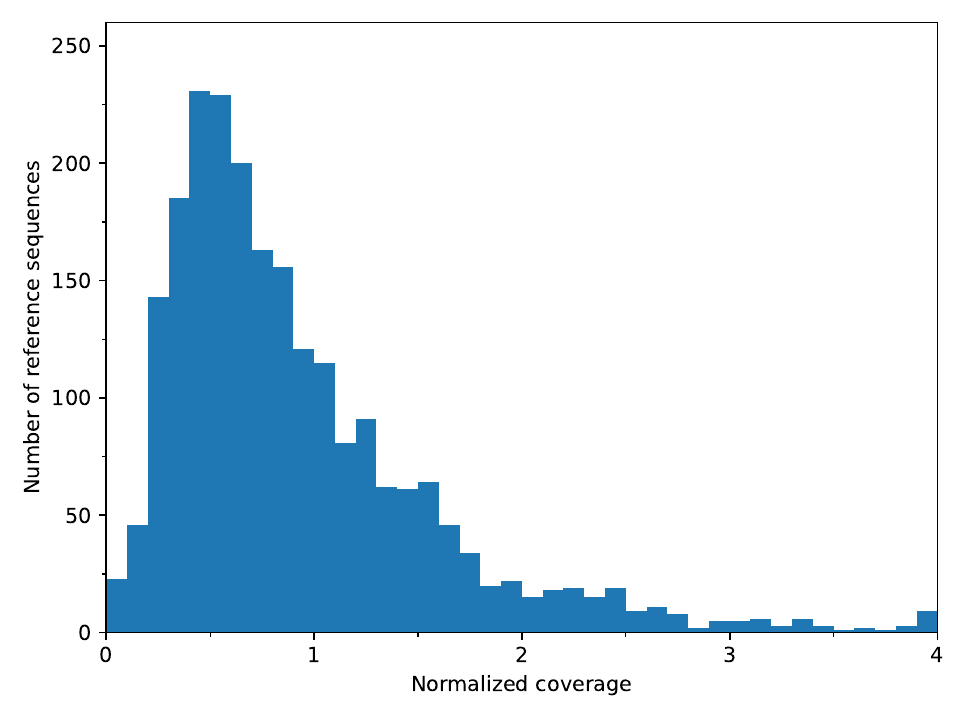}
\end{minipage}

\vspace{0.1em}
\hspace{4em}%
\begin{minipage}[c]{0.05\textwidth}
    \centering
    \rotatebox{90}{\small MGC+ (B, unfiltered)}
\end{minipage}
\begin{minipage}[c]{0.46\textwidth}
    \includegraphics[width=0.62\textwidth]{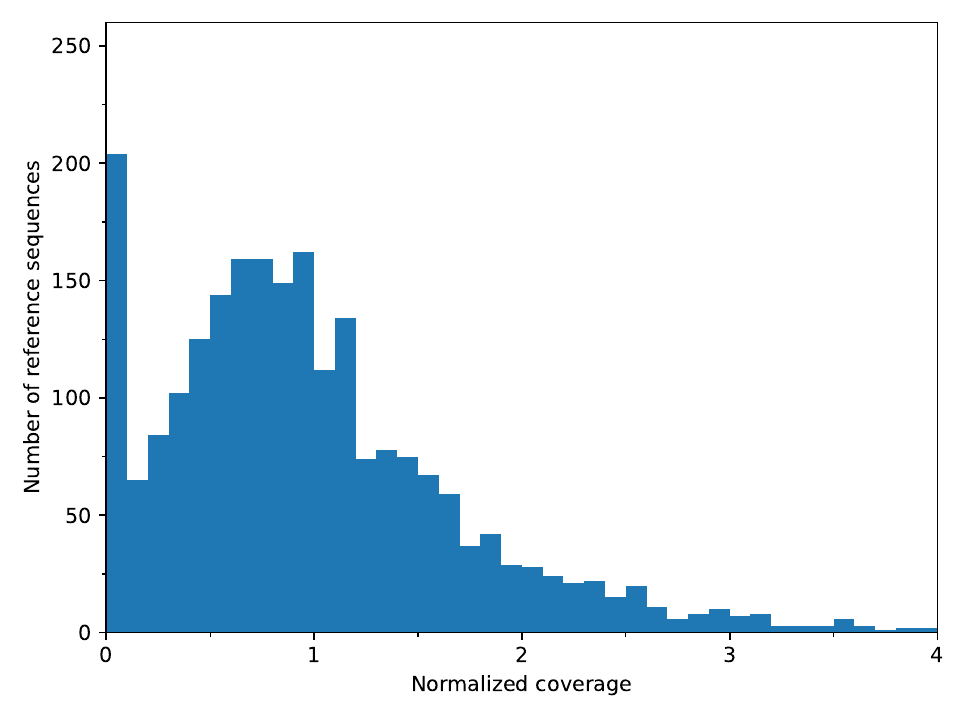}
\end{minipage}
\hspace{-4em}
\begin{minipage}[c]{0.46\textwidth}
    \includegraphics[width=0.62\textwidth]{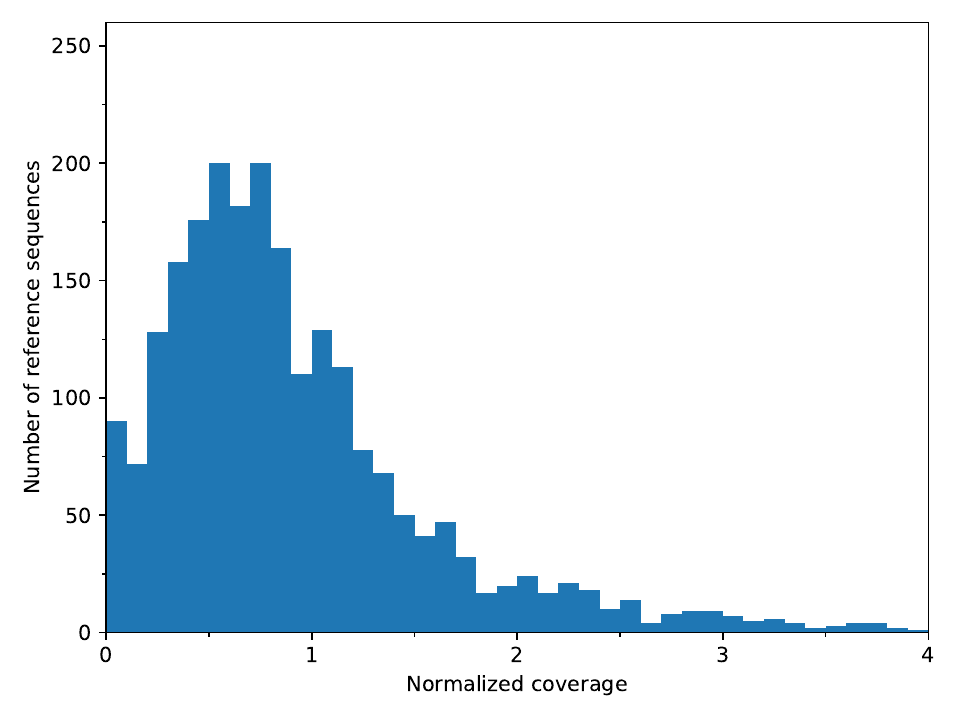}
\end{minipage}

\caption{Coverage distributions derived from the Illumina and Nanopore (dorado-sup) sequencing data in the \emph{in vitro} experiment, normalized by the mean coverage of each setting.
}

\end{figure*}

\end{document}